\documentclass[prb,twocolumn,10,aps,superscriptaddress,longbibliography]{revtex4-1}
\usepackage{mathptmx}

\usepackage[T1]{fontenc}
\usepackage[latin9]{inputenc}
\usepackage{float}
\usepackage{amsmath}
\usepackage{epsfig}
\usepackage{epstopdf}
\usepackage{amssymb}
\usepackage{array}
\usepackage{graphicx}
\usepackage{esint}
\usepackage[normalem]{ulem}
\makeatletter
\@ifundefined{textcolor}{}
{%
\definecolor{BLACK}{gray}{0}
\definecolor{WHITE}{gray}{1}
\definecolor{RED}{rgb}{1,0,0}
\definecolor{GREEN}{rgb}{0,1,0}
\definecolor{BLUE}{rgb}{0,0,1}
\definecolor{CYAN}{cmyk}{1,0,0,0}
\definecolor{MAGENTA}{cmyk}{0,1,0,0}
\definecolor{YELLOW}{cmyk}{0,0,1,0}
}
\pdfoutput=1 
\usepackage{multirow}\usepackage{dcolumn}\usepackage{bm}\usepackage{ulem}\usepackage{amsfonts}\usepackage{color}\usepackage{colordvi}\usepackage{dcolumn}\usepackage{subfigure}
\usepackage[colorlinks,bookmarks=false,citecolor=blue,linkcolor=blue,urlcolor=blue,hyperfootnotes=true]{hyperref}
\makeatother
\begin{document}
\title{Generation of pure superconducting spin current in magnetic heterostructures via non-locally induced magnetism due to Landau Fermi-liquid effects}
\date{\today}
\author{X. Montiel} 
\email{xavier.montiel@rhul.ac.uk}
\affiliation{Department of Physics, Royal Holloway, University of London, Egham, Surrey TW20 0EX, UK}
\author{M. Eschrig}
\email{matthias.eschrig@rhul.ac.uk}
\affiliation{Department of Physics, Royal Holloway, University of London, Egham, Surrey TW20 0EX, UK}
\begin{abstract}
We propose a mechanism for the generation of pure superconducting spin-current carried by equal-spin triplet Cooper pairs in a superconductor (S) sandwiched between a ferromagnet (F) and a normal metal (N$_{\rm so}$) with intrinsic spin-orbit coupling. We show that in the presence of Landau Fermi-liquid interactions the superconducting proximity effect can induce non-locally a ferromagnetic exchange field in the normal layer, which disappears above the superconducting transition temperature of the structure. 
The internal Landau Fermi-liquid exchange field leads to the onset of a spin supercurrent associated with the generation of long-range spin-triplet superconducting correlations in the trilayer. We demonstrate that the magnitude of the spin supercurrent as well as the induced magnetic order in the N$_{\rm so}$ layer depends critically on the superconducting proximity effect between the S layer and the F and N$_{\rm so}$ layers and the magnitude of the relevant Landau Fermi-liquid interaction parameter.
We investigate the effect of spin flip processes on this mechanism. Our results demonstrate the crucial role of Landau Fermi-liquid interaction in combination with spin-orbit coupling for the creation of spin supercurrent in superconducting spintronics, and
give a possible explanation of a recent experiment utilizing spin-pumping via ferromagnetic resonance [Jeon {\it et al.}, Nat. Mat. {\bf 17}, 499 (2018)].
\end{abstract}
\maketitle
\begin{section}{Introduction}
The generation of pure spin currents in superconducting devices via equal-spin Cooper pairs is one of the main challenges of superconducting spintronics.\cite{Eschrig_PhysTod2011,Eschrig_RepProgPhys2015,Robinson_Linder_2015} Contrary to singlet Cooper pairs, equal-spin Cooper pairs are triplet coherent states composed of pairs of electrons with equal spin projections on a given quantization axis and carry both charge and spin. The generation of spin-triplet correlations is a consequence of the interaction between a superconducting (S) material with a spin magnetized material, e.g. a ferromagnet (F).\cite{Izyumov2002,Eschrig_adv2004,Golubov_RevMod2004,Bergeret_RevMod2005,Buzdin_RevMod2005,Lyuksyutov2005,Eschrig_PhysTod2011,BlamireRobinson_JPhys2014,Eschrig_RepProgPhys2015,Robinson_Linder_2015,Birge_PhilTransA2018} In the vicinity of the S/F interface, the presence of an exchange field induces a spin mixing process \cite{Sauls_PRB1988,Fogelstrom00,BobkovaBobkov02,Eschrig_PRL2003,Eschrig_PhysTod2011} leading to short-range triplet correlations due to triplet Cooper pairs with zero spin projection.\cite{Golubov_RevMod2004,Buzdin_RevMod2005,Bergeret_RevMod2005,Eschrig_PhysTod2011,BlamireRobinson_JPhys2014,Eschrig_RepProgPhys2015} Equal-spin Cooper pairs are the $\pm1$ spin projection pairs (also called long-range triplet correlations) and can be produced in superconducting devices with magnetic inhomogeneities \cite{Bergeret_PRL2001,Bergeret_RevMod2005,Eschrig_PhysTod2011,Eschrig_RepProgPhys2015,Robinson_Linder_2015} of different nature: misaligned ferromagnetic magnetization,\cite{Champel_PRL2005,houzet_PRB2007,Halterman_PRB2008} magnetic domain walls \cite{Champel_PRB2005,Champel_PRB2005b,Fominov_PRB2007,Crouzy_PRB2007,Pugach_PRB2011,Kupferschmidt_PRB2009,Bergeret_PRL2001} or spin-polarized interfaces.\cite{Eschrig_PRL2003,Lofwander_PRL2010,Grein_PRL2009,Eschrig_PhysTod2011} Long-range triplets correlations also exist in S devices involving fully spin polarized materials like half-metals (HM).\cite{Eschrig_PRL2003,eschrig_NatPhys2008} or materials exhibiting spin-orbit coupling (SOC) \cite{Annunziata_PRB2012,Bergeret_PRL2013,Bergeret_PRB2014,Linder_PRB2015} Non-equilibrium spin-injection techniques in combination with direct measurement of transport properties were used to create and observe equal-spin Cooper pairs in mesoscopic devices.\cite{Hubler_PRB2010,Hubler_PRL2012,Hubler_PRB2013,quay_NatPhys2013,Wakamura_PRL2014,Beckmann_JPhysConMat2016} Investigated properties include spin and charge decoupling,\cite{quay_NatPhys2013,Wakamura_PRL2014} enhanced spin relaxation time,\cite{Poli_PRL2008,yang_NatMat2010,wakamura_NatMat2015} and a giant spin-orbit interaction\cite{Inoue_PRB2017} but a direct observation of pure spin currents carried by
equal-spin Cooper pairs has remained elusive.

A recent ferromagnetic resonance (FMR) experiment in N$_{\rm so}$/S/F/S/N$_{\rm so}$ devices, where N$_{\rm so}$ is a metallic layer exhibiting intrinsic spin-orbit coupling, could provide the first evidence of pure spin supercurrent carried by equal-spin Cooper pairs.\cite{Jeon_NatMat2018}
Precession of the ferromagnet's magnetization close to the ferromagnetic resonance induces a flow of a pure spin current from the F layer into the adjacent non-magnetic material.\cite{Tserkovnyak_RevModPhys2005,Ando_JapplPhys2011} In S/F and S/F/S devices, it has been observed \cite{Bell_PRL2008} and demonstrated \cite{Morten_EPL2008} that the amplitude of the injected spin current decreases below the superconducting critical temperature $T_{\rm c}$ because singlet Cooper pairs do not carry spin and thus lead to an effective spin-blocking. This result has been extended to the cases when the S layer is capped by a metallic spin-sink layer.\cite{Morten_EPL2008} Nevertheless, it has recently been demonstrated that an {\it increase} of the amplitude of the injected spin current takes place below $T_{\rm c}$ in Pt/Nb/Py/Nb/Pt systems where the S layer (here Nb) is capped by a metallic layer exhibiting strong spin-orbit coupling (here Pt).\cite{Jeon_NatMat2018} The increase of spin current below $T_{\rm c}$ occurs for small thicknesses of the S layer of the order of the superconducting coherence length, which emphasizes the crucial role of the S proximity effect. 
Thus it is natural to assume that this increase of the injected spin current below $T_{\rm c}$ is explained in terms of transport of spin by equal-spin Cooper pairs. For the effect to be appreciable, the triplet correlations that appear in the ferromagnetic Py region by proximity effect should be long-range, i.e. equal spin pairs with respect to the magnetization axis.

However, the underlying mechanism that explains the onset of the long-range correlations in this structure is not clear. 
The increase of the injected spin current below $T_{\rm c}$ is only observed for metallic layers formed by Pt, Ta, or W,\cite{Jeon_NatMat2018} which all exhibit strong intrinsic SOC \cite{Liu_arxiv2011,Bass_JPhysCondMatt2007,Czeschka_PRL2011,Tanaka_PRB2008} and are close to a ferromagnetic instability.\cite{Crangle_JApplPhys1965,Herrmannsdorfer_JLowTemp_1996,Konig_PRL_1999,Zhang_PRB2015} In this article, we address two questions: How can we explain the onset of equal spin Cooper pairs below $T_{\rm c}$ in such N$_{\rm so}$/S/F/S/N$_{\rm so}$ structures? What is the specific role of spin-orbit coupling and proximity to a ferromagnetic instability in the formation of such equal spin Cooper pairs?

Here, we propose a possible mechanism to explain the existence of equal-spin Cooper pairs in the experimental pentalayer. To this end, we first simplify the experimental pentalayer into an F/S/N$_{\rm so}$ trilayer (see Fig. \ref{FSNso}) where the F layer is the Py layer, the S layer is the Nb layer and the Pt layer is modelized by a metallic layer (the N$_{\rm so}$ layer) exhibiting intrinsic spin-orbit coupling and Landau Fermi-liquid corrections.\cite{Landau_JETP1956} The second simplification consists in studying the equilibrium properties of the F/S/N$_{\rm so}$ layer, leaving non-equilibrium calculation to later work. 
We also neglect the SOC in the F layer. The presence of SOC in the F layer would stabilize LR triplet correlations already in an F/S bilayer.\cite{Bergeret_PRB2014,Linder_PRB2015} However, in the FMR experiment \cite{Jeon_NatMat2018} an increase of the injected spin current below T$_c$ is associated with the use of N$_{\rm so}$ materials which exhibit strong spin-orbit coupling. In addition, the magnitude of the spin-orbit coupling in the ferromagnetic Py is much smaller than that in Nb, which in turn is much smaller than that in Pt.\cite{Morota_PRB83_174405_2011,Tsukahara_PRB89_235317_2014}
Finally, the effect is much larger in the experiment when using Pt as N$_{\rm so}$ layer than it is when using Fe$_{50}$Mn$_{50}$. As the spin diffusion length in Fe$_{50}$Mn$_{50}$ is comparable to that in Pt,\cite{ZhangPRL_113_196602_2014,YangPRB_93_094402_2016} however the intrinsic spin-orbit coupling is weaker in Fe$_{50}$Mn$_{50}$ than in Pt,\cite{ZhangPRL_113_196602_2014} this supports the idea that the effect is 
primarily a consequence of the intrinsic spin-orbit coupling in the N$_{\rm so}$ layer rather than a spin-sink effect of N$_{\rm so}$ in the presence of pre-formed LR triplet pairs.
For this reason we concentrate on a mechanism mediated entirely by the spin-orbit coupling of the N$_{\rm so}$ layer that generates and stabilizes long-range triplet correlations in such an F/S/N$_{\rm so}$ layer.
We demonstrate that the equal-spin Cooper pair channel exists in the F/S/N$_{\rm so}$ trilayer in the parameter range appropriate for the FMR experiment. 

Within our scenario, the short-range triplet correlations produced at the S/F interface decay in the S layer with the characteristic length comparable with the superconducting coherence length $\xi_{0}$ and reach the S/N$_{\rm so}$ interface. The intrinsic SOC inside the N$_{\rm so}$ layer induces a spin rotation process \cite{Robinson_Linder_2015,Bergeret_RevMod2005,Eschrig_PhysTod2011,Eschrig_RepProgPhys2015} which produces long-range triplet correlations mainly confined in the N$_{\rm so}$ layer.\cite{Bergeret_PRB2014} In addition to the SOC, the Landau FL mean fields in the N$_{\rm so}$ include an exchange field whose components are oriented along the short-range and long-range triplet correlations.\cite{Alexander_PRB1985,Sauls_PRB1988,Bratass_PRB2012}
As this induced exchange field is misaligned with the exchange field of the F layer, its onset implies the stabilization of a net spin supercurrent between in the S layer which is the signature of long-range spin-triplet correlations between the F and the N$_{\rm so}$ layer at equilibrium. The long-range triplet correlations exist only in presence of SOC in the N$_{\rm so}$ layer, but they require the Landau FL exchange field in order to spread across the entire structure; in addition the presence of the FL effects strongly amplifies these correlations.
We finally show that this new triplet channel resists to the onset of spin-flip processes. Our results provide an explanation for the stabilization of an equal-spin Cooper pair channel in the F/S/N$_{\rm so}$ layer at equilibrium which reproduces qualitatively the dependence of the injected spin current in the FMR experiment.

The paper is organized as follow: in section \ref{Theory}, we present the quasiclassical equations of superconductivity in the diffusive regime, utilizing the Usadel formalism. We also explain how we implement the spin-orbit scattering and Fermi liquid effects in the N$_{\rm so}$ layer. In section \ref{FSN}, we present the results in the F/S/N$_{\rm so}$ layer at equilibrium without spin-flip processes. Finally, we discuss the effect of spin-flip processes in section \ref{FSNspinflip}. In section \ref{Discussion}, we discuss the different parameter dependence on the F/S/N$_{\rm so}$ properties and we conclude in the section \ref{conclusion}.

\begin{figure}[!h]
\includegraphics[width=8.6cm]{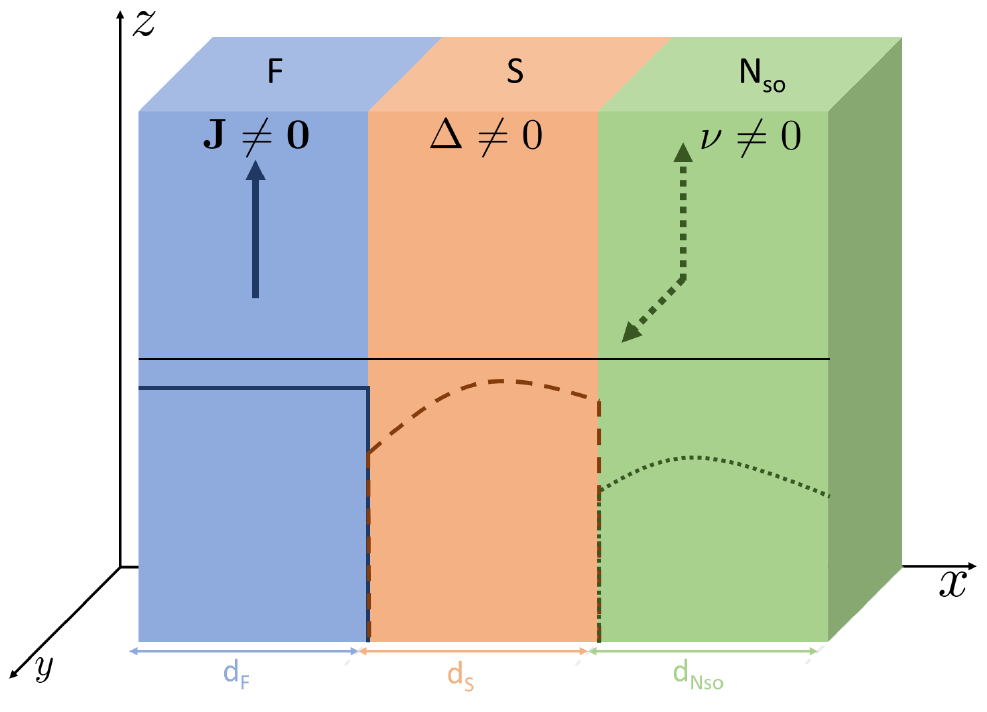}
\caption{\label{FSNso}(Color online) F/S/N$_{\rm so}$ trilayer, where $d_{S}$, $d_{F}$ and $d_{N_{\rm so}}$ refer to the S, F and N$_{\rm so}$ layer thicknesses, respectively. The trilayer is grown along the $x$-axis. The F layer exhibits a non-zero exchange field $\boldsymbol{J}$ along the $z$ axis,  
the S layer a non-zero spin-singlet superconducting order parameter $\Delta $, and the N$_{\rm so}$ layer a non-zero exchange field $\boldsymbol{\nu}$ induced non-locally from the S/F interface via Fermi liquid interactions in N$_{\rm so}$.
The magnitudes of the three order parameters are shown schematically as full, dashed, and dotted lines, respectively.}
\end{figure}

\end{section}

\begin{section}{Theoretical description}
\label{Theory}

We use the Usadel theory for diffusive superconductors \cite{Usadel_PRL1970,Belzig_SuperLattice1999} adapted for spin-polarized systems.\cite{Eschrig_RepProgPhys2015} The Usadel formalism can be deduced from the quasiclassical equations of Eilenberger \cite{Eilenberger_1968} and Larkin and Ovchinnikov \cite{larkin_JETP1969} in the diffusive limit. In the following, we focus on the equilibrium properties which are captured by the retarded Green functions. The retarded Green function $\hat{G}\left(E,\bold{R}\right)$ depends on the energy $E$ and on the spatial coordinates $\bold{R}={x,y,z}$. We define the notation $\left(\hat{...}\right)$ corresponding to quantities written in spin-dependent Nambu-Gor'kov space
(spin$\otimes$particle-hole space where $\otimes$ is the tensor product). 
In the spin$\otimes$particle-hole subspace, the Green functions are defined with respect to the Nambu spinor $\Psi=\left(\psi_{\uparrow},\psi_{\downarrow},\psi_{\uparrow}^{\dagger},\psi_{\downarrow}^{\dagger}\right)$.
The full (retarded) Green function $\hat{G}$ is then a 4$\times$4 matrix.
The internal structure of the Green function and the self-energies can be written as
\begin{equation}
\begin{array}{ccc}
\hat{G}=\left(\begin{array}{cc}
g & f\\
\widetilde{f} & \widetilde{g}
\end{array}\right) & , & \hat{\Sigma}=\left(\begin{array}{cc}
\Sigma & \Delta\\
\widetilde{\Delta} & \widetilde{\Sigma}
\end{array}\right).
\end{array}\label{eq:Gsym}
\end{equation}
We refer to the 2$\times 2$ spin subspace via the unit matrix ($\sigma_0$) and the three Pauli matrices, i.e. $\mathbf{\sigma}=\left(\sigma_{0},\sigma_{X},\sigma_{Y},\sigma_{Z}\right)$, while for the 2$\times 2$ particle-hole subspace we use the matrices $\mathbf{\tau}=\left(\tau_{0},\tau_{1},\tau_{2},\tau_{3}\right)$. Moreover, $\left(\widetilde{...}\right)$ combines complex conjugation with the transformation
$E\rightarrow-E$. $\hat{\Sigma}$ refers to the self-energies written in the Nambu-spin subspace. The self-energies $\Sigma$ and $\Delta$ are 2$\times$2 matrices in spin subspace and respectively refer to the normal and anomalous part of the self-energy.

In spin space, the Green functions can be decomposed into spin-scalar and spin-vector components using the Pauli spin matrices as basis.
The Green function then can be written in the form 
\begin{align}
\hat{G}=\left(\begin{array}{cc}
g_{s}\sigma_{0}+\boldsymbol{g}_{t}\boldsymbol{\sigma} & \left(f_{s}\sigma_{0}+\boldsymbol{f}_{t}\boldsymbol{\sigma}\right)i\sigma_{Y}\\
i\sigma_{Y}\left(\widetilde{f}_{s}\sigma_{0}-\widetilde{\boldsymbol{f}}_{t}\boldsymbol{\sigma}\right) & \sigma_{Y}\left(\widetilde{g}_{s}\sigma_{0}-\widetilde{\boldsymbol{g}}_{t}\boldsymbol{\sigma}\right)\sigma_{Y}
\end{array}\right)
\label{Green_spin}
\end{align}
where $f_{s}$ and $\boldsymbol{f}_{t}=\left(f_{t}^{X},f_{t}^{Y},f_{t}^{Z}\right)$ are singlet and triplet pair amplitudes and $g_{s}$ and $\boldsymbol{g}_{t}=\left(g_{t}^{X},g_{t}^{Y},g_{t}^{Z}\right)$ refer to spin-scalar and spin-vector part of the diagonal Green function. Here and in the following, indices $X$, $Y$, $Z$ refer to the $X$, $Y$ and $Z$ axis in spin space.

The Usadel equation for the Green function $\hat{G}\equiv \hat{G}\left(E,\bold{R}\right)$ takes the form
\begin{align}
\left[E\hat{\tau}_{3}-\hat{\Sigma},\hat{G}\right]+\frac{D}{\pi}\nabla\left(\hat{G}\nabla\hat{G}\right)=0\label{eq:UsGen}
\end{align}
where $D$ is the diffusion coefficient, $\nabla \equiv \frac{\partial}{\partial \bold{R}}$, and $\hat{\tau}_{3}$ the third Pauli matrix in the particle-hole subspace. 
In addition, the quasiclassical Green function $\hat{G}$ is required to fulfill the normalization condition $\hat{G}.\hat{G}=-\pi^{2}\hat{1}$.
Eq. (\ref{eq:UsGen}) has to be supplemented by boundary conditions.\cite{Nazarov_PRL1994,Nazarov_SuperLattMicro1999,Eschrig_NJPhys2015} For inner interfaces, we use boundary conditions appropriate for perfect transmission interfaces which describes interfaces between metals and superconductors with nearly similar electronic properties,\cite{Eschrig_PRB2005} and which take the form
\begin{align}
\begin{array}{cc}
\hat{G}^{l}=\hat{G}^{r},\qquad \displaystyle
\sigma_{l}\frac{\partial \hat{G}^{l}}{\partial x}=\sigma_{r}\frac{\partial \hat{G}^{r}}{\partial x}
\label{eq_bound}
\end{array}
\end{align}
where $l(r)$ relates to the left (right) side of the interface and $\sigma$ is the bulk conductivity.\cite{Nazarov_PRL1994,Nazarov_SuperLattMicro1999,Eschrig_NJPhys2015,Eschrig_PRB2005} For the outer interfaces, we require the charge current to vanish, which yields the boundary condition
\begin{align}
\frac{\partial \hat{G}}{\partial x}=0.
\end{align}
In the following, we study the F/S/N$_{\rm so}$ trilayer structure presented in Fig. \ref{FSNso} where we consider the $x$ axis to be the axis normal to the layers. We assume that the layers extend infinitely in the $y-z$ plane and we reduce the F/S/N trilayer problem to a one-dimensional problem where the Green functions only depend on the $x$ coordinate, $\hat{G}\left(E,\bold{R}\right)=\hat{G}\left(E,x\right)$.

\begin{subsection}{Spin orbit coupling}
Intrinsic SOC can arise due to bulk non-centrosymmetric point group symmetry\cite{Samokhin_AnnPhys2009} or due to broken inversion symmetry at interfaces in multilayer devices.\cite{Edelstein_PRB2003} Spin-orbit coupling in combination with ferromagnetic exchange splitting generates long-range spin-triplet pair correlations in superconducting devices.\cite{Bergeret_PRL2013,Bergeret_PRB2014,Linder_PRB2015} In the present case of an F/S/N$_{\rm so}$ trilayer, we include intrinsic spin-orbit effects in the N$_{\rm so}$ layer, assuming that spin-orbit processes in F and S layers are negligible.
Spin-orbit coupling allows for rotation between different spin states via two different mechanisms. The first one comes from the anisotropy of the spin-relaxation times describes by the Dyakonov-Perel tensor; the second mechanism is induced by spin precession in an inhomogeneous spin density. In superconducting devices, spin-orbit coupling allows for rotation between different spin-triplet pairing states.\cite{Bergeret_PRL2013,Bergeret_PRB2014,Linder_PRB2015}
In general, spin-orbit interaction for quasiparticles in crystalline metallic materials leads to a term 
\begin{equation}
V_{\rm SO}=-\sum_{k,\nu} v^k_F A_k^\nu \sigma_\nu
\end{equation}
with $k\in \left\{x,y,z\right\}$, 
where $\sigma_\nu $ with $\nu \in \left\{X,Y,Z\right\}$ are Pauli spin matrices, $v_F^k$ are the $k$-components of the Fermi velocity, and $A_k^\nu $ is a spin-orbit coupling tensor, which to lowest order in the momentum can be assumed to be momentum independent. It is convenient to introduce the spin-orbit field vectors $\underline{A}$, the components 
$A_k=\boldsymbol{A}_k\boldsymbol{\sigma}$ of which are 2$\times $2 spin matrices. In terms of those, the spin-orbit interaction is of the form $V_{\rm SO}=-\sum_k v_F^k A_k$.
We include spin-orbit coupling in the Usadel equations (\ref{eq:UsGen}) by substituting the standard spatial derivative by the covariant derivative \cite{Gorini_PRB2010,Bergeret_PRL2013,Bergeret_PRB2014,Linder_PRB2015}
\begin{align}
\boldsymbol{\mathfrak{D}} \rightarrow \nabla-i\left[\hat{\underline{A}},\right]
\end{align}
where $\hat{\underline{A}}$ is the spin-orbit field vector. This substitution is valid for any spin-orbit coupling linear in momentum. The SOC field $\hat{\underline{A}}$ has a vector structure in the real space and a 4$\times$4 matrix structure in spin Nambu space. The structure of $\hat{\underline{A}}$ is
\begin{align}
\begin{array}{cc}
\hat{\underline{A}}=\left(\begin{array}{cc}
\underline{A} & 0\\
0 & -\underline{A}^{*}
\end{array}\right), & \underline{A}=\left(\begin{array}{c}
A_{x}\\
A_{y}\\
A_{z}
\end{array}\right)\end{array}.
\end{align}
The three components $A_x$, $A_y$, $A_z$ can be decomposed into spin-Pauli matrices
as $A_k=\boldsymbol{A}_k\boldsymbol{\sigma}$ with real-valued vectors $\boldsymbol{A}_k$, $k\in \left\{x,y,z\right\}$.

The most common spin-orbit field types are the Rashba spin-orbit coupling \cite{Rashba_JPhys1984} (from interfacial symmetry breaking) and the Dresselhaus spin-orbit coupling \cite{Dresselhaus_PhysRev1955} (due to lack of bulk inversion symmetry). 
In order to concentrate on the most salient features of our model, 
in the following, we concentrate on a spin orbit coupling field of the form
\begin{align}
A_{x}=\left(\begin{array}{cc}
0 & \alpha\\
\alpha & 0
\end{array}\right)
\label{SOCsym}
\end{align}
assuming all the other components of the SOC field to vanish. 
This corresponds to $\boldsymbol{A}_x=(\alpha,0,0)$ with $\boldsymbol{A}_y$ and $\boldsymbol{A}_z$ vanishing.
The spin-orbit interaction in solids can be expanded around zero momentum, and for various point group symmetries a non-zero term linear in momentum appears. A full classification \cite{Samokhin_AnnPhys2009} shows that a spin-orbit coupling of the form \eqref{SOCsym} can be present in non-centrosymmetric materials with point group symmetry {\bf C}$_1(1)$, {\bf C}$_2(2)$, {\bf D}$_2(222)$, {\bf C}$_4(4)$, {\bf D}$_4(422)$, {\bf C}$_3(3)$, {\bf D}$_3(32)$, {\bf C}$_6(6)$, and {\bf D}$_6(622)$.

We do not present results here for spin-orbit coupling types with non-zero components $\boldsymbol{A}_y$ and $\boldsymbol{A}_z$, which lead to non-zero spin currents parallel to the interface.\cite{VorontsovPRL_101_127003_2008,WennerdalPRB_95_024513_2017}
The effects we discuss are, however, expected to also be present for such cases, which include other point group symmetries with components in $x$-direction, as for example {\rm O}$(432)$ (cubic), {\bf T}$(23)$ (tetrahedral), or {\bf C}$_{4v}(4mm)$ (tetragonal) symmetry. 
We are interested in spin currents perpendicular to the interface (in our geometry the $x$-direction is along the surface normal). 
As we do not wish to complicate matters by adding a spin-current component parallel to the interface, we present here results assuming the most simple form of spin-orbit coupling, Eq.~\eqref{SOCsym},  that gives a nonzero spin current in interface-normal direction.

\end{subsection}
\begin{subsection}{Riccati parameterization}
In order to solve the Usadel equations (\ref{eq:UsGen}) for the retarded Green functions respecting the normalization condition of the Green function, we use the Riccati matrices parameterization of the Green functions.\cite{Eschrig_PRB2009,Eschrig_PRB2000,Schopol_PRB1995,Nagato_JlowTempPhys1993,Nagai_JPhyssocJpn1995} The retarded Green function is parameterized in the following way:
\begin{align}
\hat{G}=- i\pi.\hat{N}.\left(\begin{array}{cc}
1+\gamma\widetilde{\gamma} & 2\gamma\\
-2\widetilde{\gamma} & -\left(1+\widetilde{\gamma}\gamma\right)
\end{array}\right)
\label{Grgam}
\end{align}
where $\gamma$ and $\tilde{\gamma}$ are matrices in the 2$\times$2 spin space and $\hat{N}$ is defined as 
\begin{align}
\hat{N}=\left(\begin{array}{cc}
\left(1-\gamma\widetilde{\gamma}\right)^{-1} & 0\\
0 & \left(1-\widetilde{\gamma}\gamma\right)^{-1}
\end{array}\right)
=\left(\begin{array}{cc}
N & 0\\
0 & \widetilde{N}
\end{array}\right).
\end{align}
The Usadel equations for the $\gamma$ matrices reads \cite{Eschrig_adv2004,Eschrig_PRB2005}
\begin{align}
&\left(\overline{\nabla}^{2}\gamma\right)+\left(\overline{\nabla}\gamma\right)\frac{\widetilde{f}}{i\pi}\left(\overline{\nabla}\gamma\right)=\nonumber\\
&\qquad =\frac{i}{D}\left[\gamma\widetilde{\Delta}\gamma-\gamma(E+\widetilde{\Sigma})-\left(E-\Sigma\right)\gamma-\Delta\right].
\label{equsgam}
\end{align}
where $\overline{\nabla}=(\overline{\partial}_{x},\overline{\partial}_{y},\overline{\partial}_{z})$ with $\overline{\partial}_{k} \gamma\equiv \partial_{k} \gamma
-i\left(A_{k}\gamma+\gamma A_{k}^{*}\right)$
is the covariant derivative and $\partial_{k}\equiv\frac{\partial}{\partial k}$ for $k\in \left\{x,y,z\right\}$. 
Note that the spatial derivatives along the $y$ and $z$ axis, $\partial_{y}$ and $\partial_{z}$, vanish because the F/S/N$_{\rm so}$ trilayer reduces to a one-dimensional problem for the case under consideration.
The equations for $\tilde{\gamma}$ can be deduced by applying the $\left(\widetilde{...}\right)$ transformation to the equations (\ref{equsgam}) .
\end{subsection}

When decomposing the Riccati amplitudes and spin-orbit fields as $\gamma=(\gamma_s + \boldsymbol{\gamma}_t \boldsymbol{\sigma} ) i\sigma_Y $, and $A_k=\boldsymbol{A}_k\boldsymbol{\sigma}$,
the covariant derivative applies to the components as 
\begin{equation}
\overline{\partial}_k \gamma = \left[\partial_k \gamma_s + (\delta_k\circ \boldsymbol{\gamma}_t) \boldsymbol{\sigma} \right] i\sigma_Y, 
\end{equation}
where we introduce the notation
$\delta_k\circ \boldsymbol{a}= \partial_k \boldsymbol{a} + 2 \boldsymbol{A}_k\times \boldsymbol{a} $ for any vector $\boldsymbol{a}$ and $k\in \left\{x,y,z\right\}$.
 In particular, the second covariant derivative is given by
\begin{align}
 \overline{\partial}_k^2\gamma=\left[\partial_k^2\gamma_s+ (\delta_k\circ (\delta_k\circ \boldsymbol{\gamma}_t))\boldsymbol{\sigma}\right]i\sigma_Y.
\label{second_deriv}
\end{align}

At the F/S interface, the boundary conditions (\ref{eq_bound}), are in the Riccati parameterization :
\begin{align}
\begin{array}{cc}
\left[\gamma\right]^{F}=\left[\gamma\right]^{S}\\
\sigma_{F}\left[\partial_{x} \gamma\right]^{F}=\sigma_{S}\left[\partial_{x}\gamma\right]^{S}
\label{eqboundgam}
\end{array}
\end{align}
where $\left[\gamma\right]^{F(S)}$ in the Riccati matrix in the F (S) side of the interface. At the S/N$_{\rm so}$ interface, the boundary conditions (\ref{eq_bound}) in the Riccati parameterization read
\begin{align}
\begin{array}{cc}
\left[\gamma\right]^{S}=\left[\gamma\right]^{N_{\rm so}}\\
\sigma_{S}\left[\partial_{x} \gamma\right]^{S}=\sigma_{N_{\rm so}} [ \overline{\partial_{x}}\gamma ]^{N_{\rm so}}
\label{eqboundgamsnso}
\end{array}
\end{align}
where $\left[\gamma\right]^{S(N_{\rm so})}$ is the Riccati matrix at the S (N$_{\rm so}$) side of the interface. The outer boundary conditions are $\left[\partial_{x} \gamma\right]^{F}=0$ on the F side and $[\overline{\partial_{x}}\gamma ]^{N_{\rm so}}=0$ on the N$_{\rm so}$ side of the trilayer.

The Usadel equation (\ref{equsgam}) and the boundary conditions (\ref{eqboundgam}) and (\ref{eqboundgamsnso}) for Riccati matrices $\gamma$ and $\tilde{\gamma}$ imply to solve a system of 8 non-linear differential equations (corresponding to the 4 spin components of each Riccati matrix $\gamma$ and $\tilde{\gamma}$). The equations (\ref{equsgam}), (\ref{eqboundgam}), and (\ref{eqboundgamsnso}) are solved numerically by using a relaxation method \cite{Numerical_Recipe} and by taking into account the self-consistency equation for the superconducting order parameter [see Eq. (\ref{Delta_self}) in section \ref{SCMF}] and the Fermi liquid order parameter [see Eq. (\ref{nu_self}) in section \ref{MF_SM}]. 
This then enables us to
calculate density of states (see section \ref{DOS}), spin-magnetization (see sections \ref{MF_SM} and \ref{MF_SMb}), pair amplitude (see section \ref{PA}) and charge and spin currents (see section \ref{Curr}) of the trilayer.

\begin{subsection}{Self-energies}
In this section, we introduce the self-energy appearing in the Usadel equations (\ref{eq:UsGen}) and (\ref{equsgam}). The total self-energy $\hat{\Sigma}$ has the form
\[
\hat{\Sigma}=\hat{\Sigma}^{\text{imp}}+\hat{\Sigma}^{\text{ex}}+\hat{\Delta}
\]
where $\hat{\Sigma}^{\text{ex}}$ describes the exchange field of a ferromagnetic layer (or alternatively, the spin-splitting produced by an external magnetic field), $\hat{\Delta}$ stands for the superconducting order parameter and $\hat{\Sigma}^{\text{imp}}$ is the self-energy produced by spin-flips in the presence of magnetic impurities
and spin-orbit scattering (see section \ref{SE_spinflip}). 

\begin{subsubsection}{Exchange field in F layer}
In order to describe the majority of minority electrons in a ferromagnet we use the exchange field $\boldsymbol{J}$, which leads to the self-energy 
\[
\hat{\Sigma}^{\text{ex}}=\left(\begin{array}{cc}
\boldsymbol{J}\boldsymbol{\sigma} & 0\\
0 & \boldsymbol{J}\boldsymbol{\sigma}^{*}
\end{array}\right)
\]
In our system (see Fig. \ref{FSNso}), we assume the exchange field in the F layer $\boldsymbol{J}$ to be constant and directed along the $z$-axis $\boldsymbol{J}=J\boldsymbol{z}$. We assume this exchange to vanish in non-ferromagnetic layers.
\end{subsubsection}
\begin{subsubsection}{Induced exchange field and the spin polarization}
\label{MF_SM}
In the N$_{\rm so}$ layer, the inclusion of the electron-electron interaction gives rise to a renormalization described by Landau Fermi liquid theory.\cite{Landau_JETP1956} The electrons and holes in a free electron gas picture are replace by electron-like and hole-like quasiparticles. The quasiparticles properties are related to the bare electronic properties through effective parameters called Landau parameters.\cite{Landau_JETP1956} The inclusion of such corrections in quasi-classical theory of superconductivity gives rise of a self-consistent exchange field produced by the onset of triplet correlations.\cite{Alexander_PRB1985,Sauls_PRB1988,Bratass_PRB2012} In the simplest case this exchange field $\boldsymbol{\nu}$ is collinear to the spin-polarization $\delta\bold{m}$.\cite{Bratass_PRB2012} In this case, the induced exchange field is given by
\begin{align}
\boldsymbol{\nu}(x)=\frac{G\delta\boldsymbol{m}(x)}{2N_{F}\mu_{B}}
\label{nu_self}
\end{align}
where $G$ is the Landau parameter, $N_{F}$ is the density of states at the Fermi level and $\mu_{B}$ the Bohr-magneton.
The spin magnetization (SM) $\delta\boldsymbol{m}$ is calculated as \cite{Sauls_PRB1988,Alexander_PRB1985,Bratass_PRB2012,Eschrig_RepProgPhys2015}
\begin{align}
\delta\boldsymbol{m}(x)=\frac{2N_{F}\mu_{B}}{1+G}\int_{-\infty}^{+\infty}\frac{dE}{2\pi}\text{Im}\left[\boldsymbol{g}_{t}\left(E,x\right)\right]\tanh\left(\frac{E}{2T}\right)
\label{magn}
\end{align}
where $\boldsymbol{g}_{t}$ is the spin-vector component written in the ${x,y,z}$ basis of the retarded Green function, see Eq.~(\ref{Green_spin}), and $T$ is the temperature.
In Usadel formalism, the Fermi liquid self-energy $\boldsymbol{\hat{\nu}}$ has the same structure as an exchange field and is given by
\[
\boldsymbol{\hat{\nu}}=\left(\begin{array}{cc}
\boldsymbol{\nu}.\boldsymbol{\sigma} & 0\\
0 & \boldsymbol{\nu}.\boldsymbol{\sigma}^\ast 
\end{array}\right)
\]
Note that the induced exchange field $\boldsymbol{\nu}$ is determined self-consistently by solving Eq.~(\ref{nu_self}) simultaneously with the Usadel equations (\ref{equsgam}). In Eq.~(\ref{nu_self}), is can be seen that the sign of the exchange field is directly related to the sign of the spin magnetization and the sign of the Landau parameter $G$. 

For a system close to a ferromagnetic instability (like Pt, W and Ta), the Landau parameter G is negative.\cite{Bratass_PRB2012} For negative G, the SM and exchange field diverges when $G\rightarrow-1$ and this divergence is known as the paramagnet instability.\cite{Bratass_PRB2012} Note that the inclusion of the exchange interaction can lead to a negative Landau parameter.\cite{Vedyaev_Phystatsolidib1975,Leiro_solidstateCommun1995,Bratass_PRB2012} The value of the Landau parameter $G$ has been calculated for light metal compounds such as Al or K \cite{Leiro_solidstateCommun1995} but, to the best of our knowledge, has not been calculated in transition metals. In materials considered as "nearly ferromagnetic" such as Pt,\cite{Crangle_JApplPhys1965,Zhang_PRB2015} we can consider $G$ to be reasonably close to $-1$. In this case, it becomes possible to induce magnetism in the N$_{\rm so}$ layer, resulting from an adjacent ferromagnet \cite{Zhang_PRB2015} or by proximity effect with a superconductor via short or long-range triplet correlations (as presented in the present paper). In the following, we assume that the Landau parameter is different from zero in the N$_{\rm so}$ layer only and vanishes in the F and S layer.
\end{subsubsection}

\begin{subsubsection}{Superconducting order parameter}
\label{SCMF}
We assume that the superconducting mean-field order parameter has only a spin-singlet  component. For this case, in the superconducting layer the SC order parameter has the form 
\[
\hat{\Delta}^{\rm R}=\left(\begin{array}{cc}
0 & \Delta^{SC}\\
\widetilde{\Delta}^{SC} & 0
\end{array}\right)
\]
where $\Delta^{SC}=i\sigma_{Y}\Delta e^{i\phi}$ with $\phi$ the superconducting phase.
The SC order parameter is fixed by the self-consistency equation 
\[
\Delta^{SC}(x)=\lambda \int_{-E_c}^{+E_c}\frac{dE}{2i\pi}f_{s}\left(E,x\right)\tanh\left(\frac{E}{2T}\right)
\]
where $f_{s}$ is the singlet part of the anomalous Green function [see (\ref{Green_spin})], $E_c$ is the technical BCS cutoff, and $\lambda$ the pairing interaction which we assume to be non-zero in the S layer only and vanishing in the F and N$_{\rm so}$ layer. We eliminate both the BCS cutoff and the pairing interaction $\lambda$ in favor of the critical temperature $T_c$ such that the self-consistency equation can be written as 
\begin{align}
\Delta^{SC}\left(x\right)=\lim_{E_{c}\rightarrow \infty}\frac{\int_{-E_{c}}^{E_{c}}\frac{dE}{2i\pi}f_{s}\left(E,x\right)\tanh\left(\frac{E}{2T}\right)}{\int_{-E_{c}}^{E_{c}}\frac{dE}{2E}\tanh\left(\frac{E}{2T}\right)+\ln\left(\frac{T}{T_{c}}\right)} .
\label{Delta_self}
\end{align}

\end{subsubsection}
\end{subsection}

\end{section}

\begin{section}{Spin-triplet correlations in F/S/N structure}
\label{FSN}
In this section, we study the onset of spin-triplet correlation in the F/S/N$_{\rm so}$ structure. 
Motivated by recent experiments we use parameters appropriate for a structure where
the F layer is permalloy Py, the S layer is Niobium Nb and the N layer is Platinum Pt. 
In the following, all the length of the layers are rescaled by the Nb coherence length $\xi_{0}=\sqrt{\frac{D}{\Delta_{0}}}=13\text{nm}$ \cite{Bell_PRL2008,Gu_PRB2002} and the energy are rescaled to $\Delta_{0}$, the bulk SC gap at zero temperature (in Nb, $\Delta_{0}=1.4\text{meV}$). In the F layer, we consider an exchange field along the z-axis whose amplitude is $J=10\Delta_{0}$.

We present results for three sets of parameters A), B) and C) which are presented in the Table \ref{Table_1}. The parameter set A) is appropriate for a trilayer in absence of spin orbit coupling in the N$_{\rm so}$ layer, which allows us to better understand the physics provided by the Fermi-liquid interactions. The parameter set B) is appropriate for the case where Fermi-liquid effects and spin-orbit scattering in the N$_{\rm so}$ layer are included and where the conductivities of the F, S and N$_{\rm so}$ layer are the same. The parameter set C) is like B), however for realistic conductivities for Py, Nb, and Pt. The comparison between the data set B) and C) will provide a better understanding of the effect of the boundary conditions. We have chosen as thicknesses of the F, S and N$_{\rm so}$ layers the ones provided in the FMR experiment.\cite{Jeon_NatMat2018}

\begin{table}
\caption{\label{Table_1} If not specified in the text, the parameters used for the calculation in the configurations A, B and C are summed up in the Table~\ref{Table_1}. Here, $d_{S}$, $d_{F}$ and $d_{N_{\rm so}}$ refer to the S, F and N$_{\rm so}$ layer thicknesses, respectively, and $\alpha$ refers to the strength of the spin orbit interaction while $T$ is the temperature, and $\Delta_{0}$ is the bulk superconducting gap. The parameters $\sigma_{F}$, $\sigma_{S}$ and $\sigma_{N_{\rm so}}$ are the bulk conductivities in the normal state in the S, F and N$_{\rm so}$ layer, respectively. In configuration C, the F layer conductivity is the one of permalloy (Py), $\sigma_{F}=\sigma_{Py}=1.72 \times10^{6} $ Sm, the S layer conductivity is the one of niobium, $\sigma_{S}=\sigma_{Nb}=6.9 \times 10^{6} $ Sm, and the N$_{\rm so}$ layer conductivity is the one of platinum, $\sigma_{N_{\rm so}}=\sigma_{Pt}=9.7\times 10^{6}$ Sm.}
\begin{tabular}{|c|c|c|c|c|c|c|c|c|}
\hline
&$d_{S}$&$d_{F}$&$d_{N_{\rm so}}$&$\alpha\xi_{0}$&$T$&$\sigma_{F}$&$\sigma_{S}$&$\sigma_{N_{\rm so}}$\\
\hline
A&$2.307\xi_{0}$&$0.231\xi_{0}$&$0.385\xi_{0}$&$0$&$0.01\Delta_{0}$&$1$&$1$&$1$\\
\hline
B&$2.307\xi_{0}$&$0.231\xi_{0}$&$0.385\xi_{0}$&$2$&$0.01\Delta_{0}$&$1$&$1$&$1$\\
\hline
C&$2.307\xi_{0}$&$0.231\xi_{0}$&$0.385\xi_{0}$&$2$&$0.01\Delta_{0}$&$\sigma_{Py}$&$\sigma_{Nb}$&$\sigma_{Pt}$\\
\hline
\end{tabular}
\end{table}

\begin{subsection}{Results}

\begin{subsubsection}{Density of states}
\label{DOS}

\begin{figure*}
\centering
\begin{minipage}{5.2cm}
a)\includegraphics[width=55mm]{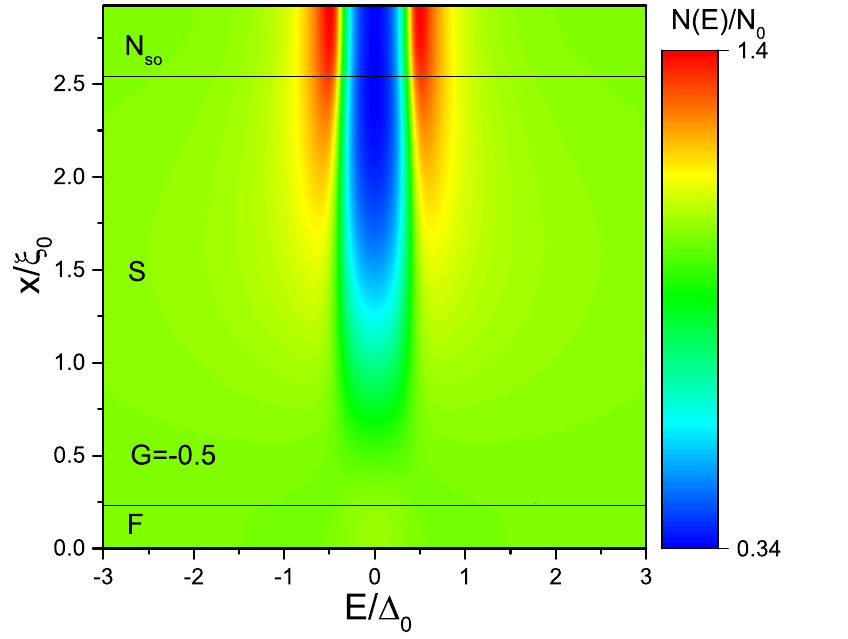}
\end{minipage}
\begin{minipage}{5.2cm}
b)\includegraphics[width=55mm]{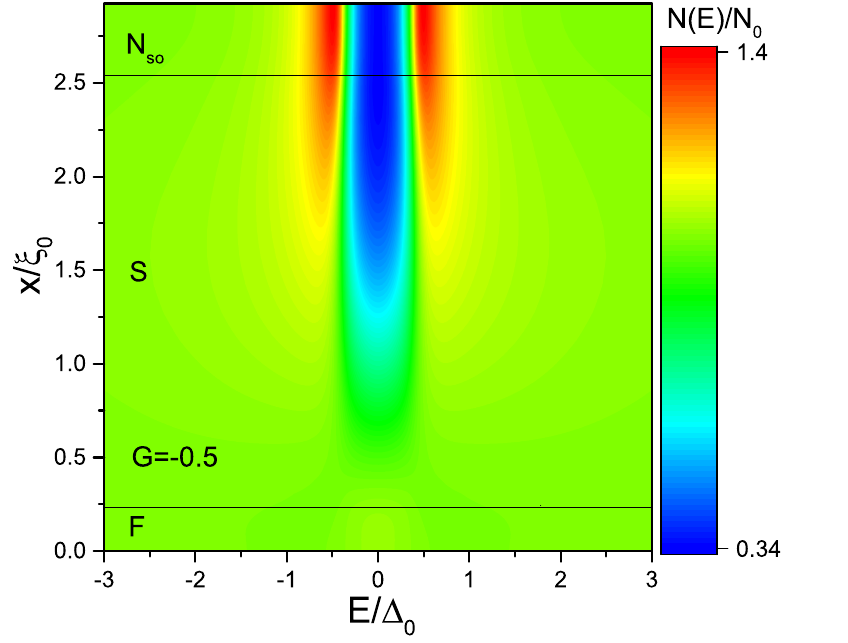}
\end{minipage}
\begin{minipage}{5.2cm}
c)\includegraphics[width=55mm]{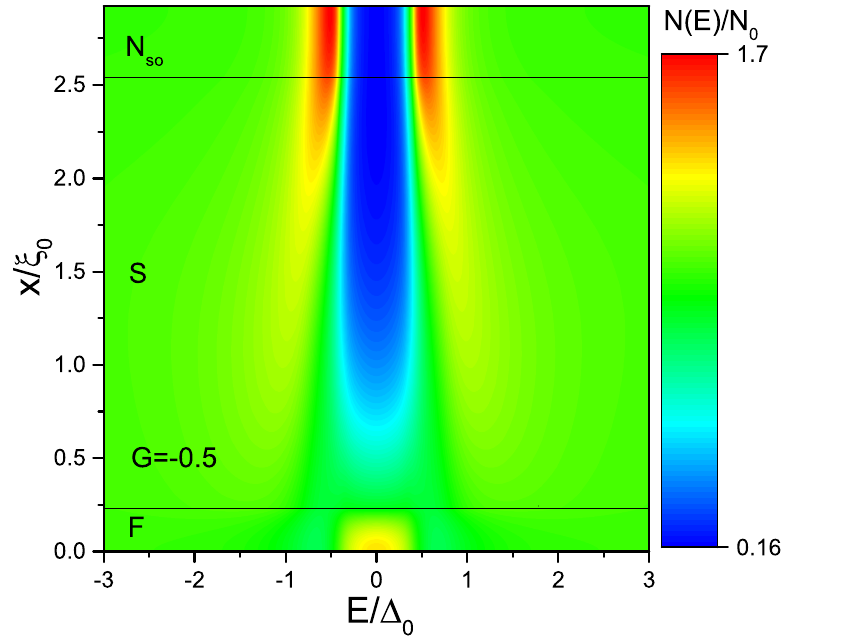}
\end{minipage}
\begin{minipage}{5.2cm}
d)\includegraphics[width=55mm]{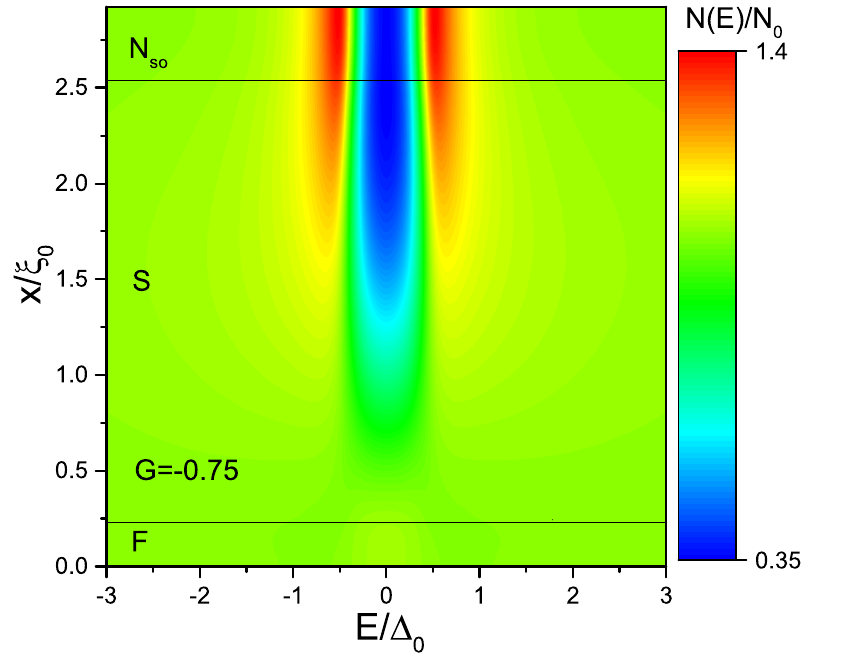}
\end{minipage}
\begin{minipage}{5.2cm}
e)\includegraphics[width=55mm]{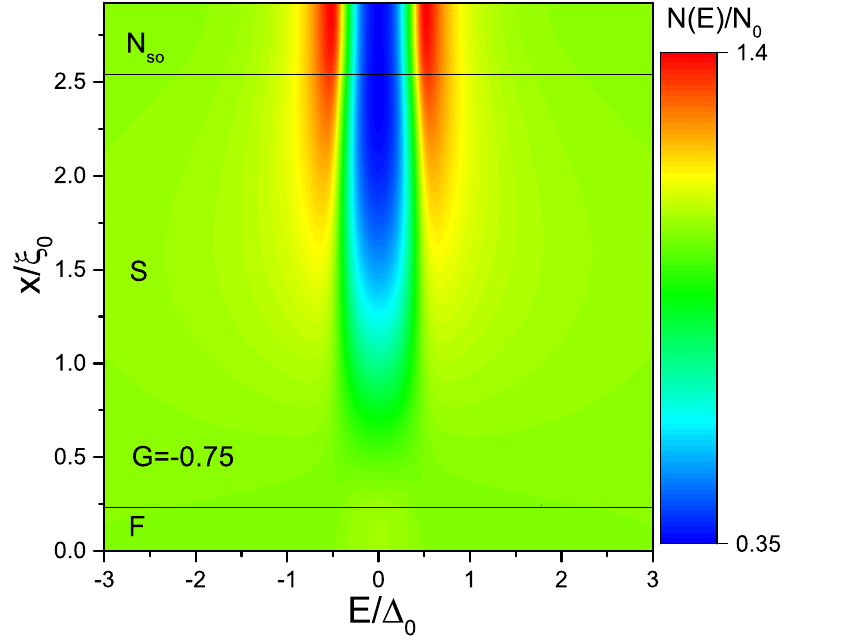}
\end{minipage}
\begin{minipage}{5.2cm}
f)\includegraphics[width=55mm]{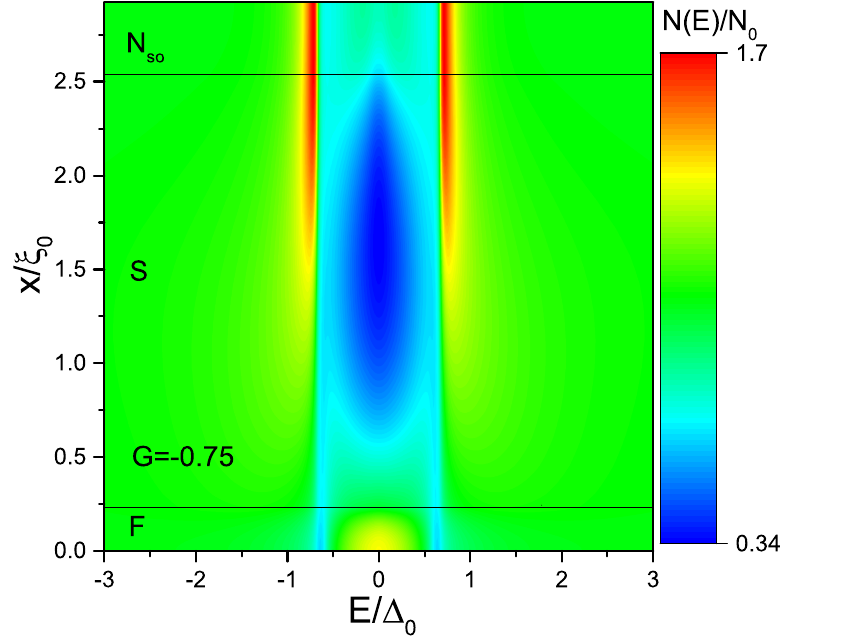}
\end{minipage}
\begin{minipage}{5.2cm}
g)\includegraphics[width=55mm]{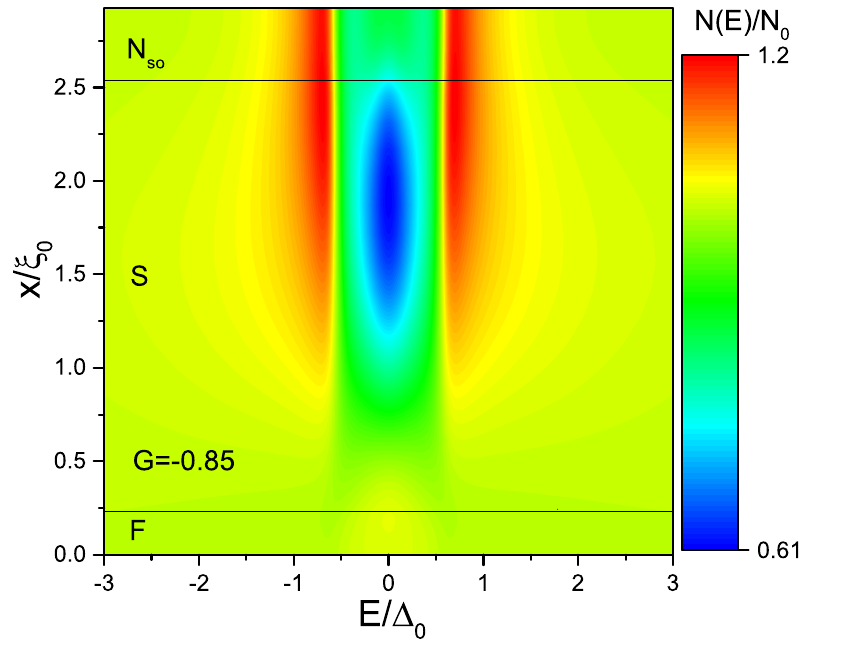}
\end{minipage}
\begin{minipage}{5.2cm}
h)\includegraphics[width=55mm]{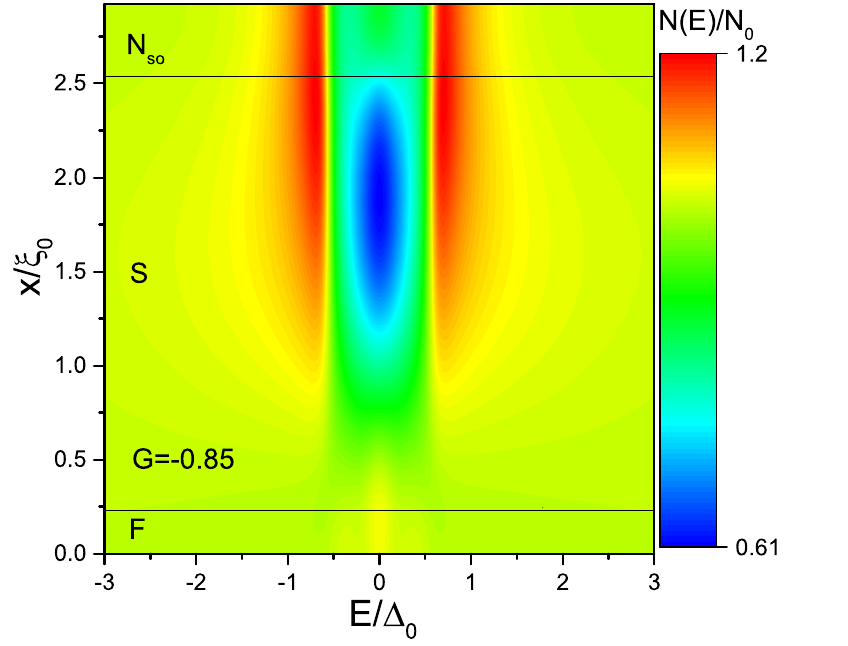}
\end{minipage}
\begin{minipage}{5.2cm}
i)\includegraphics[width=55mm]{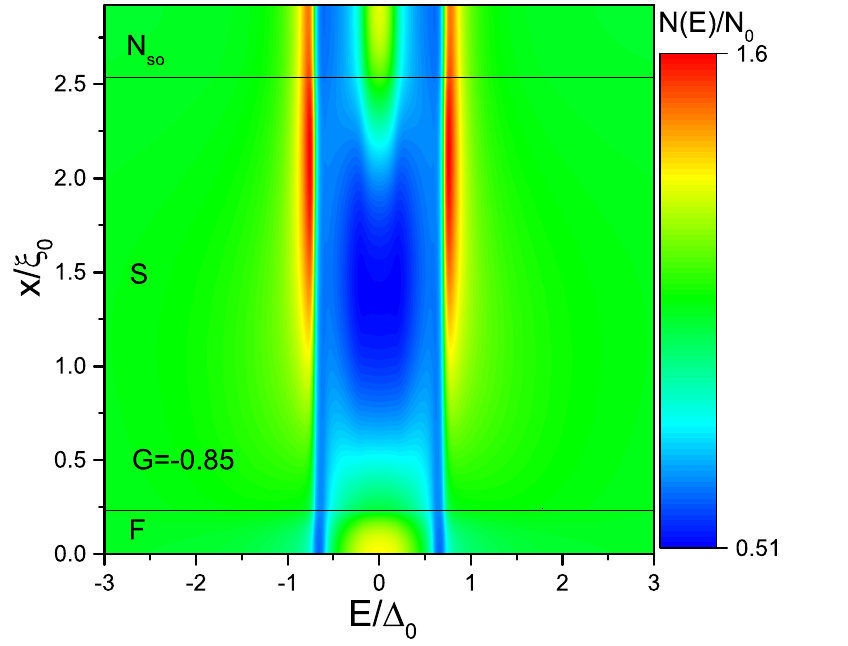}
\end{minipage}
\vspace{-2mm}
\caption{\label{Fig_DOS} (Color online) 
Density plot of the density of states as a function of energy $E$ and position $x$ in the trilayer for a), d), g) parameter set A; b), e), h) parameter set B and c), f), i) parameter set C. For each parameter set, the data are shown for three Landau parameter values, $G=-0.5$, $G=-0.75$ and $G=-0.85$. The horizontal solid lines mark the position of the F/S and the S/N$_{\rm so}$ interfaces. The proximity effect induces the opening of a minigap inside the N$_{\rm so}$ layer [a) to e)]. This gap closes for values of $G$ closer to the paramagnetic instability because of the appearance of an induced exchange field in the N$_{\rm so}$ layer [f) to i)]. A zero-energy peak appears in N$_{\rm so}$ and develops inside the S layer, i).}
\end{figure*}
We obtain the spin-resolved density of states (DOS) from the imaginary part of the normal Green function 
\[
\frac{N_{\sigma}}{N_{0}}=-\frac{1}{\pi}\text{Im}\left(G_{\sigma\sigma}\right)
\]
with $N_{0}$ the density of states at the Fermi level. The total density of states is obtained from
\begin{equation}
\frac{N}{N_{0}}=\frac{1}{2}\left(N_{\uparrow}+N_{\downarrow}\right).\label{eq:dens}
\end{equation}
In Fig.~\ref{Fig_DOS}, we present the density of states for the parameter sets A), B) and C) for three different Landau parameter values, $G=-0.5$, $G=-0.75$ and $G=-0.85$.

For Landau parameter $G=-0.5$ [see Fig.~\ref{Fig_DOS} a), b), c)], the superconducting gap opens in the SC while a minigap develops in the N$_{\rm so}$ layers
and not in the F layer. The minigap is the signature of the proximity effect and emphasizes that singlet Cooper pairs enter into the N$_{\rm so}$ layer (with a small amount of triplet Cooper pairs mixed by). In the F layer, no minigap develops because of the high amplitude of the exchange field and the presence of spin-triplet correlations.\cite{Yokoyama_PRB2006,Kawabata_JPhysSocJpn2013} Note that a small zero-energy resonance exists in the F layer as emphasized in Fig.~\ref{Fig_DOS} c).

For Landau parameter $G=-0.75$ [see Fig.~\ref{Fig_DOS} d), e), f)] we do not observe qualitative changes between parameter regime A and B [Fig.~\ref{Fig_DOS} d), e)]. However, we see that the minigap closes in the N$_{\rm so}$ layer for the parameter set C [Fig.~\ref{Fig_DOS} f)] emphasizing the onset of spin-triplet correlations and non-zero induced exchange field. The difference between the parameter sets B and C pinpoints the role of the boundary conditions. Here, the electric conductivity mismatch between the layers can induce the onset of spin-triplet correlations for a smaller value of the Landau parameter $G$.

For Landau parameter $G=-0.85$ [Fig.~\ref{Fig_DOS} g), h), i)], the minigap closes in the N$_{\rm so}$ layer for parameter set A and B [Fig.~\ref{Fig_DOS} g), h)]. For parameter set C [Fig.~\ref{Fig_DOS} i)] a zero-energy peak develops in the S layer which implies that the spin-triplet correlations are no longer confined to the N layer and are appreciable as well in the S layer. From the study of the DOS, we can deduce that spin-triplet correlations exist in the entire F/S/N$_{\rm so}$ structure, however for further information regarding the nature of these triplet correlations we need to study additional observables.
\end{subsubsection}

\begin{subsubsection}{Spin magnetization and the order parameter profile}
\label{MF_SMb}

\begin{figure*}
\centering
\begin{minipage}{5.5cm}
a)\includegraphics[width=55mm]{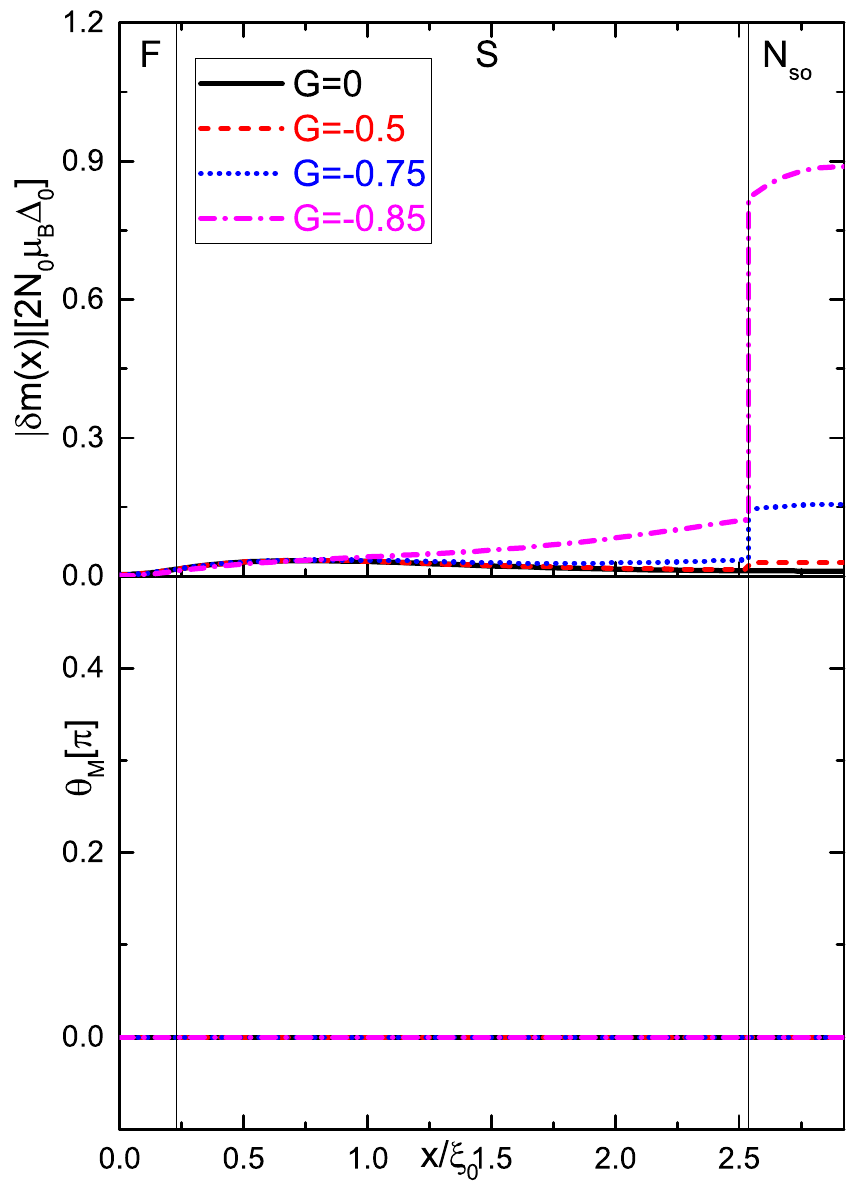}
\end{minipage}
\begin{minipage}{5.5cm}
b)\includegraphics[width=55mm]{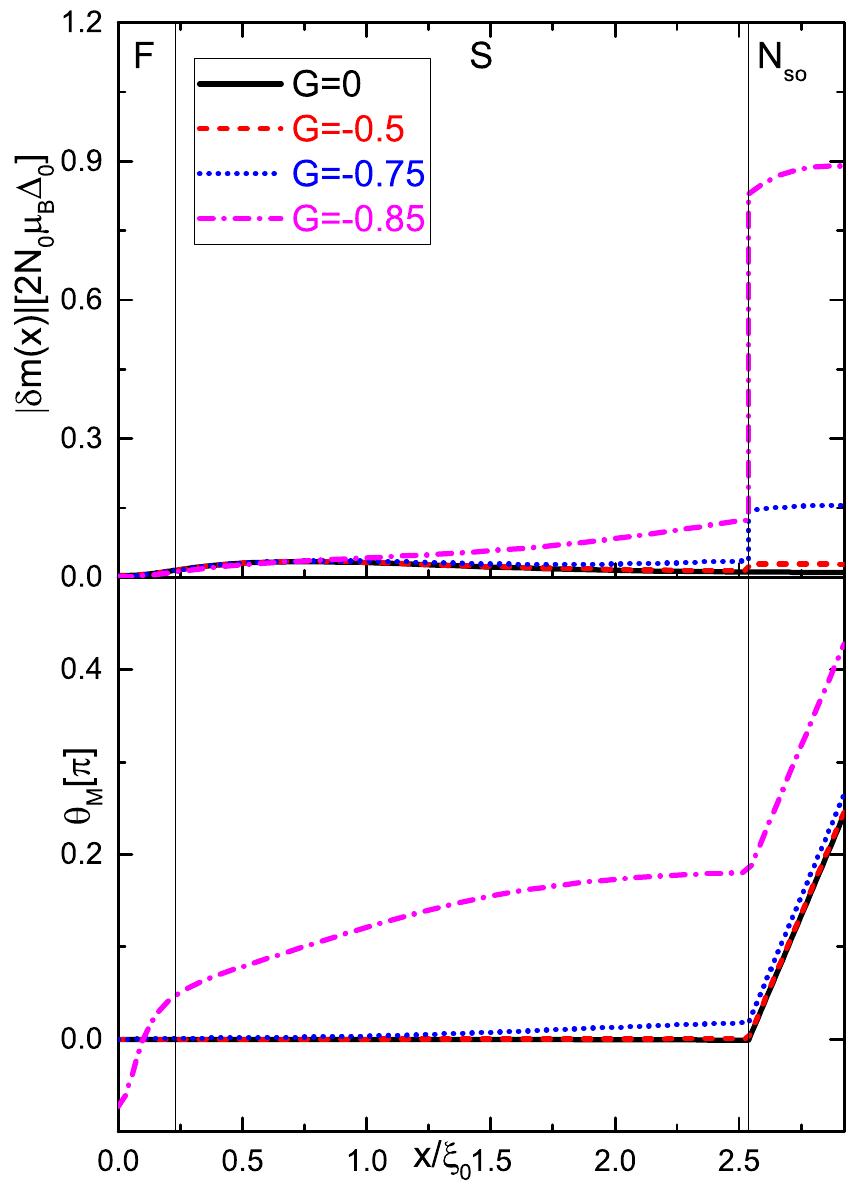}
\end{minipage}
\begin{minipage}{5.5cm}
c)\includegraphics[width=55mm]{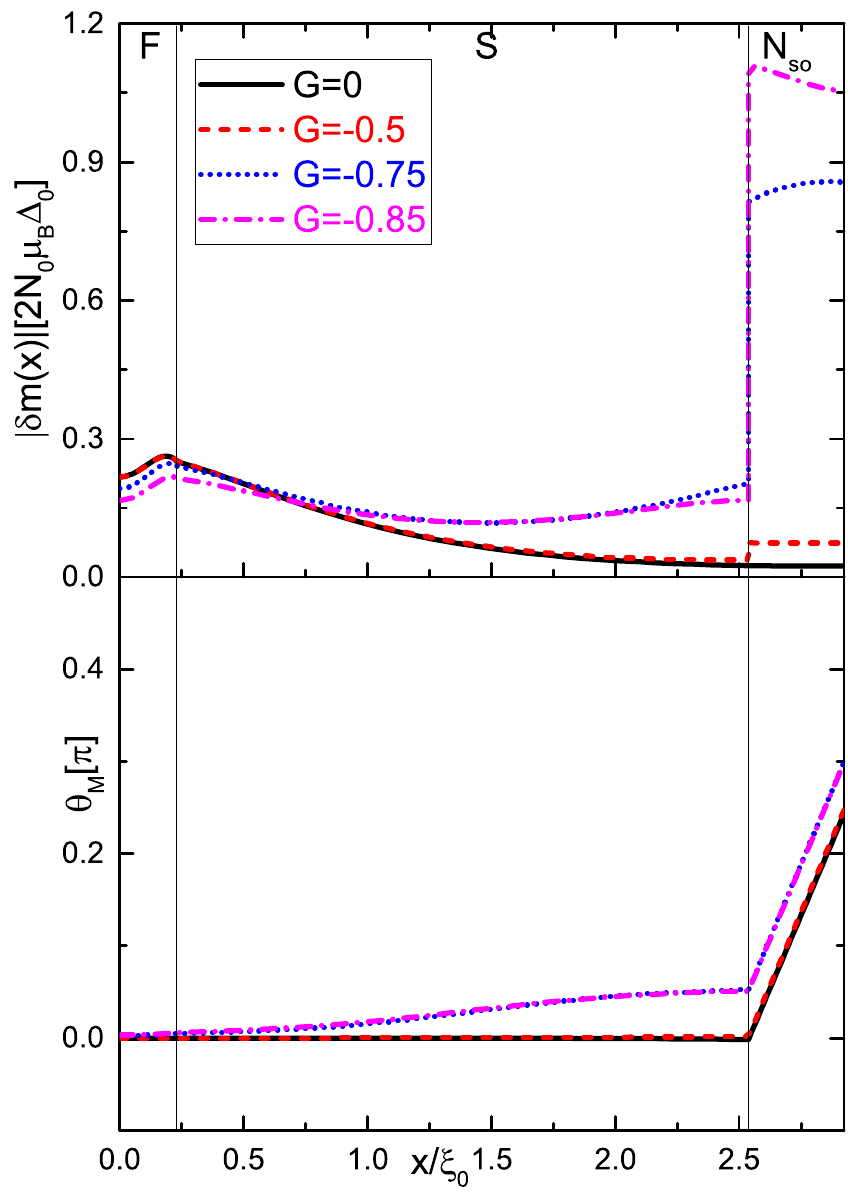}
\end{minipage}
\vspace{-2mm}
\caption{\label{Fig_SMOP} (Color online) 
Profile of the modulus of the spin magnetization, $|\delta_{m}|$, (upper panel) and the orientation angle $\theta_{M}$ (lower panel) in the F/S/N$_{\rm so}$ trilayer for parameter set A a), B b) and C c), and for Landau parameters $G=0$ (black solid line), $G=-0.5$ (red dashed line), $G=-0.75$ (blue dotted line) and $G=-0.85$ (magenta dashed-dotted line). No spin magnetization along the $x$-axis exists in the trilayer. The vertical solid lines mark the position of the F/S and the S/N$_{\rm so}$ interfaces.}
\end{figure*} 

\begin{figure*}
\centering
\begin{minipage}{5.5cm}
a)\includegraphics[width=55mm]{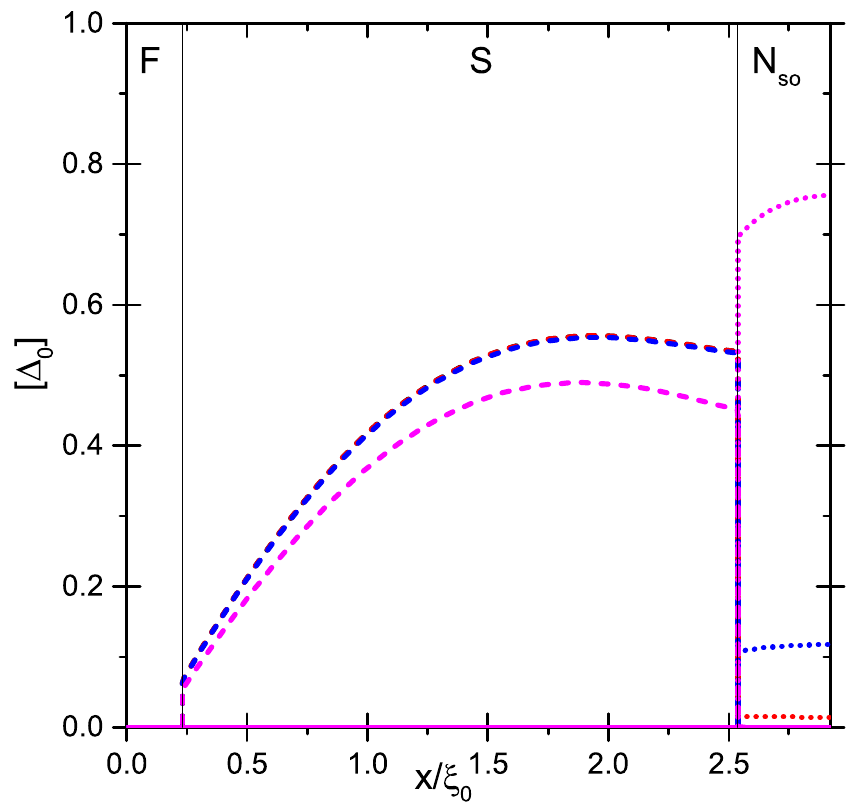}
\end{minipage}
\begin{minipage}{5.5cm}
b)\includegraphics[width=55mm]{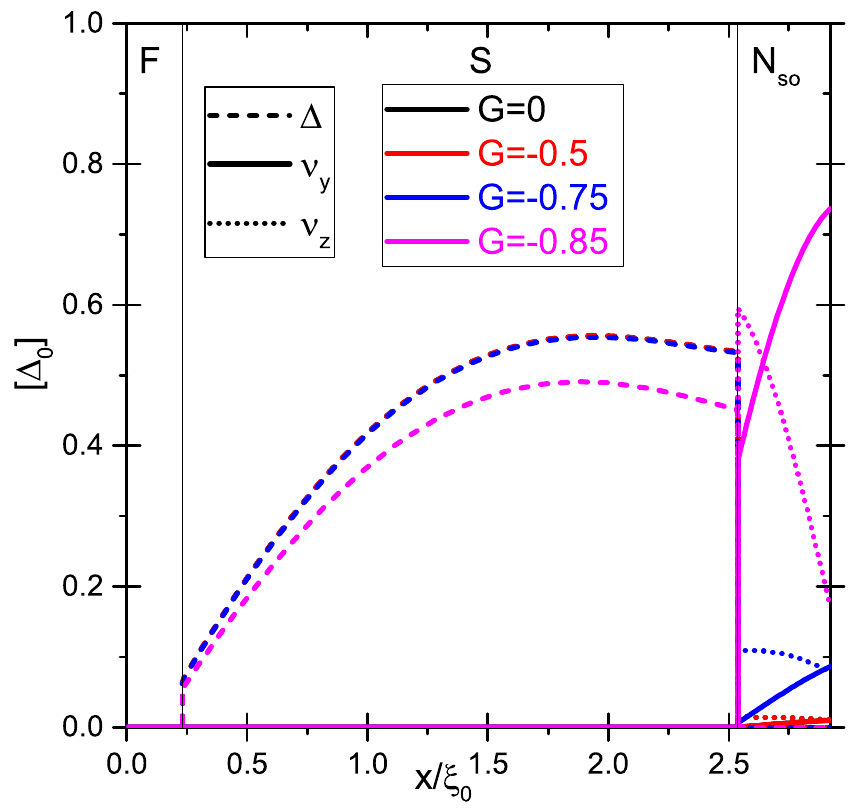}
\end{minipage}
\begin{minipage}{5.5cm}
c)\includegraphics[width=55mm]{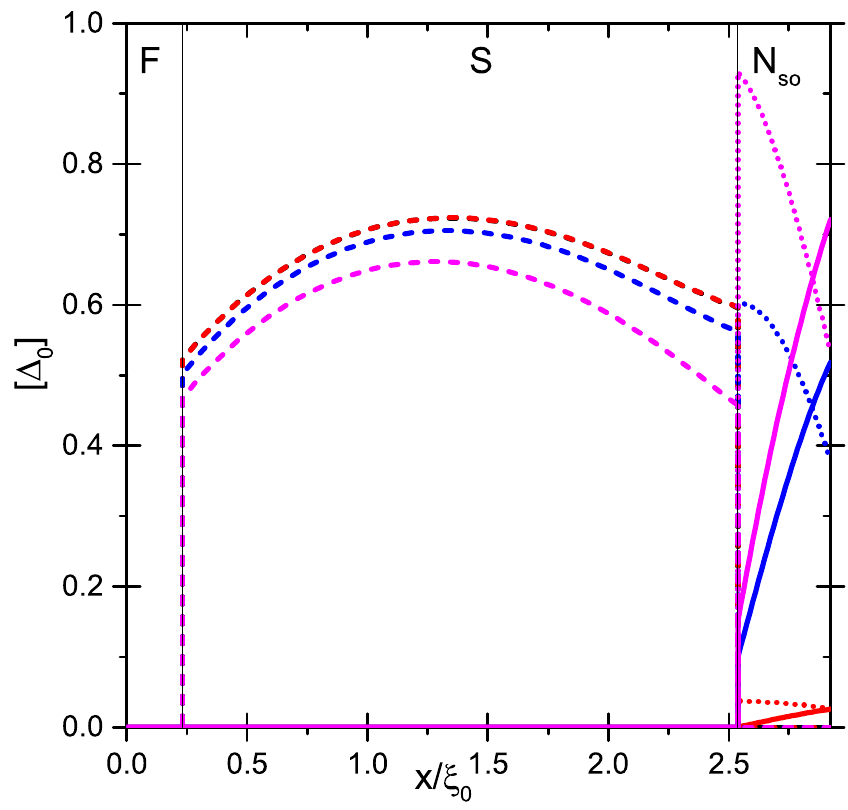}
\end{minipage}
\vspace{-2mm}
\caption{\label{Fig_MF} (Color online) 
Order parameter profiles in the F/S/N$_{\rm so}$ trilayer for parameter set A a), B b) and C c), and for Landau parameters $G=0$ (black), $G=-0.5$ (red), $G=-0.75$ (blue) and $G=-0.85$ (magenta). The SC order parameter is presented in dashed line while the exchange fields along the $y(z)$-axis are presented in solid (dotted) line. The vertical solid lines mark the position of the F/S and the S/N$_{\rm so}$ interfaces.}
\end{figure*}

The onset of spin-triplet correlations lifts the degeneracy between the spin-up and spin-down DOS which leads to a non-zero spin magnetization (SM) in superconducting devices.\cite{Eschrig_PRL2003,Champel_PRB2005b,Eschrig_RepProgPhys2015} Close to the S/F interfaces, a spin magnetization can develop inside the S layer.\cite{Sauls_PRB1988,Alexander_PRB1985,Bergeret_EPL2004,Bergeret_PRB2004} As shown in Eq.~(\ref{magn}), the spin magnetization $\delta\boldsymbol{m}(x)$ is induced by the onset of spin-triplet correlations. Fermi liquid interactions can amplify or screen the SM in the N$_{\rm so}$ layer.\cite{Sauls_PRB1988,Alexander_PRB1985}

In Fig.~\ref{Fig_SMOP} we present the SM for the three parameter sets for different magnitudes of the Landau parameter. We show the modulus of the magnetization $|\delta \bf{m}|$ and its angle $\theta_{M}$ relative to the $z$-axis which quantifies the direction of the magnetization in the $y-z$ plane. Therefore, for $\theta_{M}=0$, the SM is along the $z$-axis and if $\theta_{M}\neq0$, the SM acquires a $y$-component. Note that the direction of the SM is reversed compared to that of the ferromagnet's magnetization. This can be explained in terms of the S/F proximity effect.\cite{Bergeret_EPL2004,Bergeret_PRB2004}
The SM in the S layer decays away from the F/S interface, until it reaches the S/N$_{\rm so}$ interface. At this interface, the spin magnetization in the N$_{\rm so}$ layer is then the same as the one of the S layer. 
At $G=0$ the decay length of the SM is the coherence length in the S layer $\xi_{0}$. The onset of a non-zero Landau Fermi-liquid exchange field further amplifies the SM inside the N$_{\rm so}$ layer (see Fig. \ref{Fig_SMOP}).

In absence of SOC [Fig.~\ref{Fig_SMOP} a)], the SM only exists along the $z$-axis. With SOC, [Fig.~\ref{Fig_SMOP} b) and c)], the onset of $y$- and $z$-axis spin-triplet correlations induces the onset of $y$- and $z$-axis SM components. Therefore, the presence of SOC implies that the induced magnetization in the N$_{\rm so}$ layer is tilted compared to the F layer magnetization. The value of the induced SM at the S/N$_{\rm so}$ interface increases with the magnitude of the Landau parameter $G$. Note that the discontinuity of the SM at the S/N$_{\rm so}$ interface (despite continuous boundary conditions) is explained by the fact that the Landau parameter $G$ is non-zero in the N$_{\rm so}$ layer only and vanishes in the F and S layers. We also observe the effect of Fermi surface mismatch at the boundary on the SM profile,
where the modulus and the orientation angle differ between the case when the conductivities are the same [Fig.~\ref{Fig_SMOP} b)] and the case when the conductivities are different [Figs.~\ref{Fig_SMOP} c)]. Moreover, we note that for Landau parameter sufficiently close to the paramagnetic instability, $G=-0.75$ and $G=-0.85$, the $y$-axis SM component exists in the entire trilayer [Figs.~\ref{Fig_SMOP} b) and c)], which emphasizes the existence of long-range spin-triplet correlations in the entire trilayer.

In Fig.~\ref{Fig_MF} we show the profile of the SC gap, $\Delta(x)$, and the induced exchange field, $\nu(x)$, calculated self-consistently for the three parameter sets and various values of the Landau parameter $G$. We see that the induced exchange field appears for a non-zero value of the Landau parameter. In presence of SOC [Fig.~\ref{Fig_MF} b) and c)], the induced exchange field in the N$_{\rm so}$ layer acquires a component along the $y$ axis. This component is directly related by the onset of long-range spin-triplet correlations in the N$_{\rm so}$ layer. For increasing value of the Landau parameter $G$, the superconducting gap magnitude decreases in the S layer. The onset of the N$_{\rm so}$ exchange field implies a stronger inverse proximity effect at the S/N$_{\rm so}$ interface.

The presence of both SOC and FL corrections in the N$_{\rm so}$ layer leads to a magnetic order in the N$_{\rm so}$ layer whose magnetization direction depends on the coordinate. This magnetic order can be considered as a spiral magnetic order.\cite{Pugach_APL2018} Note that the magnetic structure of the spiral order strongly depends on the symmetry of the SOC and on the symmetry of the Landau parameters we have chosen. The S/F and S/N$_{\rm so}$ interfaces play a crucial role in the stabilization of the superconductivity and of the spin-triplet correlations. The inverse proximity effect at the S/F interface is stronger when the electrical conductivities are similar [Fig.~\ref{Fig_MF} b)] compared with the case with a conductivity mismatch [Fig.~\ref{Fig_MF} c)].
\end{subsubsection}
\begin{subsubsection}{Pair amplitude}
\label{PA}
\begin{figure*}
\centering
\begin{minipage}{5.5cm}
a)\includegraphics[width=55mm]{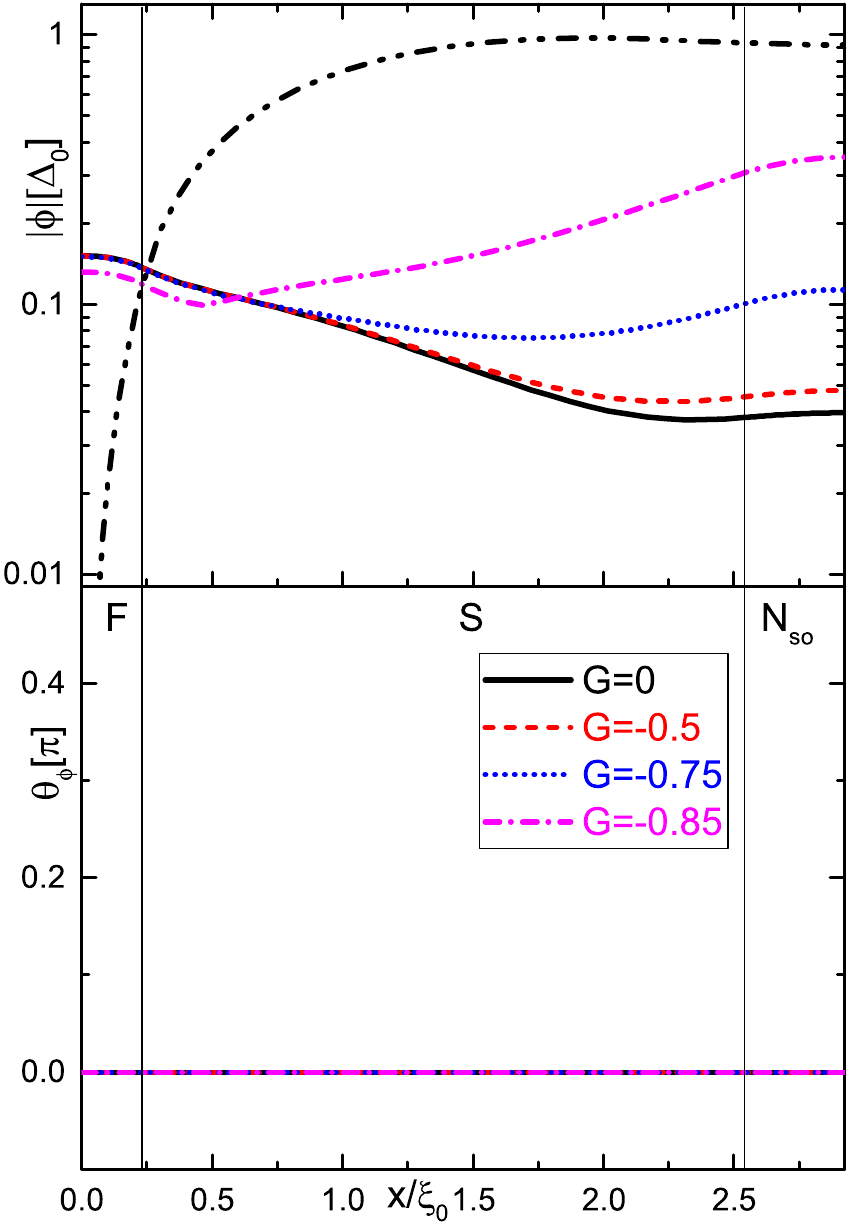}
\end{minipage}
\begin{minipage}{5.5cm}
b)\includegraphics[width=55mm]{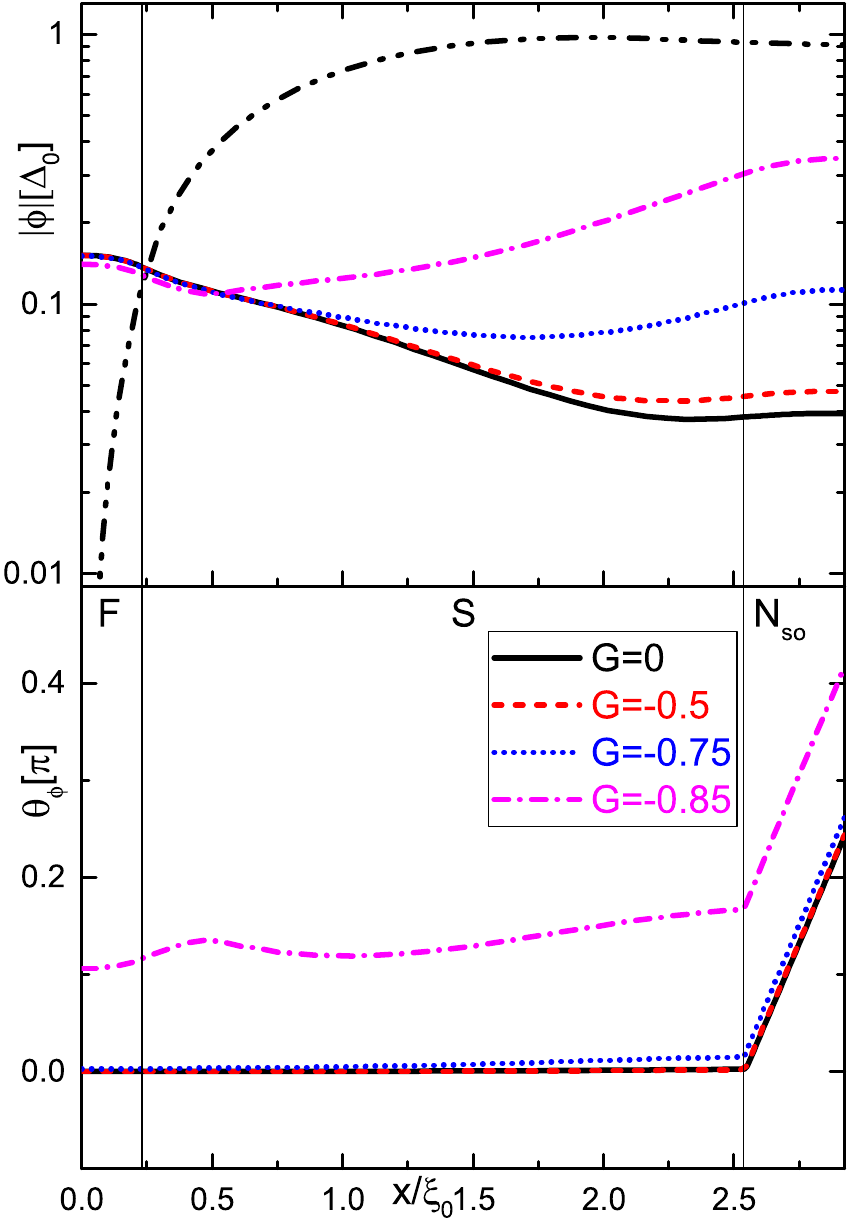}
\end{minipage}
\begin{minipage}{5.5cm}
c)\includegraphics[width=55mm]{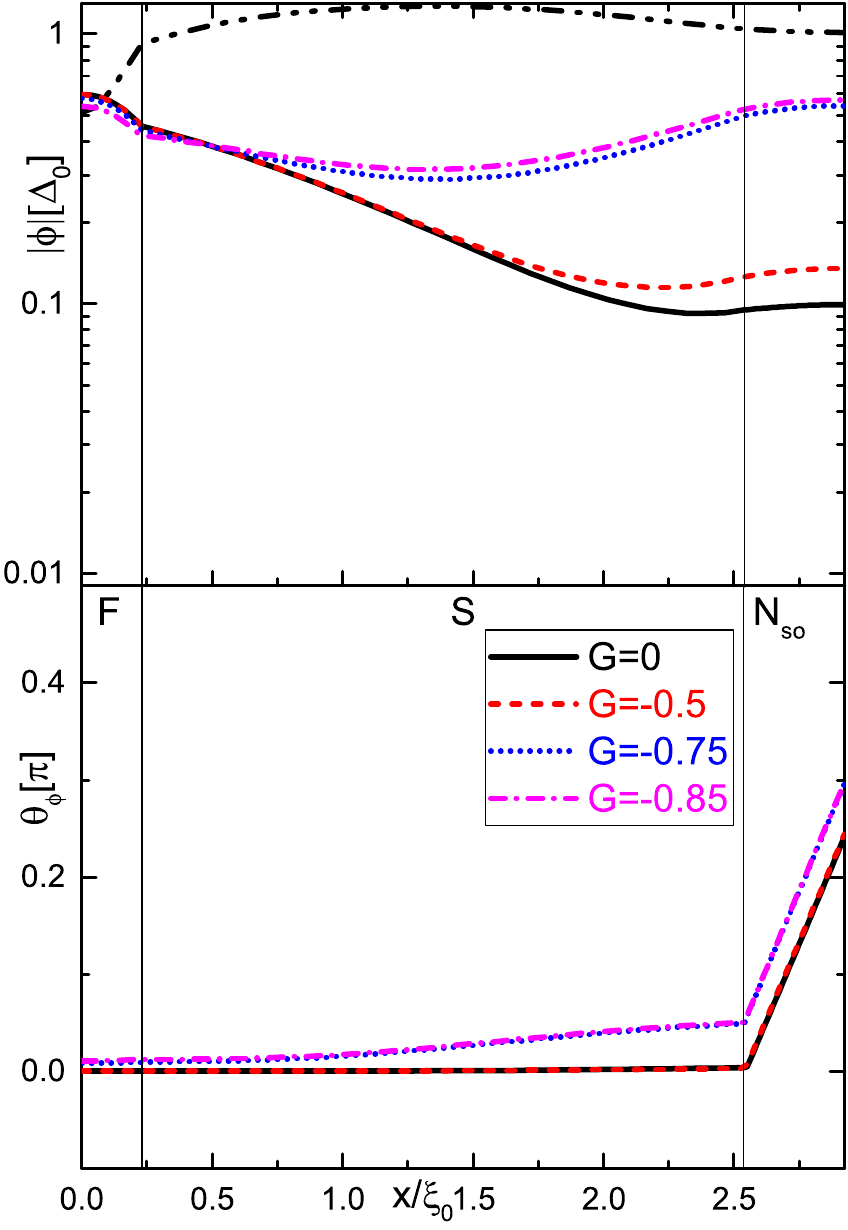}
\end{minipage}
\vspace{-2mm}
\caption{\label{Fig_PA} (Color online) 
Profile of the modulus of the triplet  pair amplitude, $|\phi_{t}|$, (upper panel) and of the orientation angle $\theta_{\phi}$ (lower panel) in the F/S/N$_{\rm so}$ trilayer for parameter set A a), B b) and C c), and for Landau parameters $G=0$ (black solid line), $G=-0.5$ (red dashed line), $G=-0.75$ (blue dotted line) and $G=-0.85$ (magenta dashed-dotted line). The modulus of the singlet pair amplitude, $|\phi_{s}|$, is weakly affected by the value of the Landau parameter and is presented in dashed-double dotted line for the case $G=0$.}
\end{figure*}

The pair amplitude reflects the strength of the SC correlations in the trilayer. The singlet and triplet pair amplitude is obtained from \cite{Champel_PRL2005}
\begin{equation}
\begin{array}{c}
\phi_{s}\left(x\right)=\int_{-\infty}^{+\infty}\frac{dE}{2i\pi}f_{s}\left(E,x\right)\tanh\left(\frac{E}{2T}\right)\\
\\
\boldsymbol{\phi}_{t}\left(x\right)=\int_{-\infty}^{+\infty}\frac{dE}{2i\pi}\boldsymbol{f}_{t}\left(E,x\right)\tanh\left(\frac{E}{2T}\right)
\end{array}.
\label{rel_PA}
\end{equation}
where $f_{s}$ and $\boldsymbol{f}_{t}$ are the singlet and triplet part of the anomalous Green function [see Eq.~(\ref{Green_spin})]. 

In Fig.~\ref{Fig_PA}, we show the pair-amplitude profile for the three parameter sets for various Landau parameters $G$.
 We present the modulus of the spin-triplet and spin-singlet pair amplitudes ($|\phi_{t}|$ and $|\phi_{s}|$), and the angle $\theta_{\phi}$ relative to the $z$-axis that quantifies the direction of the pair amplitude in the $y-z$ plane. For $\theta_{\phi}=0$, the spin-triplet pair amplitude is along the $z$-axis while if $\theta_{\phi}\neq0$, the triplet pair amplitude acquires a $y$-axis component . Therefore, $\theta_{\phi}$ quantifies the nature of spin-triplet correlations in the system. For $\theta_{\phi}=0 (\pi)$, the spin-triplet correlations are only short-range while if $\theta_{\phi}=\frac{\pi}{2}$, the spin-triplet correlations are only long-range. In the general case $0<\theta_{\phi}<\frac{\pi}{2}$, the spin-triplet correlations have both a short-range and a long-range component.

Without SOC [Fig.~\ref{Fig_PA} a)], only short-range spin-triplet Cooper pairs $\phi_{t}^{Z}$ exist in the system. The amplitude of the short-range pair amplitude is maximal in the F layer where spin-triplet pairs are produced.
With SOC, [Fig.~\ref{Fig_PA}, b) and c)], we see the onset of the long-range triplet correlations $\phi_{t}^{Y}$ which are maximal in the N$_{\rm so}$ layer [Fig.~\ref{Fig_PA} b) and c)]. The amplitude of the long-range triplet correlations increases quickly with the Landau parameter $G$: from $10^{-6}$ with $G=-0.5$ to $10^{-3}$ for $G=-0.85$. Moreover, the long-range correlations propagate in the S and the F layer with a slow spatial decay [Fig.~\ref{Fig_PA} b) and c)]. The decaying length in the S layer is the SC coherence length $\xi_{0}$, whereas it is the pair correlation length $\xi_{F}=\sqrt{D_F/2\pi T}$ in the F layer. With non-zero Landau parameter we observe an enhancement of spin-triplet correlations in the N$_{\rm so}$ layer [Fig.~\ref{Fig_PA}]. The presence of an induced exchange field in the N$_{\rm so}$ layer amplifies the spin-triplet correlations in the entire system. In Fig.~\ref{Fig_PA} b) and c), we observe that for high magnitude of the Landau parameter, $G=-0.75$ and $G=-0.85$, the long-range triplet correlations propagate in the entire trilayer. The amount of spin-triplet correlations in the trilayer differs with the proximity effect. The profile of the pair amplitude is different when the conductivities are the same [Fig.~\ref{Fig_PA} b)] compared to the case when a conductivity mismatch between the layers is present, Fig.~\ref{Fig_PA} c).
\end{subsubsection}

\begin{subsubsection}{Charge and spin current}
\label{Curr}
\begin{figure}
\centering
a)\includegraphics[width=80mm]{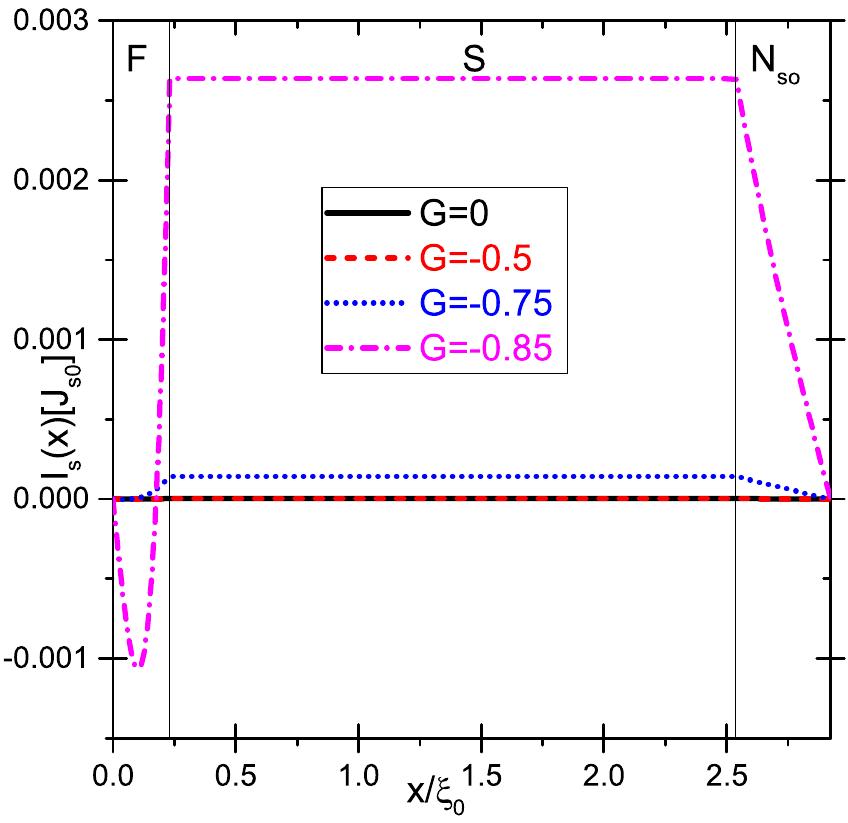}
b)\includegraphics[width=80mm]{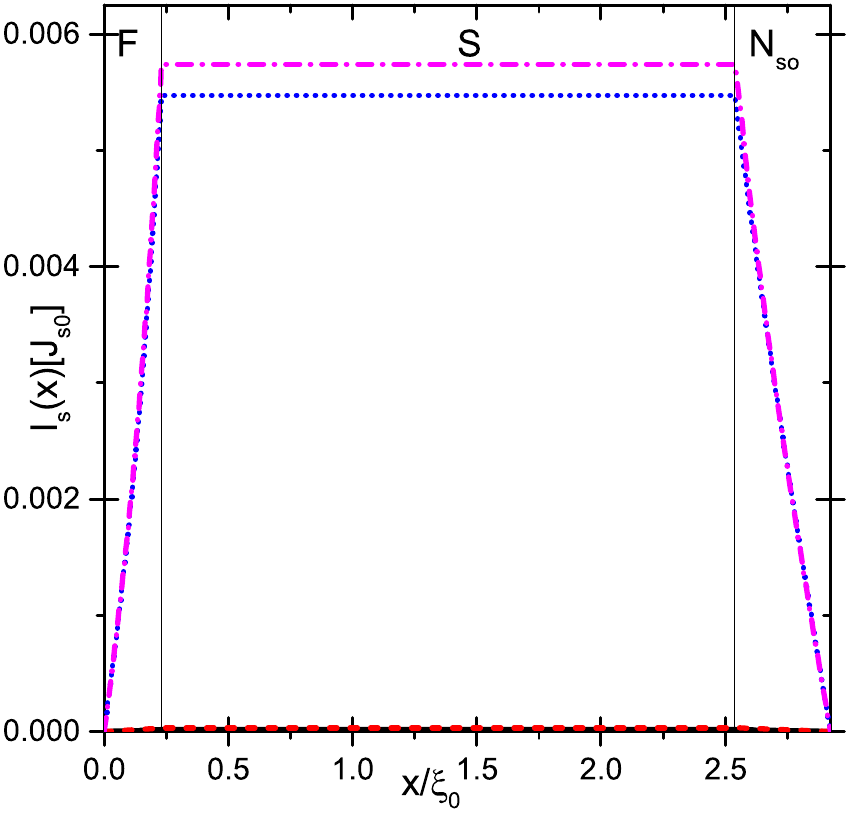}
\vspace{-2mm}
\caption{\label{Fig_IS} (Color online) 
Profile of the spin current in the F/S/N$_{\rm so}$ trilayer for parameter set B a) and C b), and for Landau parameters $G=0$ (black), $G=-0.5$ (red), $G=-0.75$ (blue) and $G=-0.85$ (magenta). The spin current varies in the F and the N$_{\rm so}$ layer while it is constant in the S layer.}
\end{figure}
One way to see and quantify long-range triplet correlations in the F/S/N$_{\rm so}$ trilayer is to calculate the spin current. For a one-dimensional system, the charge current in $x$-direction at equilibrium is given by
\begin{equation}
I_x^{c}=J_{0}\, \text{Re} \int \frac{dE}{4\pi^{2}}\text{Tr}\left(\hat{\tau}_{3}\hat{G}\partial_{x}\hat{G}\right)\tanh\left(\frac{E}{2T}\right)
\label{curr_ch_ret}
\end{equation}
where $J_{0}=-\frac{\sigma_{0}}{2e}$ is the Landau critical charge current with $\sigma_{0}=2e^{2}N_{0}D$ the electrical conductivity in the normal state. The trace Tr is taken over the 4$\times $4 matrix structure of the argument.
Correspondingly, the spin current at equilibrium is given by
\begin{equation}
\boldsymbol{I}_x^{s}=J_{s0}\, \text{Re} \int \frac{dE}{4\pi^{2}}\text{Tr}\left(\hat{\tau}_{3}
\hat{\boldsymbol{\sigma}}
\hat{G}\partial_{x}\hat{G}\right)\tanh\left(\frac{E}{2T}\right)
\label{curr_sp_ret}
\end{equation}
where $J_{s0}=\frac{\hbar}{2e}J_{0}$. In the absence of an external phase bias, the charge current vanishes in the entire structure, however a spin current can still persist. The numerically obtained spin-current profile for parameter sets B) and C) is shown in Fig.~\ref{Fig_IS}. This spin current appears because of the onset of a tilted magnetization in the N$_{\rm so}$ layer. Therefore, an equilibrium spin torque establishes between the two magnetizations. The spin current is polarized along the $x$-axis since the equilibrium spin torque is orthogonal to the plane spanned by the two magnetizations, here the $y$-$z$ plane.\cite{Waintal_PRB2002} This spin current has been predicted at zero phase difference in S/F/S/F/S junctions with tilted magnetization.\cite{Halterman_SupSciTech2016} In our case, the presence of this spin current is a signature of the presence of long-range spin-triplet correlations in the structure, i.e. it is predominantly carried by spin-polarized Cooper pairs. This spin supercurrent vanishes in the absence of spin-orbit coupling since in this case the magnetization induced in the N$_{\rm so}$ layer is collinear with the F layer one. In the presence of SOC but without an induced exchange field due to FL corrections (G=0), the spin current is restricted to the $N_{\rm so}$ region.

By decomposing the Green functions into spin-scalar and spin-vector components, see Eq.~(\ref{Green_spin}), and utilizing the notation
$\delta_x\circ \boldsymbol{a}\equiv \partial_x \boldsymbol{a}+2\boldsymbol{A}_x\times \boldsymbol{a}$ for any vector $\boldsymbol{a}$,
we rewrite the expressions (\ref{curr_ch_ret}) and (\ref{curr_sp_ret}) as \cite{Eschrig_RepProgPhys2015} 
\begin{eqnarray}
I_x^{c}&=&J_{0}\int \frac{dE}{\pi^{2}}
\text{Re}\left[\widetilde{f_{s}}\partial_{x}f_{s}-\widetilde{\boldsymbol{f_{t}}}\left(\delta_{x}\circ \boldsymbol{f_{t}}\right)\right]
\tanh\left(\frac{E}{2T}\right) 
\label{F1}
\end{eqnarray}
\begin{eqnarray}
&&\boldsymbol{I}_x^{s}=-J_{s0}\int \frac{dE}{\pi^{2}}
\text{Im}\left[\boldsymbol{g}\times(\delta_{x}\circ \boldsymbol{g})+\widetilde{\boldsymbol{f_{t}}}\times (\delta_{x}\circ \boldsymbol{f_{t}})\right] 
\tanh\left(\frac{E}{2T} \right). 
\nonumber \\
\label{F2}
\end{eqnarray}
The additional terms due to the spin-orbit field are of the form $-2\mbox{Re}[\widetilde{\boldsymbol{f_{t}}} (\boldsymbol{A}_x\times \boldsymbol{f_{t}})]$ for the charge current, and for the spin current 
$-2\mbox{Im}[\boldsymbol{g}\times (\boldsymbol{A}_x\times \boldsymbol{g})]$ and
$-2\mbox{Im}[\widetilde{\boldsymbol{f}}_t\times (\boldsymbol{A}_x\times \boldsymbol{f}_t)]$. Note that near $T_c$ the term involving $\boldsymbol{g}$ can be neglected compared to the terms involving the anomalous functions.


We notice that the charge current only depends on the presence of the anomalous Green functions $f$ and $\boldsymbol{f}_t$ which emphasizes the Cooper-pair nature of the charge Josephson current. On the other hand, the spin current depends on both normal and anomalous Green functions which emphasizes that spin can be carried by spin-triplet pairs and by quasiparticles in the S layer. 
For the special case of Eq.~(\ref{SOCsym}), we have $\boldsymbol{A}_x= (\alpha,0,0)$, and
Eqs.~\eqref{F1} and \eqref{F2} turn into
\begin{eqnarray}
I_x^{c}&=&J_{0}\int \frac{dE}{\pi^{2}}\text{Re}\left\{ \widetilde{f_{s}}\partial_{x}f_{s}-\widetilde{f}_{t}^{X}\partial_{x}f_{t}^{X}
-\widetilde{f}_{t}^{Y}\partial_{x}f_{t}^{Y}-\widetilde{f}_{t}^{Z}\partial_{x}f_{t}^{Z}
\right. \nonumber \\
&&\left.+2\alpha\left(f_{t}^{Z}\widetilde{f}_{t}^{Y}-f_{t}^{Y}\widetilde{f}_{t}^{Z}\right)\right\}\tanh\left(\frac{E}{2T}\right)  \\
I_{x}^{s,X}&=&-J_{s0}\int \frac{dE}{\pi^{2}}\text{Im}\left\{ g_{t}^{Y}\partial_{x}g_{t}^{Z}-g_{t}^{Z}\partial_{x}g_{t}^{Y}
+\widetilde{f}_{t}^{Y}\partial_{x}f_{t}^{Z}-\widetilde{f}_{t}^{Z}\partial_{x}f_{t}^{Y}\right. \nonumber \\
&+&\left. 2\alpha \left( g_{t}^{Y}g_{t}^{Y}+g_{t}^{Z}g_{t}^{Z}
+f_{t}^{Y}\widetilde{f}_{t}^{Y}+f_{t}^{Z}\widetilde{f}_{t}^{Z} \right) \right\}\tanh\left(\frac{E}{2T}\right) .
\end{eqnarray}
The spin currents $I_{x}^{s,Y}$ and $I_{x}^{s,Z}$ vanish in our case, as no $f_{t}^{X}$ or $g_{t}^{X}$ component develops.
The dominant terms are the ones proportional to $f_{t}^{Z}\widetilde{f}_{t}^{Z}$, as this is the component generated at the S/F interface. Thus, the spin current is dominantly spin-polarized in $X$-direction.
In this case, the spin current is due to equal-spin Cooper pairs existing between the S/F and the S/N$_{\rm so}$ interfaces. These equal-spin pairs exist mainly in the S layer and their intensity is expected to decrease with the S thickness.

\end{subsubsection}
\end{subsection}

\begin{subsection}{Dependence of pure spin current on model parameters}
In the last section, we have demonstrated that equal-spin Cooper pairs appear in the F/S/N$_{\rm so}$ trilayer if both SOC and FL corrections are included in the N$_{\rm so}$ layer. In this section we focus on the general variation of the spin supercurrent (which reflects the intensity of the equal spin Cooper pairs) with the various model parameters of the system.
\begin{subsubsection}{Dependence on the Landau parameter $G$}
\label{Landau_par_sec}
\begin{figure*}[htb]
\centering
\begin{minipage}{5.5cm}
a)\includegraphics[width=55mm]{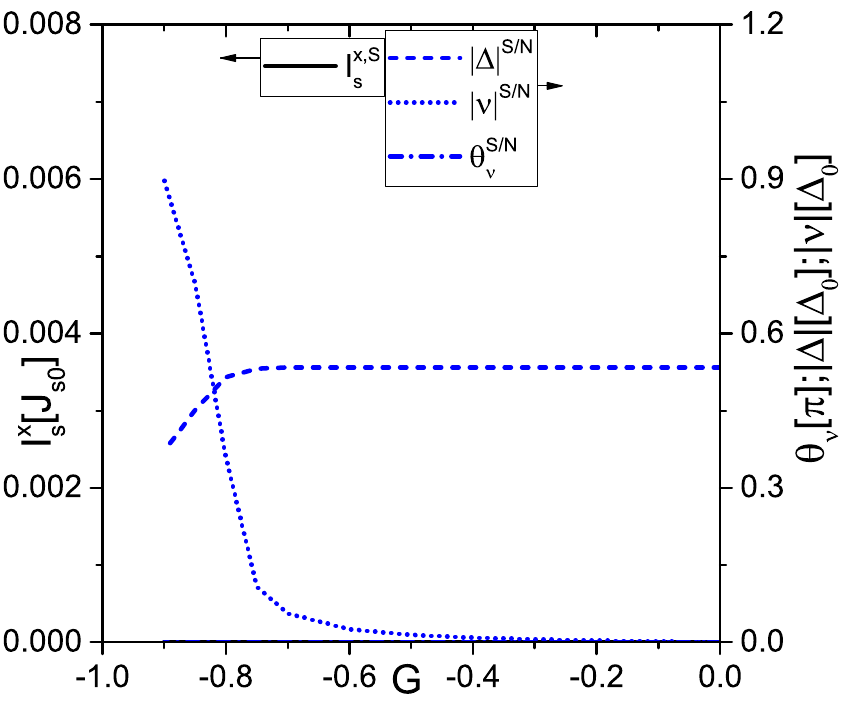}
\end{minipage}
\begin{minipage}{5.5cm}
b)\includegraphics[width=55mm]{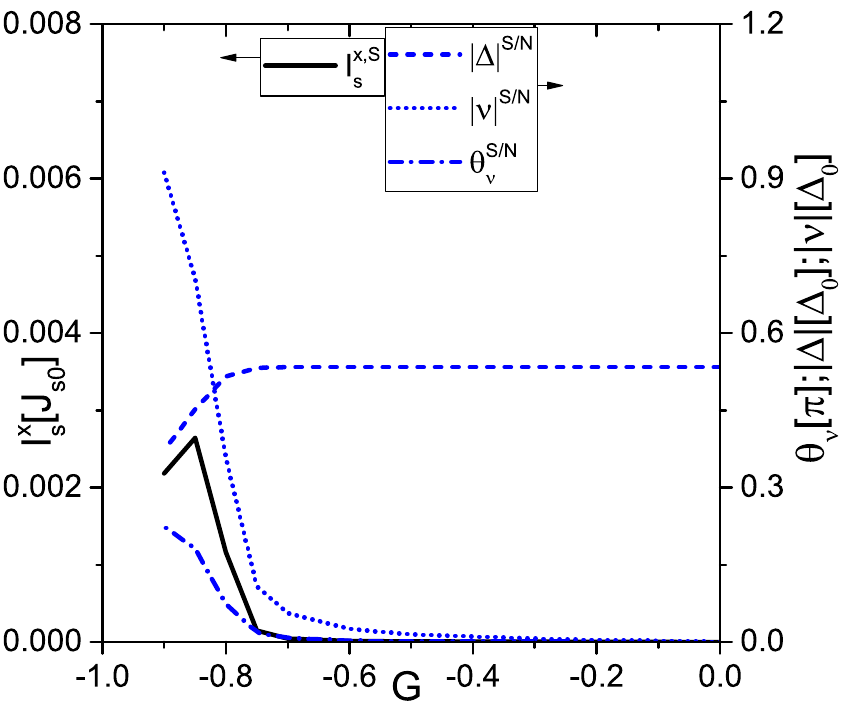}
\end{minipage}
\begin{minipage}{5.5cm}
c)\includegraphics[width=55mm]{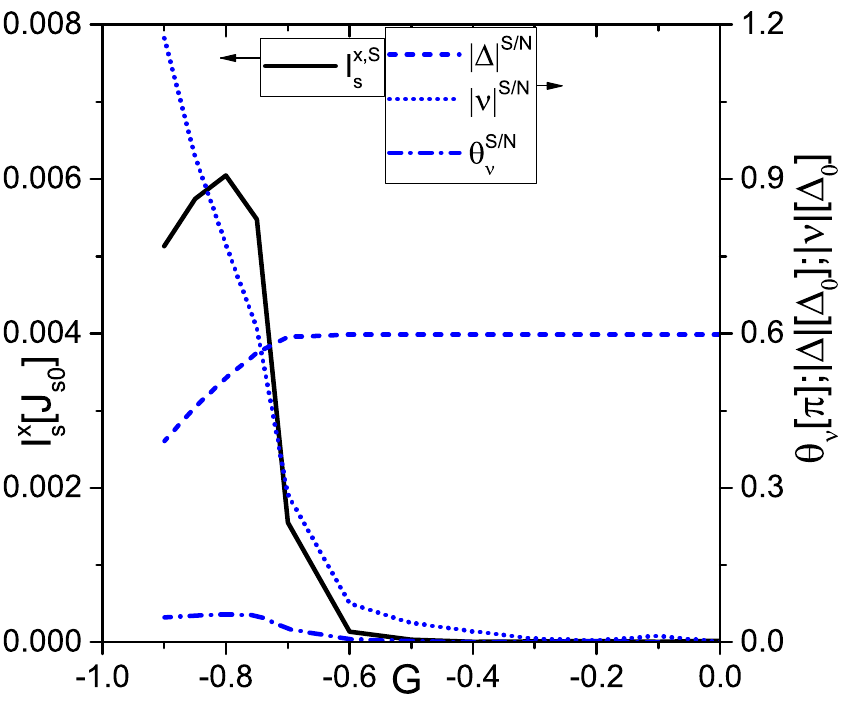}
\end{minipage}
\vspace{-2mm}
\caption{\label{G_dep}(Color online) 
Magnitude of the spin current spin-polarized along the 
$x$-axis in the S layer $I_{s}^{X}$ (solid line), magnitude of the SC order parameter $\Delta^{S/N}$ at the S/N interface (dashed line), and modulus $\nu^{S/N}$ and misorientation angle $\theta_{\nu}^{S/N}$ of the Fermi-liquid exchange field at the S/N interface (dotted and dash-dotted lines, respectively) as function of the Landau parameter for parameter set a) A, b) B and c) C. 
In the absence of spin orbit coupling, a), no spin current and $y$-axis exchange field exist in the trilayer. The intensity of the current depends on the boundary conditions b) and c). In all cases we observe a Landau parameter threshold from which the intensity of the SC order parameter decreases and the $z$-axis exchange field increases abruptly. The value of this threshold is in A) $G_{\rm th}^{\rm A}\approx 0.7$, in B) $G_{\rm th}^{B}\approx 0.7$, and in C) $G_{\rm th}^{C}\approx 0.6$.}
\end{figure*}
In Fig.~\ref{G_dep}, we present the magnitude of the spin current and of the SC order parameter at the S/N$_{\rm so}$ interface, as well as the magnitude of the Fermi liquid order parameter and the
orientation angle $\theta_{\nu}$ of the Fermi liquid exchange field 
at the S/N$_{\rm so}$ interface as a function of the Landau parameter $G$.  
The angle $\theta_{\nu}$ relative to the $z$-axis quantifies the direction of the Fermi Liquid exchange field in the $y$-$z$ plane. Therefore, for $\theta_{\nu}=0$, the exchange field $\nu$ is along the $z$-axis and if $\theta_{\nu}\neq0$, the exchange field $\nu$ acquires a $y$-axis component.
Seen from Fig.~\ref{G_dep}, no spin supercurrent exists in absence of SOC (Fig. \ref{G_dep} a)). In presence of SOC in the N$_{\rm so}$ layer [Fig.~\ref{G_dep} b) and c)], the magnitude of the spin current in the S layer increases abruptly and decreases close to $G=-1$. 
This behavior can be related to the dependence of the superconducting order parameter and the induced exchange fields at the S/N$_{\rm so}$ interface on $G$. A spin supercurrent appears when a $y$-axis exchange field appears.
There is a threshold value of the Landau parameter, $G_{\rm th}$, below which the Fermi liquid exchange field at the S/N$_{\rm so}$ interface and the magnitude of the spin current increase abruptly. The value of this threshold depends on the boundary conditions such that for parameter set A) $G_{\rm th}^{\rm A}\approx 0.7$, B) $G_{\rm th}^{B}\approx 0.7$ and C) $G_{\rm th}^{C}\approx 0.6$.

As seen from the figure, for $G>G_{\rm th}$, the SC order parameter is constant while the induced exchange fields vanish at the S/N$_{\rm so}$ interface. In this case, the is no inverse proximity effect in the structure and the triplet correlations created in the N$_{\rm so}$ layer cannot penetrate back into the S layer. Below the threshold value, $G<G_{\rm th}$, the inverse proximity effect appears which implies an onset of a non-zero exchange field at the S/N$_{\rm so}$ interface. In this regime, the spin-triplet correlations in the N$_{\rm so}$ layer can enter back into the S layer and a spin supercurrent appears. Close to the paramagnetic instability $G=-1$, the SC order parameter at the interface becomes smaller than the induced exchange field at the S/N$_{\rm so}$ interface such that the inverse proximity begins to destroy superconductivity. Consequently, the spin supercurrent intensity decreases.

The presence of the Landau parameter threshold $G_{\rm th}$ can be understood as the onset of inverse proximity effect. For $G>G_{\rm th}$, the S/N$_{\rm so}$ interface is in the rigid boundary condition regime where the inverse proximity effect in the S layer is small. This regime corresponds to $\sigma_{N_{\rm so}}/\sigma_{0}\ll\xi_{N_{\rm so}}/\xi_{0}$ where $\xi_{N_{\rm so}}$ is the coherence length in the N$_{\rm so}$ layer, i.e. this regime is expected to appear when the superconducting layer is in contact with a metal with small conductivity.

In the case without SOC [Fig.~\ref{G_dep} a)], the (zero temperature) N$_{\rm so}$ coherence length 
is given by $\xi_{N_{\rm so}}\approx\sqrt{D/|\nu|}$, where $|\nu|$ is the modulus of the induced exchange field. According to Eq.~(\ref{nu_self}), one can approximate the induced exchange field by $\boldsymbol{\nu} \approx G.\beta_{0}/(1+G)$ where $\beta_{0}$ is the integral $\beta_{0}=\int_{-\infty}^{+\infty}\frac{dE}{2\pi}\text{Im}\left(\boldsymbol{g}_{t}\left(E,x\right)\right)\tanh\left(\frac{E}{2T}\right)$ evaluated at $G=0$. From this, the N$_{\rm so}$ coherence length is expected to tend to infinity at $G=0$ and to decrease to 0 when $G\rightarrow -1$. 
The determination of the exact value of the threshold $G_{\rm th}$ from this toy calculation can be done by
assuming that the threshold between the two regimes appears approximately when $10\sigma_{N_{\rm so}}/\sigma_{0}=\xi_{N_{\rm so}}/\xi_{0}$. We deduce that $G_{\rm th}\approx -1/(1-\Gamma)$ with $\Gamma=\left(10\frac{\sigma_{N_{\rm so}}}{\sigma_{S}}\right)^{2}\frac{\beta_{0}}{\Delta_{0}}$. For parameter set A) , we find $G_{\rm th}^{\rm A}=-0.64$ and for parameter sets B and C $G_{\rm th}^{B}=-0.64$ and $G_{\rm th}^{C}=-0.48$ (with $\beta_{0}=-0.00556$). These values are comparable with self-consistently determined threshold values appearing in Fig.~\ref{G_dep}. For parameter sets B) and C), the presence of SOC can change the threshold value but we did not take these corrections into account in the toy model above.
\end{subsubsection}
\begin{subsubsection}{Dependence on the superconductor thickness}

\begin{figure}[h]
\centering
\includegraphics[width=80mm]{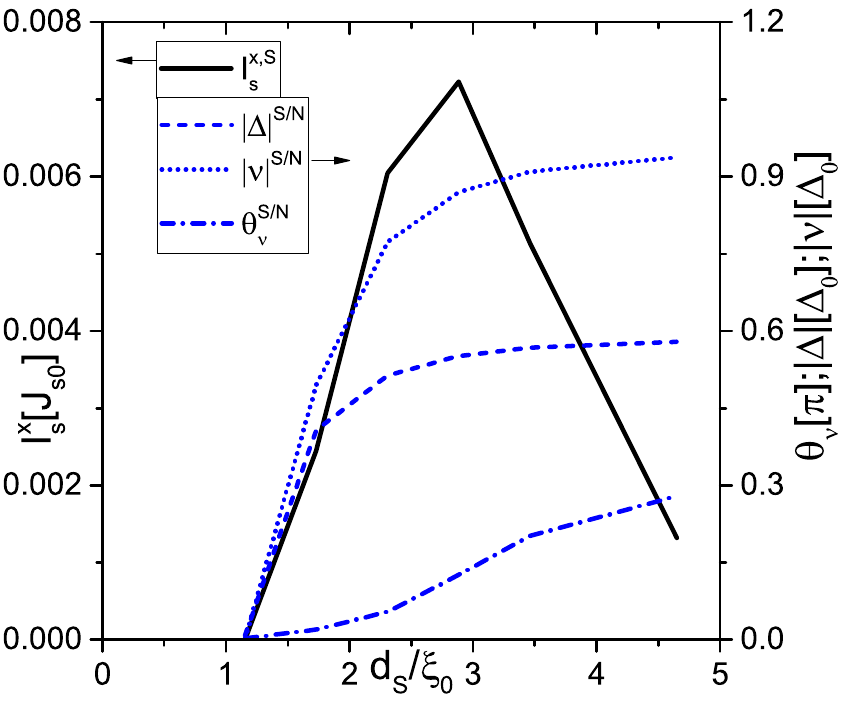}
\vspace{-2mm}
\caption{\label{thickness_dep}(Color online) 
Magnitude of the spin current polarized along the $x$-axis in the S layer, $I_{s}^{X}$ (solid line), magnitude of the SC order parameter $\Delta^{S/N}$ at the S/N interface (dashed line), and modulus $\nu^{S/N}$ and misorientation angle $\theta_{\nu}^{S/N}$ of the Fermi-liquid exchange field at the S/N interface (dotted and dash-dotted line respectively) as function of the superconducting layer thickness for a value of the Landau parameter $G=-0.85$. The other parameters are the same as in parameter set C). The results are very similar to the ones obtained with parameter set B (not shown).}
\end{figure}

The magnitude of the spin supercurrent reaches a maximum for a certain superconducting thickness as shown in Fig.~\ref{thickness_dep}. At small thicknesses, superconductivity is destroyed by the inverse proximity effect implying the vanishing of the spin supercurrent. At large thicknesses, the current intensity decreases because the spin-triplet correlations decay inside the S layer and their intensity at the S/N$_{\rm so}$ interface becomes too small. The decay length is the superconducting coherence length $\xi_{0}$. Between these two regimes, the spin current intensity reaches a maximum where the long-range triplet correlations at the S/N$_{\rm so}$ interface have a maximal intensity. 

The optimal thickness is expected to change with changing the boundary conditions. In the tunneling regime, we expect to stabilize superconductivity and consequently the spin current for smaller S layer thicknesses. Then, we expect the maximum spin current to be reached at smaller thicknesses and to decrease faster with $d_{S}$. Also, the amplitude of triplet correlations flowing out of the F layer should be smaller. 

\end{subsubsection}

\begin{subsubsection}{Dependence on the N$_{\rm so}$ layer thickness}

\begin{figure}[h]
\centering
\includegraphics[width=80mm]{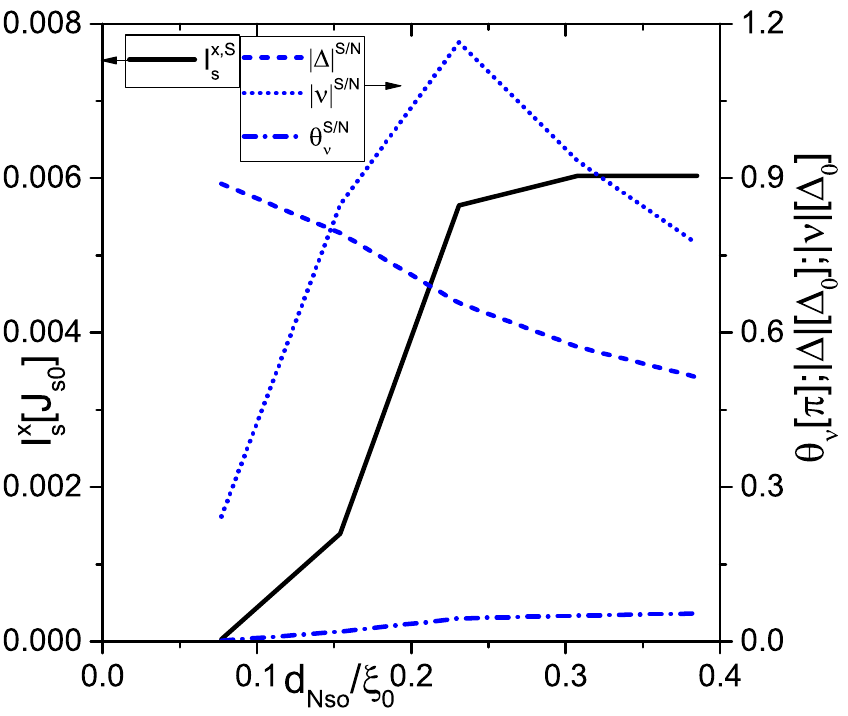}
\vspace{-2mm}
\caption{\label{Pt_layer_thick}(Color online) 
Magnitude of the spin current polarized along the $x$-axis in the S layer, $I_{s}^{X}$ (solid line), magnitude of the SC order parameter $\Delta^{S/N}$ at the S/N interface (dashed line), and modulus $\nu^{S/N}$ and misorientation angle $\theta_{\nu}^{S/N}$ of the Fermi-liquid exchange field at the S/N interface (dotted and dash-dotted line respectively) as function of 
the N$_{\rm so}$ layer thickness for a value of the Landau parameter $G=-0.8$. The other parameters are the same as in parameter set C). The results are very similar to the ones obtained with parameter set B (not shown).}
\end{figure}

As seen in Fig.~\ref{Pt_layer_thick}, the spin supercurrent intensity vanishes in absence of $N_{\rm so}$ and increases with the N$_{\rm so}$ layer thickness. At small thicknesses, the inverse proximity effect is small and the induced exchange field is small. By increasing the thickness of the N$_{\rm so}$ layer, the inverse proximity effect sets in and the SC order parameter at the S/N$_{\rm so}$ interface decreases while the induced exchange field increases. For high thicknesses, the inverse proximity effect is strong which weakens the amplitude of the induced exchange field and of the spin supercurrent amplitude. This behavior demonstrates the crucial role of spin-orbit coupling in the N$_{\rm so}$ layer and of the inverse proximity effect to stabilize the long-range spin-triplet correlations.
\end{subsubsection}

\begin{subsubsection}{Temperature dependence}

\begin{figure}[h]
\centering
\includegraphics[width=80mm]{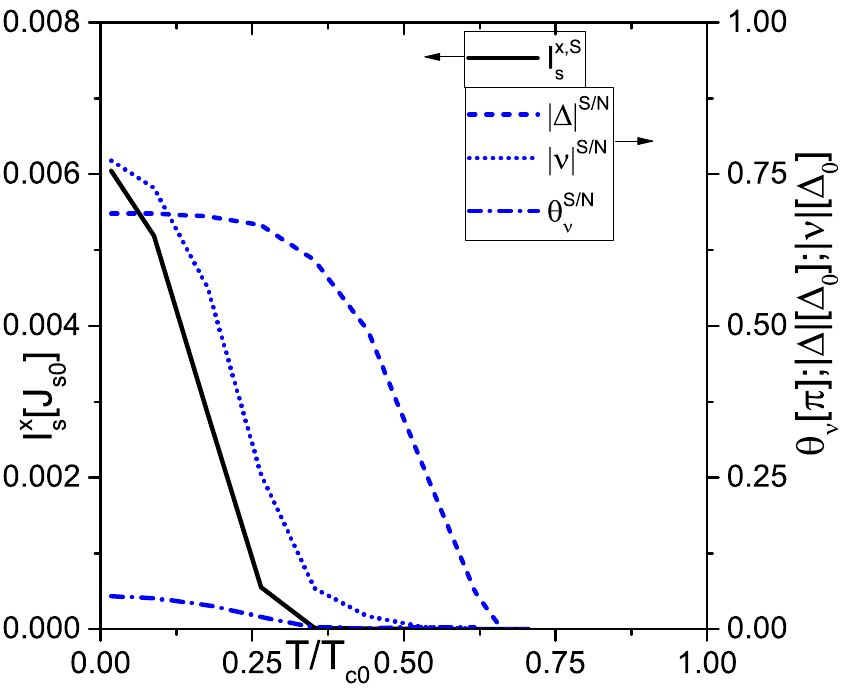}
\vspace{-2mm}
\caption{\label{temp_dep}(Color online) 
Magnitude of the spin current polarized along the $x$-axis in the S layer, $I_{s}^{X}$ (solid line), magnitude of the SC order parameter $\Delta^{S/N}$ at the S/N interface (dashed line), and modulus $\nu^{S/N}$ and misorientation angle $\theta_{\nu}^{S/N}$ of the Fermi-liquid exchange field at the S/N interface (dotted and dash-dotted line respectively) as function of 
temperature for a value of the Landau parameter $G=-0.85$. 
The other parameters are the same as in parameter set C). The results are very similar to the ones obtained with parameter set B (not shown). $T_{c0}$ is the critical temperature of a bare S layer with $\Delta_{0}=1.764T_{c0}$.}
\end{figure}

As shown in Fig.~\ref{temp_dep}, the spin current intensity appears below $T_{\rm c}$ and increases with decreasing temperature.
At low temperature, triplet correlations are maximal overall in the structure such that the current intensity is maximal. 
Close to $T_{\rm c}$, the decrease of the S order parameter amplitude implies a decrease of the proximity effect. Consequently, the induced exchange field decreases as well. As triplet correlations become negligible in this range so does the spin current intensity.
This result is in agreement with the experimental observation \cite{Jeon_NatMat2018} where the injected spin current increases below $T_{\rm c}$ and reaches a maximum at small temperature. 

\end{subsubsection}
\begin{subsubsection}{Dependence on magnitude of spin-orbit coupling}

\begin{figure}[h]
\centering
\includegraphics[width=80mm]{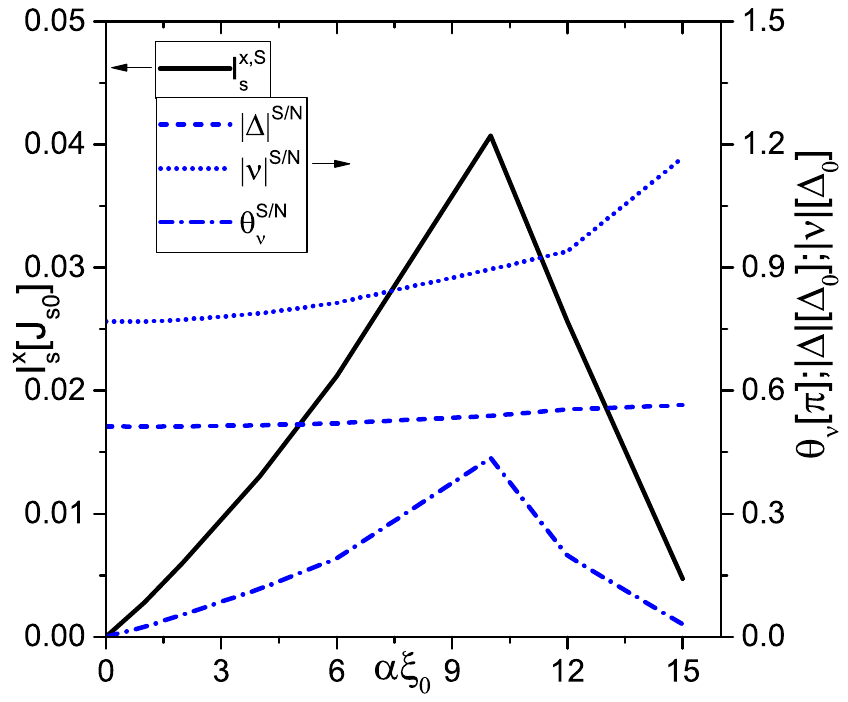}
\vspace{-2mm}
\caption{\label{SOC_dep}(Color online) 
Magnitude of the spin current polarized along the $x$-axis in the S layer, $I_{s}^{X}$ (solid line), magnitude of the SC order parameter $\Delta^{S/N}$ at the S/N interface (dashed line), and modulus $\nu^{S/N}$ and misorientation angle $\theta_{\nu}^{S/N}$ of the Fermi-liquid exchange field at the S/N interface (dotted and dash-dotted line respectively) as function of 
the SOC strength for a value of the Landau parameter $G=-0.85$. The other parameters are the same as in parameter set C). The results are very similar to the ones obtained with parameter set B (not shown).}
\end{figure}

In Fig.~\ref{SOC_dep}, we present the non-monotonic dependence of the magnitude of the spin current with the magnitude in spin-orbit coupling. The magnitude of the spin current exhibits an oscillatory behavior with spin-orbit coupling strength. 
It depends similarly on the SOC strength as the $y$-component of the spin-triplet correlations at the S/N$_{\rm so}$ interface. 
With increasing SOC the spin rotation in the N$_{\rm so}$ layer increases and the $y$-component at the interface increases and reaches a maximum value for $\alpha\xi_{0}\approx 10$ while the magnitude of the $z$-component decreases to its minimum value. 
At this point, the Fermi liquid exchange field at the S/N$_{\rm so}$ interface is oriented perpendicularly to the $z$-axis. Therefore, the spin-rotation process is maximal at the S/N$_{\rm so}$ interface and the amount of long-range triplet correlations is also maximal in this configuration.  For higher value of the SOC, the magnitude of the $y$-component decreases together with the magnitude of the spin supercurrent as the spin rotation process at the S/N$_{\rm so}$ interface is less efficient. It is natural to expect this behavior to be periodic as function of the magnitude of the SOC.
\end{subsubsection}
\end{subsection}

\end{section}

\begin{section}{Effect of spin-flip scattering}
\label{FSNspinflip}
\begin{figure}[h]
\centering
a)\includegraphics[width=80mm]{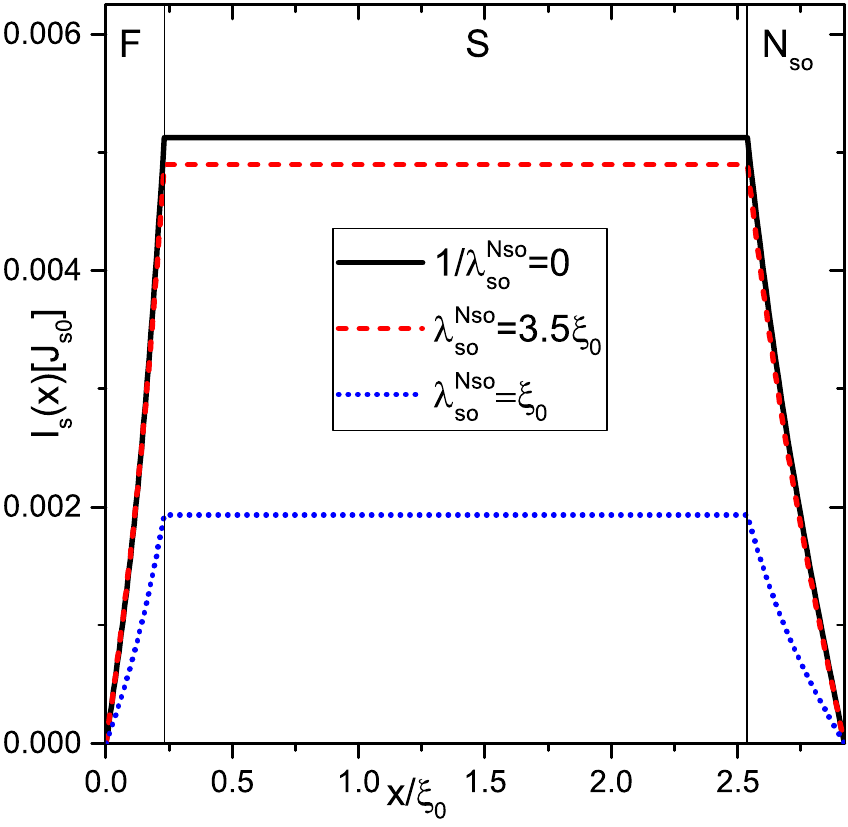}
b)\includegraphics[width=80mm]{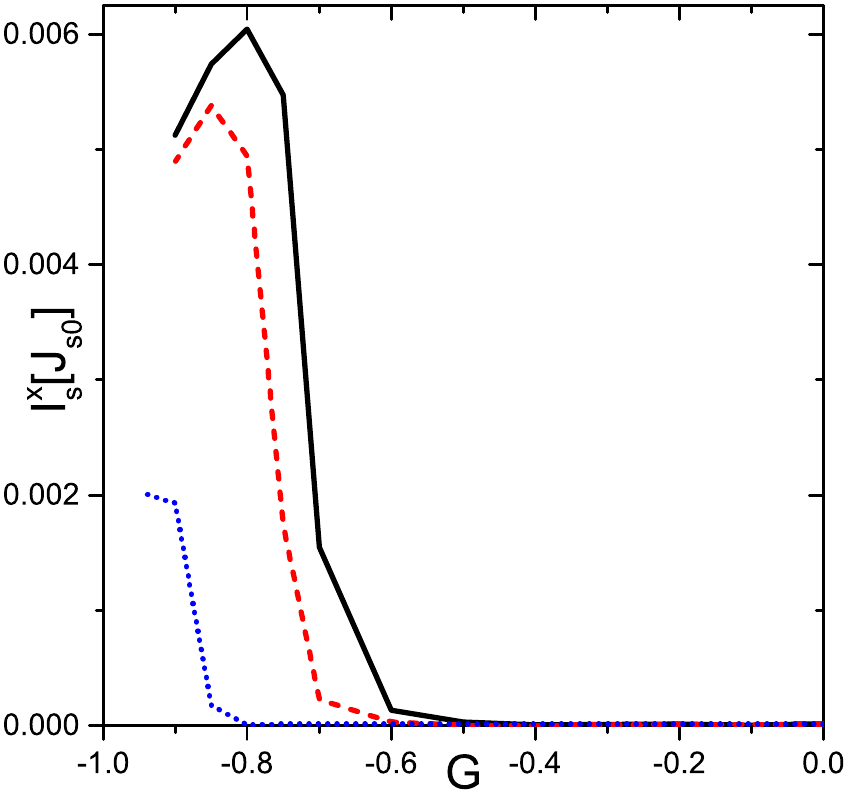}
\vspace{-2mm}
\caption{\label{G_dep_Imp}(Color online) 
In a), the profile of the magnitude of the $x$-axis polarized spin supercurrent in the F/S/N$_{\rm so}$ trilayer is shown for parameter set C) and Landau parameter $G=-0.9$. The spin diffusion length due to spin-orbit scattering, $\lambda_{\rm so}$, is as follows: in the F layer $\lambda_{\rm so}^{F}=0$ while in the the S and N$_{\rm so}$ layers $\lambda_{\rm so}^{S}=\lambda_{\rm so}^{N_{\rm so}}=0 $ (black solid line), $\lambda_{\rm so}^{S}=\lambda_{\rm so}^{N_{\rm so}}=3.5\xi_{0}$ (red solid line) and $\lambda_{\rm so}^{S}=3.5\xi_{0}$, $\lambda_{\rm so}^{N_{\rm so}}=\xi_{0}$ (blue solid line). In b), the magnitude of the spin supercurrent in the S layer is presented as a function of the Landau parameter G (the color legend is the same as in a). The results are very similar to the ones obtained for parameter set B) (not shown).}
\end{figure}
In this section, we study the effect of spin-flip processes on the spin supercurrent and on the long-range spin-triplet correlations in the F/S/N$_{\rm so}$ trilayer. Spin-flip processes are always present in real materials and can change the properties of the S and the F layer; they lead to a decrease of the magnitude of long-range and short-range correlations and may destroy superconductivity.\cite{anderson_jphyschemsol1959,Rusinov_JETP1964} Moreover, spin diffusion process are very important to describe spin pumping experiments.\cite{Morten_EPL2008}

\begin{subsection}{Theoretical implementation}
\label{SE_spinflip}
We consider two types of spin-flip processes: spin flips due to magnetic impurities \cite{anderson_jphyschemsol1959,Morten_EPL2008,Jacobsen_SciRep2016,Jacobsen_PRB2017} and spin flips due to spin-orbit scattering.\cite{Rusinov_JETP1964,Inoue_PRB2017,Morten_EPL2008,Jacobsen_SciRep2016,Jacobsen_PRB2017}. 
In the following, we discuss both processes and present the corresponding self-energies entering the Usadel equations (\ref{eq:UsGen}). 
\begin{paragraph}{Spin flips due to magnetic impurities:}
This spin flip occurs when electronic quasiparticles scatter from the impurities localized magnetic moments. This process breaks time reversal symmetry \cite{anderson_jphyschemsol1959} implying the destruction of singlet Cooper pairs, implying a decay of singlet and triplet pair correlations and a reduction of $T_{\rm c}$ \cite{anderson_jphyschemsol1959,Inoue_PRB2017}. Spin-flip scattering can be taken into account via a self-energy of the following form 
\begin{align}
\hat{\Sigma}^{\text{m}}=\frac{1}{8\tau_{m}}\hat{\boldsymbol{\tau}}.\hat{G}.\hat{\boldsymbol{\tau}}
\end{align}
where $\hat{\boldsymbol{\tau}}$ is the vector of Pauli matrices in spin-Nambu space, $\hat{\boldsymbol{\tau}}=\left(\begin{array}{cc}
\boldsymbol{\sigma} & 0\\
0 & \boldsymbol{\sigma}
\end{array}\right)$, with $\boldsymbol{\sigma}$ the vector of spin Pauli matrices. The pre-factor, $1/8\tau_{m}$, is the impurity scattering rate for and is 
related to the spin diffusion length $\lambda_{m}=\sqrt{\tau_{m}D}$.

\end{paragraph}

\begin{paragraph}{Spin flips due to spin-orbit scattering:}
Spin flip due to spin-orbit scattering happens because due to scattering of electronic quasiparticles from non-magnetic impurities in the presence of spin-orbit coupling  \cite{Rusinov_JETP1964,Inoue_PRB2017}. The strength of this process increases with the atomic number of the scattering impurity \cite{Rusinov_JETP1964,Inoue_PRB2017}.
The corresponding self-energy has in this case the form
\begin{align}
\hat{\Sigma}^{\text{SO}}=\frac{1}{8\tau_{\rm so}}\hat{\boldsymbol{\tau}}.\left(\hat{\tau}_{3}.\hat{G}.\hat{\tau}_{3}\right).\hat{\boldsymbol{\tau}}
\end{align}
where the pre-factor $1/8\tau_{\rm so}$ is the spin-orbit scattering rate and is related to a corresponding spin diffusion length $\lambda_{\rm so}=\sqrt{\tau_{\rm so}D}$. No time-reversal symmetry breaking is involved in this process. Consequently, SO scattering does not affect the singlet correlations and does not affect the critical temperature of the superconductor \cite{Rusinov_JETP1964,Inoue_PRB2017}. 
However, this scattering process induces a decay of the spin-triplet correlations in the trilayer \cite{Inoue_PRB2017,Jacobsen_SciRep2016,Jacobsen_PRB2017}.
\end{paragraph}
\begin{paragraph}{Spin-flip self-energy:}
Both $\hat{\Sigma}^{\text{m}}$ and $\hat{\Sigma}^{\text{SO}}$ create diagonal and off-diagonal contributions to the self-energy.
We combine the spin-flip self-energies as $\hat{\Sigma}^{\text{imp}}=\hat{\Sigma}^{\text{m}}+\hat{\Sigma}^{\text{SO}}$,
\begin{align}
\hat{\Sigma}^{\text{imp}}=\left(\begin{array}{cc}
\Sigma^{\text{imp}} & \Delta^{\text{imp}}\\
\widetilde{\Delta}^{\text{imp}} & \widetilde{\Sigma}^{\text{imp}}
\end{array}\right)
\end{align}
for retarded Green functions with 
\begin{align}
\begin{array}{c}
\Delta^{\text{imp}}=\left(\Gamma_{m}-\Gamma_{\rm so}\right)\left(\sigma_{X}.f.\sigma_{X}-\sigma_{Y}.f.\sigma_{Y}+\sigma_{Z}.f.\sigma_{Z}\right)\\
\Sigma^{\text{imp}}=\left(\Gamma_{m}+\Gamma_{\rm so}\right)\left(\sigma_{X}.g.\sigma_{X}+\sigma_{Y}.g.\sigma_{Y}+\sigma_{Z}.g.\sigma_{Z}\right),
\label{Self_imp}
\end{array}
\end{align}
where $\Gamma_{m}=\frac{1}{8\tau_{m}}$ and $\Gamma_{\rm so}=\frac{1}{8\tau_{\rm so}}$ are the strengths of magnetic and spin-orbit scattering, respectively. 
\end{paragraph}
\end{subsection}

\begin{subsection}{Results}
For our numerical calculations we use parameters appropriate for Py, Nb, and Pt.
In Nb, the spin diffusion length has been estimated at $\lambda^{Nb}_{sd}=48$ nm \cite{Gu_PRB2002,Morten_EPL2008}. The spin diffusion length in Pt, $\lambda_{sd}^{Pt}$, is difficult to determine unambiguously because of the presence of SOC \cite{Freeman_PRL2018}, however the range of values is $1\text{nm}<\lambda_{sd}^{Pt}<14\text{nm}$ \cite{Kurt_APL_2002,Nguyen_PRL2016,Freeman_PRL2018}. Moreover, this spin diffusion length is expected to vary with the Pt layer thickness \cite{Nguyen_PRL2016}. 
The presence of spin-orbit coupling in Pt can affect the measurements of the spin diffusion length \cite{Chen_PRL2015}. In the following, we assume that the N$_{\rm so}$ layer spin diffusion length is the same as the value for bulk Pt, $\lambda_{N_{\rm so}}=14$ nm \cite{Chen_PRL2015}. 

In the following, we focus on spin-orbit scattering and set $\tau_{m}=0$. 
This is appropriate for experiments on devices with pure chemical elements under controlled conditions which implies a small magnetic impurity scattering rate \cite{Jeon_NatMat2018}. Moreover, the strong intrinsic SOC in the Pt layer may imply that spin-orbit scattering is stronger than magnetic spin-flip scattering. 
The inclusion of magnetic impurity scattering does not qualitatively change the results we present, apart from an additional decrease of the $T_{\rm c}$ of the structure.
For simplicity, we consider spin-flip processes only inside the S and N$_{\rm so}$ layers, and neglect spin-flip processes in the F layer, $\lambda_{\rm so}^{F}=0$.

In Fig.~\ref{G_dep_Imp} a) we show the profile of the spin supercurrent in the F/S/N$_{\rm so}$ trilayer for the parameter set C. It can be seen that the spin supercurrent remains constant inside the S layer \cite{Jacobsen_SciRep2016,Jacobsen_PRB2017}, however its magnitude depends on the amount of spin-flip scattering and on the thickness of the superconducting layer.
This is due to the fact that the spin-current is produced non-locally at both interfaces of the structure and the spin needs to stay coherent between the interfaces in order for a torque to be established between the magnetizations in the adjacent materials on both sides. Thus, the magnitude of the spin supercurrent decreases with decreasing spin diffusion length. 

In Fig.~\ref{G_dep_Imp} b), we show the dependence of the magnitude of the spin supercurrent on the Landau parameter $G$ for parameter set C) and for various spin diffusion lengths in the S and N$_{\rm so}$ layers. 
The effect of spin-flip processes is very similar for parameter set B). The main effect of spin-flip scattering is to shift the threshold value $G_{\rm th}$ for the Landau parameter to values closer to the paramagnet instability. This effect is directly related to the destruction of the spin-triplet correlations in the trilayer. Consequently, the spin magnetization amplitude in the N$_{\rm so}$ layer decreases and its amplification to suitable values only occurs for higher threshold values $G_{\rm th}$. In the toy calculation presented in section \ref{Landau_par_sec}, the value of $\beta_{0}$ decreases with the spin diffusion length which implies an increase of the magnitude of $G_{\rm th}$. Moreover, the destruction of spin-triplet correlation leads a decrease of the magnitude of spin current in the S layer.

\end{subsection}

\end{section}

\begin{section}{Discussion}
\label{Discussion}
Our results provide a clear scenario to generate long-range spin-triplet correlation in F/S/N$_{\rm so}$ systems where the N$_{\rm so}$ is a normal metal with Fermi liquid interaction and intrinsic spin-orbit coupling at equilibrium. The short-range spin-triplet correlations generated at the F/S interface decays inside the S layer over the superconducting coherence length $\xi_{0}$. At the S/N$_{\rm so}$ interface and in the N$_{\rm so}$ layer, these short-range correlations are transformed by the spin rotation process induced by the SOC. The Fermi liquid interactions then induce an exchange field inside the N$_{\rm so}$ layer which (a) amplifies the long-range correlations intensity in all the F/S/N$_{\rm so}$ trilayer and (b) is misaligned with the magnetization in F. We find that these long-range correlations are more intense at small S layer thicknesses and we argue that these long-range correlations participate to the injected spin current measured in the FMR experiment \cite{Jeon_NatMat2018}.

From the equations (\ref{equsgam}) and the expression (\ref{second_deriv}), one obtains linearized Usadel equations for the singlet and triplet components which are valid close to $T_{c}$. In this regime, the relations $f\approx -2\pi i \gamma $ hold where $f$ is the anomalous Green function. The linearized Usadel equations in the N$_{\rm so}$ layer for the singlet and triplet components of the anomalous Green functions are
\begin{align}
\begin{array}{c}
\partial_x^2f_s=\frac{2i}{D}\left[\boldsymbol{f}_t \boldsymbol{\nu}-f_sE\right]\\
\partial_x^2\boldsymbol{f}_t+4\boldsymbol{A}_x\times \partial_x\boldsymbol{f}_t+4\sum_k \boldsymbol{A}_k\times\left(\boldsymbol{A}_k\times\boldsymbol{f}_t\right)=\frac{2i}{D}\left[f_s \boldsymbol{\nu}-\boldsymbol{f}_t E\right].
\end{array}
\label{Linearized_Usadel}
\end{align}
From the equation (\ref{Linearized_Usadel}), we can deduce the general condition for the SOC field symmetry that can produce long-range triplet correlations. If we consider that only $f_t^Z$ triplet component initially exists in the N$_{\rm so}$ layer (and constitutes the short-range triplet correlations), we can deduce from the term $4\boldsymbol{A}_x\times \partial_x\boldsymbol{f}_t$ in the equation (\ref{Linearized_Usadel}), that any spin-orbit field involving a nonzero component of the SOC field vector $\boldsymbol{A}_x$ can produce long-range triplet correlations $f_t^X$ and $f_t^Y$. 
From the second term $4\sum_k\left[\boldsymbol{A}_k\times\left(\boldsymbol{A}_k\times\boldsymbol{f}_t\right)\right]$  in the equation (\ref{Linearized_Usadel}), long-range triplet correlations $f_t^X$ and $f_t^Y$  can be produced from short-range triplet correlations $f_t^Z$ if the SOC field exhibits components such that $\boldsymbol{A}_y\sim (\alpha ,0,1)$ or $\sim (0,\alpha ,1)$ or $\boldsymbol{A}_z\sim (\alpha ,0,1)$ or $\sim (0,\alpha ,1)$ or linear combinations of those.
From the above conditions, one can deduce that a spin-orbit coupling of the Rashba type, which involves SOC field of the form $\boldsymbol{A}_x=\boldsymbol{0}$, $\boldsymbol{A}_y= (0 ,0,\alpha)$, $\boldsymbol{A}_z= (0,-\alpha ,0)$, 
or of a linear Dresselhaus type, which involves SOC field of the form $\boldsymbol{A}_x=\boldsymbol{0}$, $\boldsymbol{A}_y= (0 ,\beta,0)$, $\boldsymbol{A}_z= (0 ,0,-\beta)$, cannot produce LR triplet correlations in any obvious way at perfect (single crystalline) interfaces. 
Note that a bulk Dresselhaus coupling is of third order in the momentum, and a linear Dresselhaus coupling can only appear for very thin films where the $x$-component of the momentum is quantized. This is clearly not the case for the setup we consider where the N$_{\rm so}$ layer thickness is of the order of the superconducting coherence length. Therefore, although a SOC field involving both Rashba and linear Dresselhaus SOC is of the form $\boldsymbol{A}_y= (0 ,\beta,\alpha)$ and $\boldsymbol{A}_z= (0 ,-\alpha,-\beta)$ which can produce LR triplet correlations $f_t^Y$, 
this is not a likely mechanism for the FMR experiment.\cite{Jeon_NatMat2018}

In the F/S/N$_{\rm so}$ trilayer, the physics is driven by the F/S and the S/N$_{\rm so}$ interfaces. The short-range triplet correlations $f_t^Z$ are produced at the F/S interfaces while the LR correlations are produced at the S/N$_{\rm so}$ interface. Therefore, a careful study of the boundary conditions, especially at the S/N$_{\rm so}$ interface, is needed to understand the physics of the F/S/N$_{\rm so}$ trilayer. From the expression (\ref{eqboundgamsnso}), the boundary conditions at the S/N$_{\rm so}$ interface for the singlet and triplet components are given by
\begin{align}
\begin{array}{c}
\left[f_{s,t}\right]^{S}=\left[f_{s,t}\right]^{N_{\rm so}}\\
\\
\sigma_{S}\left[\partial_x f_s\right]^{S}=\sigma_{N_{\rm so}}\left[\partial_x f_s\right]^{N_{\rm so}}\\
\sigma_{S}\left[\partial_x f_t^X\right]^{S}=\sigma_{N_{\rm so}}\left[\partial_x f_t^X+2\left(A_x^Y f_t^Z-A_x^Z f_t^Y\right)\right]^{N_{\rm so}} \\
\sigma_{S}\left[\partial_x f_t^Y\right]^{S}=\sigma_{N_{\rm so}}\left[\partial_x f_t^Y+2\left(A_x^Z f_t^X-A_x^X f_t^Z\right)\right]^{N_{\rm so}}\\
\sigma_{S}\left[\partial_x f_t^Z\right]^{S}=\sigma_{N_{\rm so}}\left[\partial_x f_t^Z+2\left(A_x^X f_t^Y-A_x^Y f_t^X\right)\right]^{N_{\rm so}}
\end{array}
\label{Linearized_Bond}
\end{align}
where $\left[f_{s}\right]^{A}$ refers to the singlet anomalous Green functions on the A side of the interface and $\left[f_{t}^i\right]^{A}$ refers to the triplet anomalous Green functions in the $i$ spin direction on the A side of the interface and the SOC field vector is given by $\boldsymbol{A}_x= (A_x^X ,A_x^Y,A_x^Z)$. In our study, we consider a SOC field vector of the form $\boldsymbol{A}_x= (\alpha ,0,0)$ which simplifies the equations (\ref{Linearized_Bond}) to  
\begin{align}
\begin{array}{c}
\left[f_{s,t}\right]^{S}=\left[f_{s,t}\right]^{N_{\rm so}}\\
\\
\sigma_{S}\left[\partial_x f_s\right]^{S}=\sigma_{N_{\rm so}}\left[\partial_x f_s\right]^{N_{\rm so}}\\
\sigma_{S}\left[\partial_x f_t^X\right]^{S}=\sigma_{N_{\rm so}}\left[\partial_x f_t^X\right]^{N_{\rm so}} \\
\sigma_{S}\left[\partial_x f_t^Y\right]^{S}=\sigma_{N_{\rm so}}\left[\partial_x f_t^Y-2\alpha f_t^Z\right]^{N_{\rm so}}\\
\sigma_{S}\left[\partial_x f_t^Z\right]^{S}=\sigma_{N_{\rm so}}\left[\partial_x f_t^Z+2\alpha f_t^Y\right]^{N_{\rm so}}
\end{array}
\label{Linearized_Bond_Axx}
\end{align}
Here, we can distinguish two regimes. For a zero Landau parameter (G=0), the induced exchange field in the N$_{\rm so}$ layer vanishes and we learn from the Usadel equations in the superconducting layer that the derivative of $f_{t}^{Y}$ triplet correlations vanishes at the S/N$_{\rm so}$ interface, $\left[\partial_x f_t^Y\right]^{S}=0$. In this regime, the absence of induced exchange field in the N$_{\rm so}$ layer implies that the magnitude of $f_{t}^{Y}$ triplet correlations is negligible at the S/N$_{\rm so}$ interface, $\left[f_t^Y\right]^{S}\approx 0$. Both conditions impose a constraint on the derivative of the triplet components at the N$_{\rm so}$ interface which take the form
\begin{align}
\begin{array}{c}
\left[\partial_x f_t^Y\right]^{N_{\rm so}}=2\alpha\left[ f_t^Z\right]^{N_{\rm so}}\\
\sigma_{S}\left[\partial_x f_t^Z\right]^{S} \approx \sigma_{N_{\rm so}}\left[\partial_x f_t^Z\right]^{N_{\rm so}}
\end{array}
\label{constraint}
\end{align}
The relations (\ref{constraint}) are only valid when both the $f_t^{Y}$ triplet correlations and its derivative vanish in the superconductor at the S/N$_{\rm so}$ interface. In this regime, we observe that the $f_t^Z$ triplet component directly controls the derivative of the $f_t^Y$ component. However, these conditions does not hold if an exchange field is induced in the N$_{\rm so}$ layer. In this case, the magnitude of $f_t^{Y}$ and $f_t^{Z}$ triplet correlations are affected by the spin-mixing process. Therefore, for a non-zero Landau parameter, $G\neq 0$, we have two distinct regimes. In the regime where the inverse proximity effect is weak (for $|G|<|G_{\rm th}|$), the conditions (\ref{constraint}) still apply and the amount of LR and SR triplet correlations produced in the N$_{\rm so}$ remains small. On the other hand, for higher values of the Landau parameter, $|G|>|G_{\rm th}|$, the spin-mixing process coming from the appearance of a misaligned Landau mean field $\nu$  in the N$_{\rm so}$ layer implies that the conditions (\ref{constraint}) are no more valid. In this case, an inverse proximity effect appears and both the $f_t^{Y}$ triplet correlations and its derivative no longer vanish at the S/N$_{\rm so}$ interface. This result emphasizes the crucial role of the inverse proximity effect on the physics of the F/S/N$_{\rm so}$ trilayer.

Regarding the FMR experiment, the most straightforward explanation would be
to have a spin-orbit coupling with a nonzero component of the vector $\boldsymbol{A}_x$, which is the one relevant for current transport in $x$-direction.
One could expect to produce an out-of-plane coupling from the spin-orbit torque originating from the FMR-induced magnetization precession in the F layer \cite{Garello_nnano_2013,Hide_nnano_2014}. It has been demonstrated that a non-equilibrium situation induces such an out-of-plane component to SOC for both Rashba and Dresselhaus SOC \cite{Garello_nnano_2013,Hide_nnano_2014}. We estimate that in our case this effect would be negligibly small, due to the tiny tip angles for the precessing magnetization in the FMR experiment. 
In the FMR experiment, the spin-polarized chemical potential induced by the F layer precession is proportional to the FMR frequency $f_{\rm FMR}$ (around 20GHz). For these frequencies, FMR produces a spin-resolved chemical potential much smaller than the superconducting gap $hf_{\rm FMR}\ll\Delta_{0}$. Therefore, we expect that the main qualitative picture of the FMR experiment \cite{Jeon_NatMat2018} is captured already at the level of an equilibrium picture.

For interfacial Rashba spin-orbit interactions to be effective, the interface needs to exhibit mesoscopic facets which are misaligned within the average interface plane. 
This would then produce a non-zero component of $\boldsymbol{A}_x$.
In this case, although the Rashba SO field averaged over the interface has its spins all perpendicular to the $X$-direction, this is not the case on a scale comparable to the superconducting coherence length. It is therefore natural to assume that an out-of-plane component is present in an appreciable fraction of the interface area.
A perfect interface, on the other hand, would in this case be detrimental to the effect.
Alternatively, a bulk intrinsic spin-orbit interaction with a nonzero $\boldsymbol{A}_x$ would always be sufficient for the effect to occur.

Note that the specific form of SOC in Eq. (\ref{SOCsym}) is not the only one that can provide long-range spin-triplet correlations. For example, our results do not change if we consider a SOC of the form $\boldsymbol{A}_{x}=(0,\alpha,0)$.
In this case, the long-range spin-triplet correlations are spin-polarized along the $x$-axis and the additional spin magnetization in the N$_{\rm so}$ layer orients along the $x$-axis. With this, the spin current is spin-polarized along the $y$-axis. The 
results presented in sections \ref{FSN} and \ref{FSNspinflip} would be the same with the appropriate re-naming of spin-coordinates. 
Other forms of spin-orbit coupling could also provide a finite spin-current.

Our study provides a mechanism for generation of long-range spin-triplet correlation in an F/S/N$_{\rm so}$ trilayer. Therefore, the injection of a spin current via an F layer magnetization precession can lead to the transport of spin current via spin-triplet Cooper pairs across S. The opening of this triplet channel below $T_{\rm c}$ could explain the increase of the injected spin current in the FMR experiment in Pt/Nb/Py/Nb/Pt pentalayers \cite{Jeon_NatMat2018}. These spin-triplet correlations are strong, especially at low temperature. 

The onset of spin-triplet correlation strongly depends on the Landau parameter value $G$. 
We expect the value of the Landau parameter to be reasonably close to the paramagnet instability in Pt, Ta, W or Pd. These elements exhibit a paramagnet spin susceptibility at a low temperature that exhibit a strong Stoner enhancement ($\approx 3.9$ in Pt). This value is appropriate for bulk Pt but could be higher if the Pt is confined in a thin layer \cite{Nguyen_PRL2016}. Our calculations demonstrate that such a metal is a good candidate for the appearance of long-range triplet correlations in F/S/N$_{\rm so}$ trilayers. 
The exact value of the Landau parameter in Pt is not known and may also depend on the thickness of the Pt layer \cite{Zhang_PRB2015}. In the FMR experiment \cite{Jeon_NatMat2018}, the Pt layer is thin and the $G$ value might be enhanced above its bulk value. 
Spin-orbit scattering essentially moves the threshold for the magnitude of the Landau parameter $G$ to higher values.
The inclusion of spin-flip scattering is important to explain the physics of spin-pumping and the non-equilibrium properties of such multilayers. Here, we demonstrate that the triplet channel below $T_{\rm c}$ is not destroyed by modest amounts of spin-flip processes. As the induced magnetism in metals like Pt, Ta, W, or Pd can be enhanced below $T_{\rm c}$, the existence of such a channel is definitely a good candidate to explain the FMR experiments.


The dependence of our mechanism on the boundary conditions provides the possibility to design two experiments to test our mechanism. The first experiment should be to add an insulator at the S/F interface implying the S/F interface to be in the tunneling limit. In this limit, the magnitude of the spin mixing and the amplitude of the short-range triplet correlations should decrease in the trilayer. Therefore, our theory predicts that the effect to disappear. The second experiment would be to add a small insulating layer at the S/N$_{\rm so}$ interface only. In this case, the short-range triplet correlation produced at the S/F interface should have the same intensity as in the non-tunneling limit. However, our model predicts that the magnitude of the long-range triplet correlations should be weakened. Our predictions of the S and N$_{\rm so}$ thickness dependence of the long-range triplet correlations agrees with the thickness dependence of the injected spin current in the FMR experiment \cite{Jeon_NatMat2018}.

Moreover, our study provides a guideline for the choice of the materials composing a multilayer. Indeed, we demonstrate that heavy atom metals close to a paramagnetic instability which are subject to strong Fermi-liquid interactions and spin-orbit coupling can completely change the physics of such multilayer by inducing non-locally Fermi-liquid mean fields across superconducting spacers. 
The inclusion of heavy atom metals and their particular properties can change also the properties of Josephson junction \cite{Satchell_PRB2018}.


\end{section}

\begin{section}{Conclusion}
\label{conclusion}
We have demonstrated that spin-orbit coupling in conjunction with Fermi-liquid interactions in an N$_{\rm so}$ layer when coupled via a superconducting spacer to a ferromagnet gives rise to a non-locally induced exchange field in the normal layer that is misaligned with the magnetization in the ferromagnet, and thus leads to an equilibrium spin-torque giving rise to pure spin currents. These spin currents are carried by equal-spin triplet pairs that are long-range in the ferromagnet. 
The induced magnetism and the long-range triplet correlation are driven by the S/F proximity effect and the SOC in the N$_{\rm so}$ layer. 
Our results give a possible explanation for a recent FMR experiment in such structures \cite{Jeon_NatMat2018}.
We demonstrate that the effect survives the presence of reasonable spin-flip processes in the S and N$_{\rm so}$ layers.
Our results demonstrate that Fermi liquid interactions, which in bulk materials lead usually to renormalization on a quantitative level, can lead to drastic qualitative changes in non-local situations that dominate the physics of superconducting spintronics. We anticipate such phenomena to play an important role in the future design of superconducting spintronics devices.
\end{section}

\begin{acknowledgments}
We would like to thank M. Blamire, A. Buzdin, C. Ciccarelli, L. Cohen, K.-R. Jeon, H. Kurebayashi, and J. Robinson for fruitful discussions. This work has been supported by the EPSRC Programme Grant EP/N017242/1.
\end{acknowledgments}

\bibliographystyle{apsrev4-1}
\bibliography{Supra}

\begin{thebibliography}{104}%
\makeatletter
\providecommand \@ifxundefined [1]{%
 \@ifx{#1\undefined}
}%
\providecommand \@ifnum [1]{%
 \ifnum #1\expandafter \@firstoftwo
 \else \expandafter \@secondoftwo
 \fi
}%
\providecommand \@ifx [1]{%
 \ifx #1\expandafter \@firstoftwo
 \else \expandafter \@secondoftwo
 \fi
}%
\providecommand \natexlab [1]{#1}%
\providecommand \enquote  [1]{``#1''}%
\providecommand \bibnamefont  [1]{#1}%
\providecommand \bibfnamefont [1]{#1}%
\providecommand \citenamefont [1]{#1}%
\providecommand \href@noop [0]{\@secondoftwo}%
\providecommand \href [0]{\begingroup \@sanitize@url \@href}%
\providecommand \@href[1]{\@@startlink{#1}\@@href}%
\providecommand \@@href[1]{\endgroup#1\@@endlink}%
\providecommand \@sanitize@url [0]{\catcode `\\12\catcode `\$12\catcode
  `\&12\catcode `\#12\catcode `\^12\catcode `\_12\catcode `\%12\relax}%
\providecommand \@@startlink[1]{}%
\providecommand \@@endlink[0]{}%
\providecommand \url  [0]{\begingroup\@sanitize@url \@url }%
\providecommand \@url [1]{\endgroup\@href {#1}{\urlprefix }}%
\providecommand \urlprefix  [0]{URL }%
\providecommand \Eprint [0]{\href }%
\providecommand \doibase [0]{http://dx.doi.org/}%
\providecommand \selectlanguage [0]{\@gobble}%
\providecommand \bibinfo  [0]{\@secondoftwo}%
\providecommand \bibfield  [0]{\@secondoftwo}%
\providecommand \translation [1]{[#1]}%
\providecommand \BibitemOpen [0]{}%
\providecommand \bibitemStop [0]{}%
\providecommand \bibitemNoStop [0]{.\EOS\space}%
\providecommand \EOS [0]{\spacefactor3000\relax}%
\providecommand \BibitemShut  [1]{\csname bibitem#1\endcsname}%
\let\auto@bib@innerbib\@empty
\bibitem [{\citenamefont {Eschrig}(2011)}]{Eschrig_PhysTod2011}%
  \BibitemOpen
  \bibfield  {author} {\bibinfo {author} {\bibfnamefont {M.}~\bibnamefont
  {Eschrig}},\ }\href {https://doi.org/10.1063/1.3541944} {\bibfield  {journal}
  {\bibinfo  {journal} {Phys. Today}\ }\textbf {\bibinfo {volume} {64}},\
  \bibinfo {pages} {43} (\bibinfo {year} {2011})}\BibitemShut {NoStop}%
\bibitem [{\citenamefont {Eschrig}(2015)}]{Eschrig_RepProgPhys2015}%
  \BibitemOpen
  \bibfield  {author} {\bibinfo {author} {\bibfnamefont {M.}~\bibnamefont
  {Eschrig}},\ }\href {http://stacks.iop.org/0034-4885/78/i=10/a=104501}
  {\bibfield  {journal} {\bibinfo  {journal} {Rep. Prog. Phys.}\ }\textbf
  {\bibinfo {volume} {78}},\ \bibinfo {pages} {104501} (\bibinfo {year}
  {2015})}\BibitemShut {NoStop}%
\bibitem [{\citenamefont {Linder}\ and\ \citenamefont
  {Robinson}(2015)}]{Robinson_Linder_2015}%
  \BibitemOpen
  \bibfield  {author} {\bibinfo {author} {\bibfnamefont {J.}~\bibnamefont
  {Linder}}\ and\ \bibinfo {author} {\bibfnamefont {J.~W.~A.}\ \bibnamefont
  {Robinson}},\ }\href {http://dx.doi.org/10.1038/nphys3242} {\bibfield
  {journal} {\bibinfo  {journal} {Nat. Phys.}\ }\textbf {\bibinfo {volume}
  {11}},\ \bibinfo {pages} {307} (\bibinfo {year} {2015})}\BibitemShut
  {NoStop}%
\bibitem [{\citenamefont {Izyumov}\ \emph {et~al.}(2002)\citenamefont
  {Izyumov}, \citenamefont {Proshin},\ and\ \citenamefont
  {Khusainov}}]{Izyumov2002}%
  \BibitemOpen
  \bibfield  {author} {\bibinfo {author} {\bibfnamefont {Y.~A.}\ \bibnamefont
  {Izyumov}}, \bibinfo {author} {\bibfnamefont {Y.~N.}\ \bibnamefont
  {Proshin}}, \ and\ \bibinfo {author} {\bibfnamefont {M.~G.}\ \bibnamefont
  {Khusainov}},\ }\href {http://stacks.iop.org/1063-7869/45/i=2/a=R01}
  {\bibfield  {journal} {\bibinfo  {journal} {Physics-Uspekhi}\ }\textbf
  {\bibinfo {volume} {45}},\ \bibinfo {pages} {109} (\bibinfo {year}
  {2002})}\BibitemShut {NoStop}%
\bibitem [{\citenamefont {Eschrig}\ \emph {et~al.}(2004)\citenamefont
  {Eschrig}, \citenamefont {Kopu}, \citenamefont {Konstandin}, \citenamefont
  {Cuevas}, \citenamefont {Fogelstr{\"o}m},\ and\ \citenamefont
  {Sch{\"o}n}}]{Eschrig_adv2004}%
  \BibitemOpen
  \bibfield  {author} {\bibinfo {author} {\bibfnamefont {M.}~\bibnamefont
  {Eschrig}}, \bibinfo {author} {\bibfnamefont {J.}~\bibnamefont {Kopu}},
  \bibinfo {author} {\bibfnamefont {A.}~\bibnamefont {Konstandin}}, \bibinfo
  {author} {\bibfnamefont {J.~C.}\ \bibnamefont {Cuevas}}, \bibinfo {author}
  {\bibfnamefont {M.}~\bibnamefont {Fogelstr{\"o}m}}, \ and\ \bibinfo {author}
  {\bibfnamefont {G.}~\bibnamefont {Sch{\"o}n}},\ }in\ \href@noop {} {\emph
  {\bibinfo {booktitle} {Advances in Solid State Physics}}}\ (\bibinfo
  {publisher} {Springer},\ \bibinfo {year} {2004})\ pp.\ \bibinfo {pages}
  {533--545}\BibitemShut {NoStop}%
\bibitem [{\citenamefont {Golubov}\ \emph {et~al.}(2004)\citenamefont
  {Golubov}, \citenamefont {Kupriyanov},\ and\ \citenamefont
  {Il'ichev}}]{Golubov_RevMod2004}%
  \BibitemOpen
  \bibfield  {author} {\bibinfo {author} {\bibfnamefont {A.~A.}\ \bibnamefont
  {Golubov}}, \bibinfo {author} {\bibfnamefont {M.~Y.}\ \bibnamefont
  {Kupriyanov}}, \ and\ \bibinfo {author} {\bibfnamefont {E.}~\bibnamefont
  {Il'ichev}},\ }\href {\doibase 10.1103/RevModPhys.76.411} {\bibfield
  {journal} {\bibinfo  {journal} {Rev. Mod. Phys.}\ }\textbf {\bibinfo {volume}
  {76}},\ \bibinfo {pages} {411} (\bibinfo {year} {2004})}\BibitemShut
  {NoStop}%
\bibitem [{\citenamefont {Bergeret}\ \emph {et~al.}(2005)\citenamefont
  {Bergeret}, \citenamefont {Volkov},\ and\ \citenamefont
  {Efetov}}]{Bergeret_RevMod2005}%
  \BibitemOpen
  \bibfield  {author} {\bibinfo {author} {\bibfnamefont {F.~S.}\ \bibnamefont
  {Bergeret}}, \bibinfo {author} {\bibfnamefont {A.~F.}\ \bibnamefont
  {Volkov}}, \ and\ \bibinfo {author} {\bibfnamefont {K.~B.}\ \bibnamefont
  {Efetov}},\ }\href {\doibase 10.1103/RevModPhys.77.1321} {\bibfield
  {journal} {\bibinfo  {journal} {Rev. Mod. Phys.}\ }\textbf {\bibinfo {volume}
  {77}},\ \bibinfo {pages} {1321} (\bibinfo {year} {2005})}\BibitemShut
  {NoStop}%
\bibitem [{\citenamefont {Buzdin}(2005)}]{Buzdin_RevMod2005}%
  \BibitemOpen
  \bibfield  {author} {\bibinfo {author} {\bibfnamefont {A.~I.}\ \bibnamefont
  {Buzdin}},\ }\href {\doibase 10.1103/RevModPhys.77.935} {\bibfield  {journal}
  {\bibinfo  {journal} {Rev. Mod. Phys.}\ }\textbf {\bibinfo {volume} {77}},\
  \bibinfo {pages} {935} (\bibinfo {year} {2005})}\BibitemShut {NoStop}%
\bibitem [{\citenamefont {Lyuksyutov}\ and\ \citenamefont
  {Pokrovsky}(2005)}]{Lyuksyutov2005}%
  \BibitemOpen
  \bibfield  {author} {\bibinfo {author} {\bibfnamefont {I.~F.}\ \bibnamefont
  {Lyuksyutov}}\ and\ \bibinfo {author} {\bibfnamefont {V.~L.}\ \bibnamefont
  {Pokrovsky}},\ }\href {\doibase 10.1080/00018730500057536} {\bibfield
  {journal} {\bibinfo  {journal} {Adv. Phys.}\ }\textbf {\bibinfo {volume}
  {54}},\ \bibinfo {pages} {67} (\bibinfo {year} {2005})}\BibitemShut {NoStop}%
\bibitem [{\citenamefont {Blamire}\ and\ \citenamefont
  {Robinson}(2014)}]{BlamireRobinson_JPhys2014}%
  \BibitemOpen
  \bibfield  {author} {\bibinfo {author} {\bibfnamefont {M.~G.}\ \bibnamefont
  {Blamire}}\ and\ \bibinfo {author} {\bibfnamefont {J.~W.~A.}\ \bibnamefont
  {Robinson}},\ }\href {http://stacks.iop.org/0953-8984/26/i=45/a=453201}
  {\bibfield  {journal} {\bibinfo  {journal} {J. Phys. Condens. Matter}\
  }\textbf {\bibinfo {volume} {26}},\ \bibinfo {pages} {453201} (\bibinfo
  {year} {2014})}\BibitemShut {NoStop}%
\bibitem [{\citenamefont {Birge}(2018)}]{Birge_PhilTransA2018}%
  \BibitemOpen
  \bibfield  {author} {\bibinfo {author} {\bibfnamefont {N.~O.}\ \bibnamefont
  {Birge}},\ }\href
  {http://rsta.royalsocietypublishing.org/content/376/2125/20150150} {\bibfield
   {journal} {\bibinfo  {journal} {Philos. Trans. Royal Soc. A}\ }\textbf
  {\bibinfo {volume} {376}} (\bibinfo {year} {2018})}\BibitemShut {NoStop}%
\bibitem [{\citenamefont {Tokuyasu}\ \emph {et~al.}(1988)\citenamefont
  {Tokuyasu}, \citenamefont {Sauls},\ and\ \citenamefont
  {Rainer}}]{Sauls_PRB1988}%
  \BibitemOpen
  \bibfield  {author} {\bibinfo {author} {\bibfnamefont {T.}~\bibnamefont
  {Tokuyasu}}, \bibinfo {author} {\bibfnamefont {J.~A.}\ \bibnamefont {Sauls}},
  \ and\ \bibinfo {author} {\bibfnamefont {D.}~\bibnamefont {Rainer}},\ }\href
  {\doibase 10.1103/PhysRevB.38.8823} {\bibfield  {journal} {\bibinfo
  {journal} {Phys. Rev. B}\ }\textbf {\bibinfo {volume} {38}},\ \bibinfo
  {pages} {8823} (\bibinfo {year} {1988})}\BibitemShut {NoStop}%
\bibitem [{\citenamefont {Fogelstr\"om}(2000)}]{Fogelstrom00}%
  \BibitemOpen
  \bibfield  {author} {\bibinfo {author} {\bibfnamefont {M.}~\bibnamefont
  {Fogelstr\"om}},\ }\href {\doibase 10.1103/PhysRevB.62.11812} {\bibfield
  {journal} {\bibinfo  {journal} {Phys. Rev. B}\ }\textbf {\bibinfo {volume}
  {62}},\ \bibinfo {pages} {11812} (\bibinfo {year} {2000})}\BibitemShut
  {NoStop}%
\bibitem [{\citenamefont {Barash}\ and\ \citenamefont
  {Bobkova}(2002)}]{BobkovaBobkov02}%
  \BibitemOpen
  \bibfield  {author} {\bibinfo {author} {\bibfnamefont {Y.~S.}\ \bibnamefont
  {Barash}}\ and\ \bibinfo {author} {\bibfnamefont {I.~V.}\ \bibnamefont
  {Bobkova}},\ }\href {\doibase 10.1103/PhysRevB.65.144502} {\bibfield
  {journal} {\bibinfo  {journal} {Phys. Rev. B}\ }\textbf {\bibinfo {volume}
  {65}},\ \bibinfo {pages} {144502} (\bibinfo {year} {2002})}\BibitemShut
  {NoStop}%
\bibitem [{\citenamefont {Eschrig}\ \emph {et~al.}(2003)\citenamefont
  {Eschrig}, \citenamefont {Kopu}, \citenamefont {Cuevas},\ and\ \citenamefont
  {Sch\"on}}]{Eschrig_PRL2003}%
  \BibitemOpen
  \bibfield  {author} {\bibinfo {author} {\bibfnamefont {M.}~\bibnamefont
  {Eschrig}}, \bibinfo {author} {\bibfnamefont {J.}~\bibnamefont {Kopu}},
  \bibinfo {author} {\bibfnamefont {J.~C.}\ \bibnamefont {Cuevas}}, \ and\
  \bibinfo {author} {\bibfnamefont {G.}~\bibnamefont {Sch\"on}},\ }\href
  {\doibase 10.1103/PhysRevLett.90.137003} {\bibfield  {journal} {\bibinfo
  {journal} {Phys. Rev. Lett.}\ }\textbf {\bibinfo {volume} {90}},\ \bibinfo
  {pages} {137003} (\bibinfo {year} {2003})}\BibitemShut {NoStop}%
\bibitem [{\citenamefont {Bergeret}\ \emph {et~al.}(2001)\citenamefont
  {Bergeret}, \citenamefont {Volkov},\ and\ \citenamefont
  {Efetov}}]{Bergeret_PRL2001}%
  \BibitemOpen
  \bibfield  {author} {\bibinfo {author} {\bibfnamefont {F.~S.}\ \bibnamefont
  {Bergeret}}, \bibinfo {author} {\bibfnamefont {A.~F.}\ \bibnamefont
  {Volkov}}, \ and\ \bibinfo {author} {\bibfnamefont {K.~B.}\ \bibnamefont
  {Efetov}},\ }\href {\doibase 10.1103/PhysRevLett.86.4096} {\bibfield
  {journal} {\bibinfo  {journal} {Phys. Rev. Lett.}\ }\textbf {\bibinfo
  {volume} {86}},\ \bibinfo {pages} {4096} (\bibinfo {year}
  {2001})}\BibitemShut {NoStop}%
\bibitem [{\citenamefont {L\"ofwander}\ \emph {et~al.}(2005)\citenamefont
  {L\"ofwander}, \citenamefont {Champel}, \citenamefont {Durst},\ and\
  \citenamefont {Eschrig}}]{Champel_PRL2005}%
  \BibitemOpen
  \bibfield  {author} {\bibinfo {author} {\bibfnamefont {T.}~\bibnamefont
  {L\"ofwander}}, \bibinfo {author} {\bibfnamefont {T.}~\bibnamefont
  {Champel}}, \bibinfo {author} {\bibfnamefont {J.}~\bibnamefont {Durst}}, \
  and\ \bibinfo {author} {\bibfnamefont {M.}~\bibnamefont {Eschrig}},\ }\href
  {\doibase 10.1103/PhysRevLett.95.187003} {\bibfield  {journal} {\bibinfo
  {journal} {Phys. Rev. Lett.}\ }\textbf {\bibinfo {volume} {95}},\ \bibinfo
  {pages} {187003} (\bibinfo {year} {2005})}\BibitemShut {NoStop}%
\bibitem [{\citenamefont {Houzet}\ and\ \citenamefont
  {Buzdin}(2007)}]{houzet_PRB2007}%
  \BibitemOpen
  \bibfield  {author} {\bibinfo {author} {\bibfnamefont {M.}~\bibnamefont
  {Houzet}}\ and\ \bibinfo {author} {\bibfnamefont {A.~I.}\ \bibnamefont
  {Buzdin}},\ }\href {\doibase 10.1103/PhysRevB.76.060504} {\bibfield
  {journal} {\bibinfo  {journal} {Phys. Rev. B}\ }\textbf {\bibinfo {volume}
  {76}},\ \bibinfo {pages} {060504} (\bibinfo {year} {2007})}\BibitemShut
  {NoStop}%
\bibitem [{\citenamefont {Halterman}\ \emph {et~al.}(2008)\citenamefont
  {Halterman}, \citenamefont {Valls},\ and\ \citenamefont
  {Barsic}}]{Halterman_PRB2008}%
  \BibitemOpen
  \bibfield  {author} {\bibinfo {author} {\bibfnamefont {K.}~\bibnamefont
  {Halterman}}, \bibinfo {author} {\bibfnamefont {O.~T.}\ \bibnamefont
  {Valls}}, \ and\ \bibinfo {author} {\bibfnamefont {P.~H.}\ \bibnamefont
  {Barsic}},\ }\href {\doibase 10.1103/PhysRevB.77.174511} {\bibfield
  {journal} {\bibinfo  {journal} {Phys. Rev. B}\ }\textbf {\bibinfo {volume}
  {77}},\ \bibinfo {pages} {174511} (\bibinfo {year} {2008})}\BibitemShut
  {NoStop}%
\bibitem [{\citenamefont {Champel}\ and\ \citenamefont
  {Eschrig}(2005{\natexlab{a}})}]{Champel_PRB2005}%
  \BibitemOpen
  \bibfield  {author} {\bibinfo {author} {\bibfnamefont {T.}~\bibnamefont
  {Champel}}\ and\ \bibinfo {author} {\bibfnamefont {M.}~\bibnamefont
  {Eschrig}},\ }\href {\doibase 10.1103/PhysRevB.71.220506} {\bibfield
  {journal} {\bibinfo  {journal} {Phys. Rev. B}\ }\textbf {\bibinfo {volume}
  {71}},\ \bibinfo {pages} {220506} (\bibinfo {year}
  {2005}{\natexlab{a}})}\BibitemShut {NoStop}%
\bibitem [{\citenamefont {Champel}\ and\ \citenamefont
  {Eschrig}(2005{\natexlab{b}})}]{Champel_PRB2005b}%
  \BibitemOpen
  \bibfield  {author} {\bibinfo {author} {\bibfnamefont {T.}~\bibnamefont
  {Champel}}\ and\ \bibinfo {author} {\bibfnamefont {M.}~\bibnamefont
  {Eschrig}},\ }\href {\doibase 10.1103/PhysRevB.72.054523} {\bibfield
  {journal} {\bibinfo  {journal} {Phys. Rev. B}\ }\textbf {\bibinfo {volume}
  {72}},\ \bibinfo {pages} {054523} (\bibinfo {year}
  {2005}{\natexlab{b}})}\BibitemShut {NoStop}%
\bibitem [{\citenamefont {Fominov}\ \emph {et~al.}(2007)\citenamefont
  {Fominov}, \citenamefont {Volkov},\ and\ \citenamefont
  {Efetov}}]{Fominov_PRB2007}%
  \BibitemOpen
  \bibfield  {author} {\bibinfo {author} {\bibfnamefont {Y.~V.}\ \bibnamefont
  {Fominov}}, \bibinfo {author} {\bibfnamefont {A.~F.}\ \bibnamefont {Volkov}},
  \ and\ \bibinfo {author} {\bibfnamefont {K.~B.}\ \bibnamefont {Efetov}},\
  }\href {\doibase 10.1103/PhysRevB.75.104509} {\bibfield  {journal} {\bibinfo
  {journal} {Phys. Rev. B}\ }\textbf {\bibinfo {volume} {75}},\ \bibinfo
  {pages} {104509} (\bibinfo {year} {2007})}\BibitemShut {NoStop}%
\bibitem [{\citenamefont {Crouzy}\ \emph {et~al.}(2007)\citenamefont {Crouzy},
  \citenamefont {Tollis},\ and\ \citenamefont {Ivanov}}]{Crouzy_PRB2007}%
  \BibitemOpen
  \bibfield  {author} {\bibinfo {author} {\bibfnamefont {B.}~\bibnamefont
  {Crouzy}}, \bibinfo {author} {\bibfnamefont {S.}~\bibnamefont {Tollis}}, \
  and\ \bibinfo {author} {\bibfnamefont {D.~A.}\ \bibnamefont {Ivanov}},\
  }\href {\doibase 10.1103/PhysRevB.76.134502} {\bibfield  {journal} {\bibinfo
  {journal} {Phys. Rev. B}\ }\textbf {\bibinfo {volume} {76}},\ \bibinfo
  {pages} {134502} (\bibinfo {year} {2007})}\BibitemShut {NoStop}%
\bibitem [{\citenamefont {Buzdin}\ \emph {et~al.}(2011)\citenamefont {Buzdin},
  \citenamefont {Mel'nikov},\ and\ \citenamefont {Pugach}}]{Pugach_PRB2011}%
  \BibitemOpen
  \bibfield  {author} {\bibinfo {author} {\bibfnamefont {A.~I.}\ \bibnamefont
  {Buzdin}}, \bibinfo {author} {\bibfnamefont {A.~S.}\ \bibnamefont
  {Mel'nikov}}, \ and\ \bibinfo {author} {\bibfnamefont {N.~G.}\ \bibnamefont
  {Pugach}},\ }\href {\doibase 10.1103/PhysRevB.83.144515} {\bibfield
  {journal} {\bibinfo  {journal} {Phys. Rev. B}\ }\textbf {\bibinfo {volume}
  {83}},\ \bibinfo {pages} {144515} (\bibinfo {year} {2011})}\BibitemShut
  {NoStop}%
\bibitem [{\citenamefont {Kupferschmidt}\ and\ \citenamefont
  {Brouwer}(2009)}]{Kupferschmidt_PRB2009}%
  \BibitemOpen
  \bibfield  {author} {\bibinfo {author} {\bibfnamefont {J.~N.}\ \bibnamefont
  {Kupferschmidt}}\ and\ \bibinfo {author} {\bibfnamefont {P.~W.}\ \bibnamefont
  {Brouwer}},\ }\href {\doibase 10.1103/PhysRevB.80.214537} {\bibfield
  {journal} {\bibinfo  {journal} {Phys. Rev. B}\ }\textbf {\bibinfo {volume}
  {80}},\ \bibinfo {pages} {214537} (\bibinfo {year} {2009})}\BibitemShut
  {NoStop}%
\bibitem [{\citenamefont {L\"ofwander}\ \emph {et~al.}(2010)\citenamefont
  {L\"ofwander}, \citenamefont {Grein},\ and\ \citenamefont
  {Eschrig}}]{Lofwander_PRL2010}%
  \BibitemOpen
  \bibfield  {author} {\bibinfo {author} {\bibfnamefont {T.}~\bibnamefont
  {L\"ofwander}}, \bibinfo {author} {\bibfnamefont {R.}~\bibnamefont {Grein}},
  \ and\ \bibinfo {author} {\bibfnamefont {M.}~\bibnamefont {Eschrig}},\ }\href
  {\doibase 10.1103/PhysRevLett.105.207001} {\bibfield  {journal} {\bibinfo
  {journal} {Phys. Rev. Lett.}\ }\textbf {\bibinfo {volume} {105}},\ \bibinfo
  {pages} {207001} (\bibinfo {year} {2010})}\BibitemShut {NoStop}%
\bibitem [{\citenamefont {Grein}\ \emph {et~al.}(2009)\citenamefont {Grein},
  \citenamefont {Eschrig}, \citenamefont {Metalidis},\ and\ \citenamefont
  {Sch\"on}}]{Grein_PRL2009}%
  \BibitemOpen
  \bibfield  {author} {\bibinfo {author} {\bibfnamefont {R.}~\bibnamefont
  {Grein}}, \bibinfo {author} {\bibfnamefont {M.}~\bibnamefont {Eschrig}},
  \bibinfo {author} {\bibfnamefont {G.}~\bibnamefont {Metalidis}}, \ and\
  \bibinfo {author} {\bibfnamefont {G.}~\bibnamefont {Sch\"on}},\ }\href
  {\doibase 10.1103/PhysRevLett.102.227005} {\bibfield  {journal} {\bibinfo
  {journal} {Phys. Rev. Lett.}\ }\textbf {\bibinfo {volume} {102}},\ \bibinfo
  {pages} {227005} (\bibinfo {year} {2009})}\BibitemShut {NoStop}%
\bibitem [{\citenamefont {Eschrig}\ and\ \citenamefont
  {L{\"o}fwander}(2008)}]{eschrig_NatPhys2008}%
  \BibitemOpen
  \bibfield  {author} {\bibinfo {author} {\bibfnamefont {M.}~\bibnamefont
  {Eschrig}}\ and\ \bibinfo {author} {\bibfnamefont {T.}~\bibnamefont
  {L{\"o}fwander}},\ }\href {\doibase 10.1038/nphys831} {\bibfield  {journal}
  {\bibinfo  {journal} {Nat. Phys.}\ }\textbf {\bibinfo {volume} {4}},\
  \bibinfo {pages} {138} (\bibinfo {year} {2008})}\BibitemShut {NoStop}%
\bibitem [{\citenamefont {Annunziata}\ \emph {et~al.}(2012)\citenamefont
  {Annunziata}, \citenamefont {Manske},\ and\ \citenamefont
  {Linder}}]{Annunziata_PRB2012}%
  \BibitemOpen
  \bibfield  {author} {\bibinfo {author} {\bibfnamefont {G.}~\bibnamefont
  {Annunziata}}, \bibinfo {author} {\bibfnamefont {D.}~\bibnamefont {Manske}},
  \ and\ \bibinfo {author} {\bibfnamefont {J.}~\bibnamefont {Linder}},\ }\href
  {\doibase 10.1103/PhysRevB.86.174514} {\bibfield  {journal} {\bibinfo
  {journal} {Phys. Rev. B}\ }\textbf {\bibinfo {volume} {86}},\ \bibinfo
  {pages} {174514} (\bibinfo {year} {2012})}\BibitemShut {NoStop}%
\bibitem [{\citenamefont {Bergeret}\ and\ \citenamefont
  {Tokatly}(2013)}]{Bergeret_PRL2013}%
  \BibitemOpen
  \bibfield  {author} {\bibinfo {author} {\bibfnamefont {F.~S.}\ \bibnamefont
  {Bergeret}}\ and\ \bibinfo {author} {\bibfnamefont {I.~V.}\ \bibnamefont
  {Tokatly}},\ }\href {\doibase 10.1103/PhysRevLett.110.117003} {\bibfield
  {journal} {\bibinfo  {journal} {Phys. Rev. Lett.}\ }\textbf {\bibinfo
  {volume} {110}},\ \bibinfo {pages} {117003} (\bibinfo {year}
  {2013})}\BibitemShut {NoStop}%
\bibitem [{\citenamefont {Bergeret}\ and\ \citenamefont
  {Tokatly}(2014)}]{Bergeret_PRB2014}%
  \BibitemOpen
  \bibfield  {author} {\bibinfo {author} {\bibfnamefont {F.~S.}\ \bibnamefont
  {Bergeret}}\ and\ \bibinfo {author} {\bibfnamefont {I.~V.}\ \bibnamefont
  {Tokatly}},\ }\href {\doibase 10.1103/PhysRevB.89.134517} {\bibfield
  {journal} {\bibinfo  {journal} {Phys. Rev. B}\ }\textbf {\bibinfo {volume}
  {89}},\ \bibinfo {pages} {134517} (\bibinfo {year} {2014})}\BibitemShut
  {NoStop}%
\bibitem [{\citenamefont {Jacobsen}\ \emph {et~al.}(2015)\citenamefont
  {Jacobsen}, \citenamefont {Ouassou},\ and\ \citenamefont
  {Linder}}]{Linder_PRB2015}%
  \BibitemOpen
  \bibfield  {author} {\bibinfo {author} {\bibfnamefont {S.~H.}\ \bibnamefont
  {Jacobsen}}, \bibinfo {author} {\bibfnamefont {J.~A.}\ \bibnamefont
  {Ouassou}}, \ and\ \bibinfo {author} {\bibfnamefont {J.}~\bibnamefont
  {Linder}},\ }\href {\doibase 10.1103/PhysRevB.92.024510} {\bibfield
  {journal} {\bibinfo  {journal} {Phys. Rev. B}\ }\textbf {\bibinfo {volume}
  {92}},\ \bibinfo {pages} {024510} (\bibinfo {year} {2015})}\BibitemShut
  {NoStop}%
\bibitem [{\citenamefont {H\"ubler}\ \emph {et~al.}(2010)\citenamefont
  {H\"ubler}, \citenamefont {Lemyre}, \citenamefont {Beckmann},\ and\
  \citenamefont {v.~L\"ohneysen}}]{Hubler_PRB2010}%
  \BibitemOpen
  \bibfield  {author} {\bibinfo {author} {\bibfnamefont {F.}~\bibnamefont
  {H\"ubler}}, \bibinfo {author} {\bibfnamefont {J.~C.}\ \bibnamefont
  {Lemyre}}, \bibinfo {author} {\bibfnamefont {D.}~\bibnamefont {Beckmann}}, \
  and\ \bibinfo {author} {\bibfnamefont {H.}~\bibnamefont {v.~L\"ohneysen}},\
  }\href {\doibase 10.1103/PhysRevB.81.184524} {\bibfield  {journal} {\bibinfo
  {journal} {Phys. Rev. B}\ }\textbf {\bibinfo {volume} {81}},\ \bibinfo
  {pages} {184524} (\bibinfo {year} {2010})}\BibitemShut {NoStop}%
\bibitem [{\citenamefont {H\"ubler}\ \emph {et~al.}(2012)\citenamefont
  {H\"ubler}, \citenamefont {Wolf}, \citenamefont {Beckmann},\ and\
  \citenamefont {v.~L\"ohneysen}}]{Hubler_PRL2012}%
  \BibitemOpen
  \bibfield  {author} {\bibinfo {author} {\bibfnamefont {F.}~\bibnamefont
  {H\"ubler}}, \bibinfo {author} {\bibfnamefont {M.~J.}\ \bibnamefont {Wolf}},
  \bibinfo {author} {\bibfnamefont {D.}~\bibnamefont {Beckmann}}, \ and\
  \bibinfo {author} {\bibfnamefont {H.}~\bibnamefont {v.~L\"ohneysen}},\ }\href
  {\doibase 10.1103/PhysRevLett.109.207001} {\bibfield  {journal} {\bibinfo
  {journal} {Phys. Rev. Lett.}\ }\textbf {\bibinfo {volume} {109}},\ \bibinfo
  {pages} {207001} (\bibinfo {year} {2012})}\BibitemShut {NoStop}%
\bibitem [{\citenamefont {Wolf}\ \emph {et~al.}(2013)\citenamefont {Wolf},
  \citenamefont {H\"ubler}, \citenamefont {Kolenda}, \citenamefont
  {v.~L\"ohneysen},\ and\ \citenamefont {Beckmann}}]{Hubler_PRB2013}%
  \BibitemOpen
  \bibfield  {author} {\bibinfo {author} {\bibfnamefont {M.~J.}\ \bibnamefont
  {Wolf}}, \bibinfo {author} {\bibfnamefont {F.}~\bibnamefont {H\"ubler}},
  \bibinfo {author} {\bibfnamefont {S.}~\bibnamefont {Kolenda}}, \bibinfo
  {author} {\bibfnamefont {H.}~\bibnamefont {v.~L\"ohneysen}}, \ and\ \bibinfo
  {author} {\bibfnamefont {D.}~\bibnamefont {Beckmann}},\ }\href {\doibase
  10.1103/PhysRevB.87.024517} {\bibfield  {journal} {\bibinfo  {journal} {Phys.
  Rev. B}\ }\textbf {\bibinfo {volume} {87}},\ \bibinfo {pages} {024517}
  (\bibinfo {year} {2013})}\BibitemShut {NoStop}%
\bibitem [{\citenamefont {Quay}\ \emph {et~al.}(2013)\citenamefont {Quay},
  \citenamefont {Chevallier}, \citenamefont {Bena},\ and\ \citenamefont
  {Aprili}}]{quay_NatPhys2013}%
  \BibitemOpen
  \bibfield  {author} {\bibinfo {author} {\bibfnamefont {C.}~\bibnamefont
  {Quay}}, \bibinfo {author} {\bibfnamefont {D.}~\bibnamefont {Chevallier}},
  \bibinfo {author} {\bibfnamefont {C.}~\bibnamefont {Bena}}, \ and\ \bibinfo
  {author} {\bibfnamefont {M.}~\bibnamefont {Aprili}},\ }\href {\doibase
  10.1038/nphys2518} {\bibfield  {journal} {\bibinfo  {journal} {Nat. Phys.}\
  }\textbf {\bibinfo {volume} {9}},\ \bibinfo {pages} {84} (\bibinfo {year}
  {2013})}\BibitemShut {NoStop}%
\bibitem [{\citenamefont {Wakamura}\ \emph {et~al.}(2014)\citenamefont
  {Wakamura}, \citenamefont {Hasegawa}, \citenamefont {Ohnishi}, \citenamefont
  {Niimi},\ and\ \citenamefont {Otani}}]{Wakamura_PRL2014}%
  \BibitemOpen
  \bibfield  {author} {\bibinfo {author} {\bibfnamefont {T.}~\bibnamefont
  {Wakamura}}, \bibinfo {author} {\bibfnamefont {N.}~\bibnamefont {Hasegawa}},
  \bibinfo {author} {\bibfnamefont {K.}~\bibnamefont {Ohnishi}}, \bibinfo
  {author} {\bibfnamefont {Y.}~\bibnamefont {Niimi}}, \ and\ \bibinfo {author}
  {\bibfnamefont {Y.}~\bibnamefont {Otani}},\ }\href {\doibase
  10.1103/PhysRevLett.112.036602} {\bibfield  {journal} {\bibinfo  {journal}
  {Phys. Rev. Lett.}\ }\textbf {\bibinfo {volume} {112}},\ \bibinfo {pages}
  {036602} (\bibinfo {year} {2014})}\BibitemShut {NoStop}%
\bibitem [{\citenamefont {Beckmann}(2016)}]{Beckmann_JPhysConMat2016}%
  \BibitemOpen
  \bibfield  {author} {\bibinfo {author} {\bibfnamefont {D.}~\bibnamefont
  {Beckmann}},\ }\href {http://stacks.iop.org/0953-8984/28/i=16/a=163001}
  {\bibfield  {journal} {\bibinfo  {journal} {J. Phys.: Condens. Matter}\
  }\textbf {\bibinfo {volume} {28}},\ \bibinfo {pages} {163001} (\bibinfo
  {year} {2016})}\BibitemShut {NoStop}%
\bibitem [{\citenamefont {Poli}\ \emph {et~al.}(2008)\citenamefont {Poli},
  \citenamefont {Morten}, \citenamefont {Urech}, \citenamefont {Brataas},
  \citenamefont {Haviland},\ and\ \citenamefont {Korenivski}}]{Poli_PRL2008}%
  \BibitemOpen
  \bibfield  {author} {\bibinfo {author} {\bibfnamefont {N.}~\bibnamefont
  {Poli}}, \bibinfo {author} {\bibfnamefont {J.~P.}\ \bibnamefont {Morten}},
  \bibinfo {author} {\bibfnamefont {M.}~\bibnamefont {Urech}}, \bibinfo
  {author} {\bibfnamefont {A.}~\bibnamefont {Brataas}}, \bibinfo {author}
  {\bibfnamefont {D.~B.}\ \bibnamefont {Haviland}}, \ and\ \bibinfo {author}
  {\bibfnamefont {V.}~\bibnamefont {Korenivski}},\ }\href {\doibase
  10.1103/PhysRevLett.100.136601} {\bibfield  {journal} {\bibinfo  {journal}
  {Phys. Rev. Lett.}\ }\textbf {\bibinfo {volume} {100}},\ \bibinfo {pages}
  {136601} (\bibinfo {year} {2008})}\BibitemShut {NoStop}%
\bibitem [{\citenamefont {Yang}\ \emph {et~al.}(2010)\citenamefont {Yang},
  \citenamefont {Yang}, \citenamefont {Takahashi}, \citenamefont {Maekawa},\
  and\ \citenamefont {Parkin}}]{yang_NatMat2010}%
  \BibitemOpen
  \bibfield  {author} {\bibinfo {author} {\bibfnamefont {H.}~\bibnamefont
  {Yang}}, \bibinfo {author} {\bibfnamefont {S.-H.}\ \bibnamefont {Yang}},
  \bibinfo {author} {\bibfnamefont {S.}~\bibnamefont {Takahashi}}, \bibinfo
  {author} {\bibfnamefont {S.}~\bibnamefont {Maekawa}}, \ and\ \bibinfo
  {author} {\bibfnamefont {S.~S.}\ \bibnamefont {Parkin}},\ }\href {\doibase
  10.1038/nmat2781} {\bibfield  {journal} {\bibinfo  {journal} {Nat. Mater.}\
  }\textbf {\bibinfo {volume} {9}},\ \bibinfo {pages} {586} (\bibinfo {year}
  {2010})}\BibitemShut {NoStop}%
\bibitem [{\citenamefont {Wakamura}\ \emph {et~al.}(2015)\citenamefont
  {Wakamura}, \citenamefont {Akaike}, \citenamefont {Omori}, \citenamefont
  {Niimi}, \citenamefont {Takahashi}, \citenamefont {Fujimaki}, \citenamefont
  {Maekawa},\ and\ \citenamefont {Otani}}]{wakamura_NatMat2015}%
  \BibitemOpen
  \bibfield  {author} {\bibinfo {author} {\bibfnamefont {T.}~\bibnamefont
  {Wakamura}}, \bibinfo {author} {\bibfnamefont {H.}~\bibnamefont {Akaike}},
  \bibinfo {author} {\bibfnamefont {Y.}~\bibnamefont {Omori}}, \bibinfo
  {author} {\bibfnamefont {Y.}~\bibnamefont {Niimi}}, \bibinfo {author}
  {\bibfnamefont {S.}~\bibnamefont {Takahashi}}, \bibinfo {author}
  {\bibfnamefont {A.}~\bibnamefont {Fujimaki}}, \bibinfo {author}
  {\bibfnamefont {S.}~\bibnamefont {Maekawa}}, \ and\ \bibinfo {author}
  {\bibfnamefont {Y.}~\bibnamefont {Otani}},\ }\href {\doibase
  10.1038/nmat4276} {\bibfield  {journal} {\bibinfo  {journal} {Nat. Mater.}\
  }\textbf {\bibinfo {volume} {14}},\ \bibinfo {pages} {675} (\bibinfo {year}
  {2015})}\BibitemShut {NoStop}%
\bibitem [{\citenamefont {Inoue}\ \emph {et~al.}(2017)\citenamefont {Inoue},
  \citenamefont {Ichioka},\ and\ \citenamefont {Adachi}}]{Inoue_PRB2017}%
  \BibitemOpen
  \bibfield  {author} {\bibinfo {author} {\bibfnamefont {M.}~\bibnamefont
  {Inoue}}, \bibinfo {author} {\bibfnamefont {M.}~\bibnamefont {Ichioka}}, \
  and\ \bibinfo {author} {\bibfnamefont {H.}~\bibnamefont {Adachi}},\ }\href
  {\doibase 10.1103/PhysRevB.96.024414} {\bibfield  {journal} {\bibinfo
  {journal} {Phys. Rev. B}\ }\textbf {\bibinfo {volume} {96}},\ \bibinfo
  {pages} {024414} (\bibinfo {year} {2017})}\BibitemShut {NoStop}%
\bibitem [{\citenamefont {Jeon}\ \emph {et~al.}(2018)\citenamefont {Jeon},
  \citenamefont {Ciccarelli}, \citenamefont {Ferguson}, \citenamefont
  {Kurebayashi}, \citenamefont {Cohen}, \citenamefont {Montiel}, \citenamefont
  {Eschrig}, \citenamefont {Robinson},\ and\ \citenamefont
  {Blamire}}]{Jeon_NatMat2018}%
  \BibitemOpen
  \bibfield  {author} {\bibinfo {author} {\bibfnamefont {K.-R.}\ \bibnamefont
  {Jeon}}, \bibinfo {author} {\bibfnamefont {C.}~\bibnamefont {Ciccarelli}},
  \bibinfo {author} {\bibfnamefont {A.~J.}\ \bibnamefont {Ferguson}}, \bibinfo
  {author} {\bibfnamefont {H.}~\bibnamefont {Kurebayashi}}, \bibinfo {author}
  {\bibfnamefont {L.~F.}\ \bibnamefont {Cohen}}, \bibinfo {author}
  {\bibfnamefont {X.}~\bibnamefont {Montiel}}, \bibinfo {author} {\bibfnamefont
  {M.}~\bibnamefont {Eschrig}}, \bibinfo {author} {\bibfnamefont {J.~W.~A.}\
  \bibnamefont {Robinson}}, \ and\ \bibinfo {author} {\bibfnamefont {M.~G.}\
  \bibnamefont {Blamire}},\ }\href {\doibase 10.1038/s41563-018-0058-9}
  {\bibfield  {journal} {\bibinfo  {journal} {Nat. Mater.}\ }\textbf {\bibinfo
  {volume} {17}},\ \bibinfo {pages} {499} (\bibinfo {year} {2018})}\BibitemShut
  {NoStop}%
\bibitem [{\citenamefont {Tserkovnyak}\ \emph {et~al.}(2005)\citenamefont
  {Tserkovnyak}, \citenamefont {Brataas}, \citenamefont {Bauer},\ and\
  \citenamefont {Halperin}}]{Tserkovnyak_RevModPhys2005}%
  \BibitemOpen
  \bibfield  {author} {\bibinfo {author} {\bibfnamefont {Y.}~\bibnamefont
  {Tserkovnyak}}, \bibinfo {author} {\bibfnamefont {A.}~\bibnamefont
  {Brataas}}, \bibinfo {author} {\bibfnamefont {G.~E.~W.}\ \bibnamefont
  {Bauer}}, \ and\ \bibinfo {author} {\bibfnamefont {B.~I.}\ \bibnamefont
  {Halperin}},\ }\href {\doibase 10.1103/RevModPhys.77.1375} {\bibfield
  {journal} {\bibinfo  {journal} {Rev. Mod. Phys.}\ }\textbf {\bibinfo {volume}
  {77}},\ \bibinfo {pages} {1375} (\bibinfo {year} {2005})}\BibitemShut
  {NoStop}%
\bibitem [{\citenamefont {Ando}\ \emph {et~al.}(2011)\citenamefont {Ando},
  \citenamefont {Takahashi}, \citenamefont {Ieda}, \citenamefont {Kajiwara},
  \citenamefont {Nakayama}, \citenamefont {Yoshino}, \citenamefont {Harii},
  \citenamefont {Fujikawa}, \citenamefont {Matsuo}, \citenamefont {Maekawa},\
  and\ \citenamefont {Saitoh}}]{Ando_JapplPhys2011}%
  \BibitemOpen
  \bibfield  {author} {\bibinfo {author} {\bibfnamefont {K.}~\bibnamefont
  {Ando}}, \bibinfo {author} {\bibfnamefont {S.}~\bibnamefont {Takahashi}},
  \bibinfo {author} {\bibfnamefont {J.}~\bibnamefont {Ieda}}, \bibinfo {author}
  {\bibfnamefont {Y.}~\bibnamefont {Kajiwara}}, \bibinfo {author}
  {\bibfnamefont {H.}~\bibnamefont {Nakayama}}, \bibinfo {author}
  {\bibfnamefont {T.}~\bibnamefont {Yoshino}}, \bibinfo {author} {\bibfnamefont
  {K.}~\bibnamefont {Harii}}, \bibinfo {author} {\bibfnamefont
  {Y.}~\bibnamefont {Fujikawa}}, \bibinfo {author} {\bibfnamefont
  {M.}~\bibnamefont {Matsuo}}, \bibinfo {author} {\bibfnamefont
  {S.}~\bibnamefont {Maekawa}}, \ and\ \bibinfo {author} {\bibfnamefont
  {E.}~\bibnamefont {Saitoh}},\ }\href {\doibase 10.1063/1.3587173} {\bibfield
  {journal} {\bibinfo  {journal} {J. Appl. Phys.}\ }\textbf {\bibinfo {volume}
  {109}},\ \bibinfo {pages} {103913} (\bibinfo {year} {2011})}\BibitemShut
  {NoStop}%
\bibitem [{\citenamefont {Bell}\ \emph {et~al.}(2008)\citenamefont {Bell},
  \citenamefont {Milikisyants}, \citenamefont {Huber},\ and\ \citenamefont
  {Aarts}}]{Bell_PRL2008}%
  \BibitemOpen
  \bibfield  {author} {\bibinfo {author} {\bibfnamefont {C.}~\bibnamefont
  {Bell}}, \bibinfo {author} {\bibfnamefont {S.}~\bibnamefont {Milikisyants}},
  \bibinfo {author} {\bibfnamefont {M.}~\bibnamefont {Huber}}, \ and\ \bibinfo
  {author} {\bibfnamefont {J.}~\bibnamefont {Aarts}},\ }\href {\doibase
  10.1103/PhysRevLett.100.047002} {\bibfield  {journal} {\bibinfo  {journal}
  {Phys. Rev. Lett.}\ }\textbf {\bibinfo {volume} {100}},\ \bibinfo {pages}
  {047002} (\bibinfo {year} {2008})}\BibitemShut {NoStop}%
\bibitem [{\citenamefont {Morten}\ \emph {et~al.}(2008)\citenamefont {Morten},
  \citenamefont {Brataas}, \citenamefont {Bauer}, \citenamefont {Belzig},\ and\
  \citenamefont {Tserkovnyak}}]{Morten_EPL2008}%
  \BibitemOpen
  \bibfield  {author} {\bibinfo {author} {\bibfnamefont {J.~P.}\ \bibnamefont
  {Morten}}, \bibinfo {author} {\bibfnamefont {A.}~\bibnamefont {Brataas}},
  \bibinfo {author} {\bibfnamefont {G.~E.~W.}\ \bibnamefont {Bauer}}, \bibinfo
  {author} {\bibfnamefont {W.}~\bibnamefont {Belzig}}, \ and\ \bibinfo {author}
  {\bibfnamefont {Y.}~\bibnamefont {Tserkovnyak}},\ }\href
  {http://stacks.iop.org/0295-5075/84/i=5/a=57008} {\bibfield  {journal}
  {\bibinfo  {journal} {Europhys. Lett.}\ }\textbf {\bibinfo {volume} {84}},\
  \bibinfo {pages} {57008} (\bibinfo {year} {2008})}\BibitemShut {NoStop}%
\bibitem [{\citenamefont {{Liu}}\ \emph {et~al.}(2011)\citenamefont {{Liu}},
  \citenamefont {{Buhrman}},\ and\ \citenamefont {{Ralph}}}]{Liu_arxiv2011}%
  \BibitemOpen
  \bibfield  {author} {\bibinfo {author} {\bibfnamefont {L.}~\bibnamefont
  {{Liu}}}, \bibinfo {author} {\bibfnamefont {R.~A.}\ \bibnamefont
  {{Buhrman}}}, \ and\ \bibinfo {author} {\bibfnamefont {D.~C.}\ \bibnamefont
  {{Ralph}}},\ }\href {http://adsabs.harvard.edu/abs/2011arXiv1111.3702L}
  {\bibfield  {journal} {\bibinfo  {journal} {arXiv:1111.3702}\ } (\bibinfo
  {year} {2011})}\BibitemShut {NoStop}%
\bibitem [{\citenamefont {Bass}\ and\ \citenamefont
  {Jr}(2007)}]{Bass_JPhysCondMatt2007}%
  \BibitemOpen
  \bibfield  {author} {\bibinfo {author} {\bibfnamefont {J.}~\bibnamefont
  {Bass}}\ and\ \bibinfo {author} {\bibfnamefont {W.~P.~P.}\ \bibnamefont
  {Jr}},\ }\href {http://stacks.iop.org/0953-8984/19/i=18/a=183201} {\bibfield
  {journal} {\bibinfo  {journal} {J. Phys.: Condens. Matter}\ }\textbf
  {\bibinfo {volume} {19}},\ \bibinfo {pages} {183201} (\bibinfo {year}
  {2007})}\BibitemShut {NoStop}%
\bibitem [{\citenamefont {Czeschka}\ \emph {et~al.}(2011)\citenamefont
  {Czeschka}, \citenamefont {Dreher}, \citenamefont {Brandt}, \citenamefont
  {Weiler}, \citenamefont {Althammer}, \citenamefont {Imort}, \citenamefont
  {Reiss}, \citenamefont {Thomas}, \citenamefont {Schoch}, \citenamefont
  {Limmer}, \citenamefont {Huebl}, \citenamefont {Gross},\ and\ \citenamefont
  {Goennenwein}}]{Czeschka_PRL2011}%
  \BibitemOpen
  \bibfield  {author} {\bibinfo {author} {\bibfnamefont {F.~D.}\ \bibnamefont
  {Czeschka}}, \bibinfo {author} {\bibfnamefont {L.}~\bibnamefont {Dreher}},
  \bibinfo {author} {\bibfnamefont {M.~S.}\ \bibnamefont {Brandt}}, \bibinfo
  {author} {\bibfnamefont {M.}~\bibnamefont {Weiler}}, \bibinfo {author}
  {\bibfnamefont {M.}~\bibnamefont {Althammer}}, \bibinfo {author}
  {\bibfnamefont {I.-M.}\ \bibnamefont {Imort}}, \bibinfo {author}
  {\bibfnamefont {G.}~\bibnamefont {Reiss}}, \bibinfo {author} {\bibfnamefont
  {A.}~\bibnamefont {Thomas}}, \bibinfo {author} {\bibfnamefont
  {W.}~\bibnamefont {Schoch}}, \bibinfo {author} {\bibfnamefont
  {W.}~\bibnamefont {Limmer}}, \bibinfo {author} {\bibfnamefont
  {H.}~\bibnamefont {Huebl}}, \bibinfo {author} {\bibfnamefont
  {R.}~\bibnamefont {Gross}}, \ and\ \bibinfo {author} {\bibfnamefont
  {S.~T.~B.}\ \bibnamefont {Goennenwein}},\ }\href {\doibase
  10.1103/PhysRevLett.107.046601} {\bibfield  {journal} {\bibinfo  {journal}
  {Phys. Rev. Lett.}\ }\textbf {\bibinfo {volume} {107}},\ \bibinfo {pages}
  {046601} (\bibinfo {year} {2011})}\BibitemShut {NoStop}%
\bibitem [{\citenamefont {Tanaka}\ \emph {et~al.}(2008)\citenamefont {Tanaka},
  \citenamefont {Kontani}, \citenamefont {Naito}, \citenamefont {Naito},
  \citenamefont {Hirashima}, \citenamefont {Yamada},\ and\ \citenamefont
  {Inoue}}]{Tanaka_PRB2008}%
  \BibitemOpen
  \bibfield  {author} {\bibinfo {author} {\bibfnamefont {T.}~\bibnamefont
  {Tanaka}}, \bibinfo {author} {\bibfnamefont {H.}~\bibnamefont {Kontani}},
  \bibinfo {author} {\bibfnamefont {M.}~\bibnamefont {Naito}}, \bibinfo
  {author} {\bibfnamefont {T.}~\bibnamefont {Naito}}, \bibinfo {author}
  {\bibfnamefont {D.~S.}\ \bibnamefont {Hirashima}}, \bibinfo {author}
  {\bibfnamefont {K.}~\bibnamefont {Yamada}}, \ and\ \bibinfo {author}
  {\bibfnamefont {J.}~\bibnamefont {Inoue}},\ }\href {\doibase
  10.1103/PhysRevB.77.165117} {\bibfield  {journal} {\bibinfo  {journal} {Phys.
  Rev. B}\ }\textbf {\bibinfo {volume} {77}},\ \bibinfo {pages} {165117}
  (\bibinfo {year} {2008})}\BibitemShut {NoStop}%
\bibitem [{\citenamefont {Crangle}\ and\ \citenamefont
  {Scott}(1965)}]{Crangle_JApplPhys1965}%
  \BibitemOpen
  \bibfield  {author} {\bibinfo {author} {\bibfnamefont {J.}~\bibnamefont
  {Crangle}}\ and\ \bibinfo {author} {\bibfnamefont {W.~R.}\ \bibnamefont
  {Scott}},\ }\href {\doibase 10.1063/1.1714264} {\bibfield  {journal}
  {\bibinfo  {journal} {J. Appl. Phys.}\ }\textbf {\bibinfo {volume} {36}},\
  \bibinfo {pages} {921} (\bibinfo {year} {1965})}\BibitemShut {NoStop}%
\bibitem [{\citenamefont {Herrmannsd{\"o}rfer}\ \emph
  {et~al.}(1996)\citenamefont {Herrmannsd{\"o}rfer}, \citenamefont {Rehmann},
  \citenamefont {Wendler},\ and\ \citenamefont
  {Pobell}}]{Herrmannsdorfer_JLowTemp_1996}%
  \BibitemOpen
  \bibfield  {author} {\bibinfo {author} {\bibfnamefont {T.}~\bibnamefont
  {Herrmannsd{\"o}rfer}}, \bibinfo {author} {\bibfnamefont {S.}~\bibnamefont
  {Rehmann}}, \bibinfo {author} {\bibfnamefont {W.}~\bibnamefont {Wendler}}, \
  and\ \bibinfo {author} {\bibfnamefont {F.}~\bibnamefont {Pobell}},\ }\href
  {\doibase 10.1007/BF00754089} {\bibfield  {journal} {\bibinfo  {journal} {J.
  Low Temp. Phys.}\ }\textbf {\bibinfo {volume} {104}},\ \bibinfo {pages} {49}
  (\bibinfo {year} {1996})}\BibitemShut {NoStop}%
\bibitem [{\citenamefont {K\"onig}\ \emph {et~al.}(1999)\citenamefont
  {K\"onig}, \citenamefont {Schindler},\ and\ \citenamefont
  {Herrmannsd\"orfer}}]{Konig_PRL_1999}%
  \BibitemOpen
  \bibfield  {author} {\bibinfo {author} {\bibfnamefont {R.}~\bibnamefont
  {K\"onig}}, \bibinfo {author} {\bibfnamefont {A.}~\bibnamefont {Schindler}},
  \ and\ \bibinfo {author} {\bibfnamefont {T.}~\bibnamefont
  {Herrmannsd\"orfer}},\ }\href {\doibase 10.1103/PhysRevLett.82.4528}
  {\bibfield  {journal} {\bibinfo  {journal} {Phys. Rev. Lett.}\ }\textbf
  {\bibinfo {volume} {82}},\ \bibinfo {pages} {4528} (\bibinfo {year}
  {1999})}\BibitemShut {NoStop}%
\bibitem [{\citenamefont {Zhang}\ \emph {et~al.}(2015)\citenamefont {Zhang},
  \citenamefont {Jungfleisch}, \citenamefont {Jiang}, \citenamefont {Liu},
  \citenamefont {Pearson}, \citenamefont {Velthuis}, \citenamefont {Hoffmann},
  \citenamefont {Freimuth},\ and\ \citenamefont {Mokrousov}}]{Zhang_PRB2015}%
  \BibitemOpen
  \bibfield  {author} {\bibinfo {author} {\bibfnamefont {W.}~\bibnamefont
  {Zhang}}, \bibinfo {author} {\bibfnamefont {M.~B.}\ \bibnamefont
  {Jungfleisch}}, \bibinfo {author} {\bibfnamefont {W.}~\bibnamefont {Jiang}},
  \bibinfo {author} {\bibfnamefont {Y.}~\bibnamefont {Liu}}, \bibinfo {author}
  {\bibfnamefont {J.~E.}\ \bibnamefont {Pearson}}, \bibinfo {author}
  {\bibfnamefont {S.~G. E.~t.}\ \bibnamefont {Velthuis}}, \bibinfo {author}
  {\bibfnamefont {A.}~\bibnamefont {Hoffmann}}, \bibinfo {author}
  {\bibfnamefont {F.}~\bibnamefont {Freimuth}}, \ and\ \bibinfo {author}
  {\bibfnamefont {Y.}~\bibnamefont {Mokrousov}},\ }\href {\doibase
  10.1103/PhysRevB.91.115316} {\bibfield  {journal} {\bibinfo  {journal} {Phys.
  Rev. B}\ }\textbf {\bibinfo {volume} {91}},\ \bibinfo {pages} {115316}
  (\bibinfo {year} {2015})}\BibitemShut {NoStop}%
\bibitem [{\citenamefont {Landau}(1956)}]{Landau_JETP1956}%
  \BibitemOpen
  \bibfield  {author} {\bibinfo {author} {\bibfnamefont {L.}~\bibnamefont
  {Landau}},\ }\href {http://www.jetp.ac.ru/cgi-bin/e/index/e/3/6/p920?a=list}
  {\bibfield  {journal} {\bibinfo  {journal} {Sov. Phys. JETP}\ }\textbf
  {\bibinfo {volume} {3}},\ \bibinfo {pages} {920} (\bibinfo {year}
  {1956})}\BibitemShut {NoStop}%
\bibitem [{\citenamefont {Morota}\ \emph {et~al.}(2011)\citenamefont {Morota},
  \citenamefont {Niimi}, \citenamefont {Ohnishi}, \citenamefont {Wei},
  \citenamefont {Tanaka}, \citenamefont {Kontani}, \citenamefont {Kimura},\
  and\ \citenamefont {Otani}}]{Morota_PRB83_174405_2011}%
  \BibitemOpen
  \bibfield  {author} {\bibinfo {author} {\bibfnamefont {M.}~\bibnamefont
  {Morota}}, \bibinfo {author} {\bibfnamefont {Y.}~\bibnamefont {Niimi}},
  \bibinfo {author} {\bibfnamefont {K.}~\bibnamefont {Ohnishi}}, \bibinfo
  {author} {\bibfnamefont {D.~H.}\ \bibnamefont {Wei}}, \bibinfo {author}
  {\bibfnamefont {T.}~\bibnamefont {Tanaka}}, \bibinfo {author} {\bibfnamefont
  {H.}~\bibnamefont {Kontani}}, \bibinfo {author} {\bibfnamefont
  {T.}~\bibnamefont {Kimura}}, \ and\ \bibinfo {author} {\bibfnamefont
  {Y.}~\bibnamefont {Otani}},\ }\href {\doibase 10.1103/PhysRevB.83.174405}
  {\bibfield  {journal} {\bibinfo  {journal} {Phys. Rev. B}\ }\textbf {\bibinfo
  {volume} {83}},\ \bibinfo {pages} {174405} (\bibinfo {year}
  {2011})}\BibitemShut {NoStop}%
\bibitem [{\citenamefont {Tsukahara}\ \emph {et~al.}(2014)\citenamefont
  {Tsukahara}, \citenamefont {Ando}, \citenamefont {Kitamura}, \citenamefont
  {Emoto}, \citenamefont {Shikoh}, \citenamefont {Delmo}, \citenamefont
  {Shinjo},\ and\ \citenamefont {Shiraishi}}]{Tsukahara_PRB89_235317_2014}%
  \BibitemOpen
  \bibfield  {author} {\bibinfo {author} {\bibfnamefont {A.}~\bibnamefont
  {Tsukahara}}, \bibinfo {author} {\bibfnamefont {Y.}~\bibnamefont {Ando}},
  \bibinfo {author} {\bibfnamefont {Y.}~\bibnamefont {Kitamura}}, \bibinfo
  {author} {\bibfnamefont {H.}~\bibnamefont {Emoto}}, \bibinfo {author}
  {\bibfnamefont {E.}~\bibnamefont {Shikoh}}, \bibinfo {author} {\bibfnamefont
  {M.~P.}\ \bibnamefont {Delmo}}, \bibinfo {author} {\bibfnamefont
  {T.}~\bibnamefont {Shinjo}}, \ and\ \bibinfo {author} {\bibfnamefont
  {M.}~\bibnamefont {Shiraishi}},\ }\href {\doibase 10.1103/PhysRevB.89.235317}
  {\bibfield  {journal} {\bibinfo  {journal} {Phys. Rev. B}\ }\textbf {\bibinfo
  {volume} {89}},\ \bibinfo {pages} {235317} (\bibinfo {year}
  {2014})}\BibitemShut {NoStop}%
\bibitem [{\citenamefont {Zhang}\ \emph {et~al.}(2014)\citenamefont {Zhang},
  \citenamefont {Jungfleisch}, \citenamefont {Jiang}, \citenamefont {Pearson},
  \citenamefont {Hoffmann}, \citenamefont {Freimuth},\ and\ \citenamefont
  {Mokrousov}}]{ZhangPRL_113_196602_2014}%
  \BibitemOpen
  \bibfield  {author} {\bibinfo {author} {\bibfnamefont {W.}~\bibnamefont
  {Zhang}}, \bibinfo {author} {\bibfnamefont {M.~B.}\ \bibnamefont
  {Jungfleisch}}, \bibinfo {author} {\bibfnamefont {W.}~\bibnamefont {Jiang}},
  \bibinfo {author} {\bibfnamefont {J.~E.}\ \bibnamefont {Pearson}}, \bibinfo
  {author} {\bibfnamefont {A.}~\bibnamefont {Hoffmann}}, \bibinfo {author}
  {\bibfnamefont {F.}~\bibnamefont {Freimuth}}, \ and\ \bibinfo {author}
  {\bibfnamefont {Y.}~\bibnamefont {Mokrousov}},\ }\href {\doibase
  10.1103/PhysRevLett.113.196602} {\bibfield  {journal} {\bibinfo  {journal}
  {Phys. Rev. Lett.}\ }\textbf {\bibinfo {volume} {113}},\ \bibinfo {pages}
  {196602} (\bibinfo {year} {2014})}\BibitemShut {NoStop}%
\bibitem [{\citenamefont {Yang}\ \emph {et~al.}(2016)\citenamefont {Yang},
  \citenamefont {Xu}, \citenamefont {Zhang}, \citenamefont {Wang},
  \citenamefont {Zhang}, \citenamefont {Li}, \citenamefont {Mirshekarloo},
  \citenamefont {Yao},\ and\ \citenamefont {Wu}}]{YangPRB_93_094402_2016}%
  \BibitemOpen
  \bibfield  {author} {\bibinfo {author} {\bibfnamefont {Y.}~\bibnamefont
  {Yang}}, \bibinfo {author} {\bibfnamefont {Y.}~\bibnamefont {Xu}}, \bibinfo
  {author} {\bibfnamefont {X.}~\bibnamefont {Zhang}}, \bibinfo {author}
  {\bibfnamefont {Y.}~\bibnamefont {Wang}}, \bibinfo {author} {\bibfnamefont
  {S.}~\bibnamefont {Zhang}}, \bibinfo {author} {\bibfnamefont {R.-W.}\
  \bibnamefont {Li}}, \bibinfo {author} {\bibfnamefont {M.~S.}\ \bibnamefont
  {Mirshekarloo}}, \bibinfo {author} {\bibfnamefont {K.}~\bibnamefont {Yao}}, \
  and\ \bibinfo {author} {\bibfnamefont {Y.}~\bibnamefont {Wu}},\ }\href
  {\doibase 10.1103/PhysRevB.93.094402} {\bibfield  {journal} {\bibinfo
  {journal} {Phys. Rev. B}\ }\textbf {\bibinfo {volume} {93}},\ \bibinfo
  {pages} {094402} (\bibinfo {year} {2016})}\BibitemShut {NoStop}%
\bibitem [{\citenamefont {Alexander}\ \emph {et~al.}(1985)\citenamefont
  {Alexander}, \citenamefont {Orlando}, \citenamefont {Rainer},\ and\
  \citenamefont {Tedrow}}]{Alexander_PRB1985}%
  \BibitemOpen
  \bibfield  {author} {\bibinfo {author} {\bibfnamefont {J.~A.~X.}\
  \bibnamefont {Alexander}}, \bibinfo {author} {\bibfnamefont {T.~P.}\
  \bibnamefont {Orlando}}, \bibinfo {author} {\bibfnamefont {D.}~\bibnamefont
  {Rainer}}, \ and\ \bibinfo {author} {\bibfnamefont {P.~M.}\ \bibnamefont
  {Tedrow}},\ }\href {\doibase 10.1103/PhysRevB.31.5811} {\bibfield  {journal}
  {\bibinfo  {journal} {Phys. Rev. B}\ }\textbf {\bibinfo {volume} {31}},\
  \bibinfo {pages} {5811} (\bibinfo {year} {1985})}\BibitemShut {NoStop}%
\bibitem [{\citenamefont {Mal'shukov}\ and\ \citenamefont
  {Brataas}(2012)}]{Bratass_PRB2012}%
  \BibitemOpen
  \bibfield  {author} {\bibinfo {author} {\bibfnamefont {A.~G.}\ \bibnamefont
  {Mal'shukov}}\ and\ \bibinfo {author} {\bibfnamefont {A.}~\bibnamefont
  {Brataas}},\ }\href {\doibase 10.1103/PhysRevB.86.094517} {\bibfield
  {journal} {\bibinfo  {journal} {Phys. Rev. B}\ }\textbf {\bibinfo {volume}
  {86}},\ \bibinfo {pages} {094517} (\bibinfo {year} {2012})}\BibitemShut
  {NoStop}%
\bibitem [{\citenamefont {Usadel}(1970)}]{Usadel_PRL1970}%
  \BibitemOpen
  \bibfield  {author} {\bibinfo {author} {\bibfnamefont {K.~D.}\ \bibnamefont
  {Usadel}},\ }\href {\doibase 10.1103/PhysRevLett.25.507} {\bibfield
  {journal} {\bibinfo  {journal} {Phys. Rev. Lett.}\ }\textbf {\bibinfo
  {volume} {25}},\ \bibinfo {pages} {507} (\bibinfo {year} {1970})}\BibitemShut
  {NoStop}%
\bibitem [{\citenamefont {Belzig}\ \emph {et~al.}(1999)\citenamefont {Belzig},
  \citenamefont {Wilhelm}, \citenamefont {Bruder}, \citenamefont {Schön},\
  and\ \citenamefont {Zaikin}}]{Belzig_SuperLattice1999}%
  \BibitemOpen
  \bibfield  {author} {\bibinfo {author} {\bibfnamefont {W.}~\bibnamefont
  {Belzig}}, \bibinfo {author} {\bibfnamefont {F.~K.}\ \bibnamefont {Wilhelm}},
  \bibinfo {author} {\bibfnamefont {C.}~\bibnamefont {Bruder}}, \bibinfo
  {author} {\bibfnamefont {G.}~\bibnamefont {Schön}}, \ and\ \bibinfo {author}
  {\bibfnamefont {A.~D.}\ \bibnamefont {Zaikin}},\ }\href {\doibase
  https://doi.org/10.1006/spmi.1999.0710} {\bibfield  {journal} {\bibinfo
  {journal} {Superlattices Microstruc.}\ }\textbf {\bibinfo {volume} {25}},\
  \bibinfo {pages} {1251 } (\bibinfo {year} {1999})}\BibitemShut {NoStop}%
\bibitem [{\citenamefont {Eilenberger}(1968)}]{Eilenberger_1968}%
  \BibitemOpen
  \bibfield  {author} {\bibinfo {author} {\bibfnamefont {G.}~\bibnamefont
  {Eilenberger}},\ }\href {\doibase 10.1007/BF01379803} {\bibfield  {journal}
  {\bibinfo  {journal} {Z. Phys.}\ }\textbf {\bibinfo {volume} {214}},\
  \bibinfo {pages} {195} (\bibinfo {year} {1968})}\BibitemShut {NoStop}%
\bibitem [{\citenamefont {Larkin}\ and\ \citenamefont
  {Ovchinnikov}(1969)}]{larkin_JETP1969}%
  \BibitemOpen
  \bibfield  {author} {\bibinfo {author} {\bibfnamefont {A.}~\bibnamefont
  {Larkin}}\ and\ \bibinfo {author} {\bibfnamefont {Y.~N.}\ \bibnamefont
  {Ovchinnikov}},\ }\href
  {http://www.jetp.ac.ru/cgi-bin/e/index/e/28/6/p1200?a=list} {\bibfield
  {journal} {\bibinfo  {journal} {Sov Phys JETP}\ }\textbf {\bibinfo {volume}
  {28}},\ \bibinfo {pages} {1200} (\bibinfo {year} {1969})}\BibitemShut
  {NoStop}%
\bibitem [{\citenamefont {Nazarov}(1994)}]{Nazarov_PRL1994}%
  \BibitemOpen
  \bibfield  {author} {\bibinfo {author} {\bibfnamefont {Y.~V.}\ \bibnamefont
  {Nazarov}},\ }\href {\doibase 10.1103/PhysRevLett.73.1420} {\bibfield
  {journal} {\bibinfo  {journal} {Phys. Rev. Lett.}\ }\textbf {\bibinfo
  {volume} {73}},\ \bibinfo {pages} {1420} (\bibinfo {year}
  {1994})}\BibitemShut {NoStop}%
\bibitem [{\citenamefont {Nazarov}(1999)}]{Nazarov_SuperLattMicro1999}%
  \BibitemOpen
  \bibfield  {author} {\bibinfo {author} {\bibfnamefont {Y.~V.}\ \bibnamefont
  {Nazarov}},\ }\href {\doibase https://doi.org/10.1006/spmi.1999.0738}
  {\bibfield  {journal} {\bibinfo  {journal} {Superlattices Microstruc.}\
  }\textbf {\bibinfo {volume} {25}},\ \bibinfo {pages} {1221 } (\bibinfo {year}
  {1999})}\BibitemShut {NoStop}%
\bibitem [{\citenamefont {Eschrig}\ \emph {et~al.}(2015)\citenamefont
  {Eschrig}, \citenamefont {Cottet}, \citenamefont {Belzig},\ and\
  \citenamefont {Linder}}]{Eschrig_NJPhys2015}%
  \BibitemOpen
  \bibfield  {author} {\bibinfo {author} {\bibfnamefont {M.}~\bibnamefont
  {Eschrig}}, \bibinfo {author} {\bibfnamefont {A.}~\bibnamefont {Cottet}},
  \bibinfo {author} {\bibfnamefont {W.}~\bibnamefont {Belzig}}, \ and\ \bibinfo
  {author} {\bibfnamefont {J.}~\bibnamefont {Linder}},\ }\href
  {http://stacks.iop.org/1367-2630/17/i=8/a=083037} {\bibfield  {journal}
  {\bibinfo  {journal} {New J. Phys.}\ }\textbf {\bibinfo {volume} {17}},\
  \bibinfo {pages} {083037} (\bibinfo {year} {2015})}\BibitemShut {NoStop}%
\bibitem [{\citenamefont {Konstandin}\ \emph {et~al.}(2005)\citenamefont
  {Konstandin}, \citenamefont {Kopu},\ and\ \citenamefont
  {Eschrig}}]{Eschrig_PRB2005}%
  \BibitemOpen
  \bibfield  {author} {\bibinfo {author} {\bibfnamefont {A.}~\bibnamefont
  {Konstandin}}, \bibinfo {author} {\bibfnamefont {J.}~\bibnamefont {Kopu}}, \
  and\ \bibinfo {author} {\bibfnamefont {M.}~\bibnamefont {Eschrig}},\ }\href
  {\doibase 10.1103/PhysRevB.72.140501} {\bibfield  {journal} {\bibinfo
  {journal} {Phys. Rev. B}\ }\textbf {\bibinfo {volume} {72}},\ \bibinfo
  {pages} {140501} (\bibinfo {year} {2005})}\BibitemShut {NoStop}%
\bibitem [{\citenamefont {Samokhin}(2009)}]{Samokhin_AnnPhys2009}%
  \BibitemOpen
  \bibfield  {author} {\bibinfo {author} {\bibfnamefont {K.}~\bibnamefont
  {Samokhin}},\ }\href {\doibase https://doi.org/10.1016/j.aop.2009.08.008}
  {\bibfield  {journal} {\bibinfo  {journal} {Ann. Phys.}\ }\textbf {\bibinfo
  {volume} {324}},\ \bibinfo {pages} {2385 } (\bibinfo {year}
  {2009})}\BibitemShut {NoStop}%
\bibitem [{\citenamefont {Edelstein}(2003)}]{Edelstein_PRB2003}%
  \BibitemOpen
  \bibfield  {author} {\bibinfo {author} {\bibfnamefont {V.~M.}\ \bibnamefont
  {Edelstein}},\ }\href {\doibase 10.1103/PhysRevB.67.020505} {\bibfield
  {journal} {\bibinfo  {journal} {Phys. Rev. B}\ }\textbf {\bibinfo {volume}
  {67}},\ \bibinfo {pages} {020505} (\bibinfo {year} {2003})}\BibitemShut
  {NoStop}%
\bibitem [{\citenamefont {Gorini}\ \emph {et~al.}(2010)\citenamefont {Gorini},
  \citenamefont {Schwab}, \citenamefont {Raimondi},\ and\ \citenamefont
  {Shelankov}}]{Gorini_PRB2010}%
  \BibitemOpen
  \bibfield  {author} {\bibinfo {author} {\bibfnamefont {C.}~\bibnamefont
  {Gorini}}, \bibinfo {author} {\bibfnamefont {P.}~\bibnamefont {Schwab}},
  \bibinfo {author} {\bibfnamefont {R.}~\bibnamefont {Raimondi}}, \ and\
  \bibinfo {author} {\bibfnamefont {A.~L.}\ \bibnamefont {Shelankov}},\ }\href
  {\doibase 10.1103/PhysRevB.82.195316} {\bibfield  {journal} {\bibinfo
  {journal} {Phys. Rev. B}\ }\textbf {\bibinfo {volume} {82}},\ \bibinfo
  {pages} {195316} (\bibinfo {year} {2010})}\BibitemShut {NoStop}%
\bibitem [{\citenamefont {Bychkov}\ and\ \citenamefont
  {Rashba}(1984)}]{Rashba_JPhys1984}%
  \BibitemOpen
  \bibfield  {author} {\bibinfo {author} {\bibfnamefont {Y.~A.}\ \bibnamefont
  {Bychkov}}\ and\ \bibinfo {author} {\bibfnamefont {E.~I.}\ \bibnamefont
  {Rashba}},\ }\href {http://stacks.iop.org/0022-3719/17/i=33/a=015} {\bibfield
   {journal} {\bibinfo  {journal} {J. Phys. C Solid State Phys.}\ }\textbf
  {\bibinfo {volume} {17}},\ \bibinfo {pages} {6039} (\bibinfo {year}
  {1984})}\BibitemShut {NoStop}%
\bibitem [{\citenamefont {Dresselhaus}(1955)}]{Dresselhaus_PhysRev1955}%
  \BibitemOpen
  \bibfield  {author} {\bibinfo {author} {\bibfnamefont {G.}~\bibnamefont
  {Dresselhaus}},\ }\href {\doibase 10.1103/PhysRev.100.580} {\bibfield
  {journal} {\bibinfo  {journal} {Phys. Rev.}\ }\textbf {\bibinfo {volume}
  {100}},\ \bibinfo {pages} {580} (\bibinfo {year} {1955})}\BibitemShut
  {NoStop}%
\bibitem [{\citenamefont {Vorontsov}\ \emph {et~al.}(2008)\citenamefont
  {Vorontsov}, \citenamefont {Vekhter},\ and\ \citenamefont
  {Eschrig}}]{VorontsovPRL_101_127003_2008}%
  \BibitemOpen
  \bibfield  {author} {\bibinfo {author} {\bibfnamefont {A.~B.}\ \bibnamefont
  {Vorontsov}}, \bibinfo {author} {\bibfnamefont {I.}~\bibnamefont {Vekhter}},
  \ and\ \bibinfo {author} {\bibfnamefont {M.}~\bibnamefont {Eschrig}},\ }\href
  {\doibase 10.1103/PhysRevLett.101.127003} {\bibfield  {journal} {\bibinfo
  {journal} {Phys. Rev. Lett.}\ }\textbf {\bibinfo {volume} {101}},\ \bibinfo
  {pages} {127003} (\bibinfo {year} {2008})}\BibitemShut {NoStop}%
\bibitem [{\citenamefont {Wennerdal}\ and\ \citenamefont
  {Eschrig}(2017)}]{WennerdalPRB_95_024513_2017}%
  \BibitemOpen
  \bibfield  {author} {\bibinfo {author} {\bibfnamefont {N.}~\bibnamefont
  {Wennerdal}}\ and\ \bibinfo {author} {\bibfnamefont {M.}~\bibnamefont
  {Eschrig}},\ }\href {\doibase 10.1103/PhysRevB.95.024513} {\bibfield
  {journal} {\bibinfo  {journal} {Phys. Rev. B}\ }\textbf {\bibinfo {volume}
  {95}},\ \bibinfo {pages} {024513} (\bibinfo {year} {2017})}\BibitemShut
  {NoStop}%
\bibitem [{\citenamefont {Eschrig}(2009)}]{Eschrig_PRB2009}%
  \BibitemOpen
  \bibfield  {author} {\bibinfo {author} {\bibfnamefont {M.}~\bibnamefont
  {Eschrig}},\ }\href {\doibase 10.1103/PhysRevB.80.134511} {\bibfield
  {journal} {\bibinfo  {journal} {Phys. Rev. B}\ }\textbf {\bibinfo {volume}
  {80}},\ \bibinfo {pages} {134511} (\bibinfo {year} {2009})}\BibitemShut
  {NoStop}%
\bibitem [{\citenamefont {Eschrig}(2000)}]{Eschrig_PRB2000}%
  \BibitemOpen
  \bibfield  {author} {\bibinfo {author} {\bibfnamefont {M.}~\bibnamefont
  {Eschrig}},\ }\href {\doibase 10.1103/PhysRevB.61.9061} {\bibfield  {journal}
  {\bibinfo  {journal} {Phys. Rev. B}\ }\textbf {\bibinfo {volume} {61}},\
  \bibinfo {pages} {9061} (\bibinfo {year} {2000})}\BibitemShut {NoStop}%
\bibitem [{\citenamefont {Schopohl}\ and\ \citenamefont
  {Maki}(1995)}]{Schopol_PRB1995}%
  \BibitemOpen
  \bibfield  {author} {\bibinfo {author} {\bibfnamefont {N.}~\bibnamefont
  {Schopohl}}\ and\ \bibinfo {author} {\bibfnamefont {K.}~\bibnamefont
  {Maki}},\ }\href {\doibase 10.1103/PhysRevB.52.490} {\bibfield  {journal}
  {\bibinfo  {journal} {Phys. Rev. B}\ }\textbf {\bibinfo {volume} {52}},\
  \bibinfo {pages} {490} (\bibinfo {year} {1995})}\BibitemShut {NoStop}%
\bibitem [{\citenamefont {Nagato}\ \emph {et~al.}(1993)\citenamefont {Nagato},
  \citenamefont {Nagai},\ and\ \citenamefont {Hara}}]{Nagato_JlowTempPhys1993}%
  \BibitemOpen
  \bibfield  {author} {\bibinfo {author} {\bibfnamefont {Y.}~\bibnamefont
  {Nagato}}, \bibinfo {author} {\bibfnamefont {K.}~\bibnamefont {Nagai}}, \
  and\ \bibinfo {author} {\bibfnamefont {J.}~\bibnamefont {Hara}},\ }\href
  {\doibase 10.1007/BF00682280} {\bibfield  {journal} {\bibinfo  {journal} {J.
  Low Temp. Phys.}\ }\textbf {\bibinfo {volume} {93}},\ \bibinfo {pages} {33}
  (\bibinfo {year} {1993})}\BibitemShut {NoStop}%
\bibitem [{\citenamefont {Higashitani}\ and\ \citenamefont
  {Nagai}(1995)}]{Nagai_JPhyssocJpn1995}%
  \BibitemOpen
  \bibfield  {author} {\bibinfo {author} {\bibfnamefont {S.}~\bibnamefont
  {Higashitani}}\ and\ \bibinfo {author} {\bibfnamefont {K.}~\bibnamefont
  {Nagai}},\ }\href {\doibase 10.1143/JPSJ.64.549} {\bibfield  {journal}
  {\bibinfo  {journal} {J. Phys. Soc. Jpn.}\ }\textbf {\bibinfo {volume}
  {64}},\ \bibinfo {pages} {549} (\bibinfo {year} {1995})}\BibitemShut
  {NoStop}%
\bibitem [{\citenamefont {Press}\ \emph {et~al.}(1992)\citenamefont {Press},
  \citenamefont {Flannery}, \citenamefont {Teukolsky},\ and\ \citenamefont
  {Vetterling}}]{Numerical_Recipe}%
  \BibitemOpen
  \bibfield  {author} {\bibinfo {author} {\bibfnamefont {W.~H.}\ \bibnamefont
  {Press}}, \bibinfo {author} {\bibfnamefont {B.}~\bibnamefont {Flannery}},
  \bibinfo {author} {\bibfnamefont {S.}~\bibnamefont {Teukolsky}}, \ and\
  \bibinfo {author} {\bibfnamefont {W.}~\bibnamefont {Vetterling}},\
  }\href@noop {} {\enquote {\bibinfo {title} {Numerical recipe, the art of
  scientific computation},}\ } (\bibinfo {year} {1992})\BibitemShut {NoStop}%
\bibitem [{\citenamefont {Vedyaev}\ \emph {et~al.}(1975)\citenamefont
  {Vedyaev}, \citenamefont {Kondorskii},\ and\ \citenamefont
  {Mizie}}]{Vedyaev_Phystatsolidib1975}%
  \BibitemOpen
  \bibfield  {author} {\bibinfo {author} {\bibfnamefont {A.~V.}\ \bibnamefont
  {Vedyaev}}, \bibinfo {author} {\bibfnamefont {E.~I.}\ \bibnamefont
  {Kondorskii}}, \ and\ \bibinfo {author} {\bibfnamefont {E.}~\bibnamefont
  {Mizie}},\ }\href {\doibase 10.1002/pssb.2220720122} {\bibfield  {journal}
  {\bibinfo  {journal} {Phys. Status Solidi (b)}\ }\textbf {\bibinfo {volume}
  {72}},\ \bibinfo {pages} {205} (\bibinfo {year} {1975})}\BibitemShut
  {NoStop}%
\bibitem [{\citenamefont {Leiro}(1995)}]{Leiro_solidstateCommun1995}%
  \BibitemOpen
  \bibfield  {author} {\bibinfo {author} {\bibfnamefont {J.}~\bibnamefont
  {Leiro}},\ }\href {\doibase https://doi.org/10.1016/0038-1098(94)00824-8}
  {\bibfield  {journal} {\bibinfo  {journal} {Solid State Commun.}\ }\textbf
  {\bibinfo {volume} {93}},\ \bibinfo {pages} {953 } (\bibinfo {year}
  {1995})}\BibitemShut {NoStop}%
\bibitem [{\citenamefont {Gu}\ \emph {et~al.}(2002)\citenamefont {Gu},
  \citenamefont {Caballero}, \citenamefont {Slater}, \citenamefont {Loloee},\
  and\ \citenamefont {Pratt}}]{Gu_PRB2002}%
  \BibitemOpen
  \bibfield  {author} {\bibinfo {author} {\bibfnamefont {J.~Y.}\ \bibnamefont
  {Gu}}, \bibinfo {author} {\bibfnamefont {J.~A.}\ \bibnamefont {Caballero}},
  \bibinfo {author} {\bibfnamefont {R.~D.}\ \bibnamefont {Slater}}, \bibinfo
  {author} {\bibfnamefont {R.}~\bibnamefont {Loloee}}, \ and\ \bibinfo {author}
  {\bibfnamefont {W.~P.}\ \bibnamefont {Pratt}},\ }\href {\doibase
  10.1103/PhysRevB.66.140507} {\bibfield  {journal} {\bibinfo  {journal} {Phys.
  Rev. B}\ }\textbf {\bibinfo {volume} {66}},\ \bibinfo {pages} {140507}
  (\bibinfo {year} {2002})}\BibitemShut {NoStop}%
\bibitem [{\citenamefont {Yokoyama}\ \emph {et~al.}(2006)\citenamefont
  {Yokoyama}, \citenamefont {Tanaka},\ and\ \citenamefont
  {Golubov}}]{Yokoyama_PRB2006}%
  \BibitemOpen
  \bibfield  {author} {\bibinfo {author} {\bibfnamefont {T.}~\bibnamefont
  {Yokoyama}}, \bibinfo {author} {\bibfnamefont {Y.}~\bibnamefont {Tanaka}}, \
  and\ \bibinfo {author} {\bibfnamefont {A.~A.}\ \bibnamefont {Golubov}},\
  }\href {\doibase 10.1103/PhysRevB.73.094501} {\bibfield  {journal} {\bibinfo
  {journal} {Phys. Rev. B}\ }\textbf {\bibinfo {volume} {73}},\ \bibinfo
  {pages} {094501} (\bibinfo {year} {2006})}\BibitemShut {NoStop}%
\bibitem [{\citenamefont {Kawabata}\ \emph {et~al.}(2013)\citenamefont
  {Kawabata}, \citenamefont {Asano}, \citenamefont {Tanaka},\ and\
  \citenamefont {Golubov}}]{Kawabata_JPhysSocJpn2013}%
  \BibitemOpen
  \bibfield  {author} {\bibinfo {author} {\bibfnamefont {S.}~\bibnamefont
  {Kawabata}}, \bibinfo {author} {\bibfnamefont {Y.}~\bibnamefont {Asano}},
  \bibinfo {author} {\bibfnamefont {Y.}~\bibnamefont {Tanaka}}, \ and\ \bibinfo
  {author} {\bibfnamefont {A.~A.}\ \bibnamefont {Golubov}},\ }\href {\doibase
  10.7566/JPSJ.82.124702} {\bibfield  {journal} {\bibinfo  {journal} {J. Phys.
  Soc. Jpn.}\ }\textbf {\bibinfo {volume} {82}},\ \bibinfo {pages} {124702}
  (\bibinfo {year} {2013})}\BibitemShut {NoStop}%
\bibitem [{\citenamefont {Bergeret}\ \emph
  {et~al.}(2004{\natexlab{a}})\citenamefont {Bergeret}, \citenamefont
  {Volkov},\ and\ \citenamefont {Efetov}}]{Bergeret_EPL2004}%
  \BibitemOpen
  \bibfield  {author} {\bibinfo {author} {\bibfnamefont {F.~S.}\ \bibnamefont
  {Bergeret}}, \bibinfo {author} {\bibfnamefont {A.~F.}\ \bibnamefont
  {Volkov}}, \ and\ \bibinfo {author} {\bibfnamefont {K.~B.}\ \bibnamefont
  {Efetov}},\ }\href {http://stacks.iop.org/0295-5075/66/i=1/a=111} {\bibfield
  {journal} {\bibinfo  {journal} {Europhys. Lett.}\ }\textbf {\bibinfo {volume}
  {66}},\ \bibinfo {pages} {111} (\bibinfo {year}
  {2004}{\natexlab{a}})}\BibitemShut {NoStop}%
\bibitem [{\citenamefont {Bergeret}\ \emph
  {et~al.}(2004{\natexlab{b}})\citenamefont {Bergeret}, \citenamefont
  {Volkov},\ and\ \citenamefont {Efetov}}]{Bergeret_PRB2004}%
  \BibitemOpen
  \bibfield  {author} {\bibinfo {author} {\bibfnamefont {F.~S.}\ \bibnamefont
  {Bergeret}}, \bibinfo {author} {\bibfnamefont {A.~F.}\ \bibnamefont
  {Volkov}}, \ and\ \bibinfo {author} {\bibfnamefont {K.~B.}\ \bibnamefont
  {Efetov}},\ }\href {\doibase 10.1103/PhysRevB.69.174504} {\bibfield
  {journal} {\bibinfo  {journal} {Phys. Rev. B}\ }\textbf {\bibinfo {volume}
  {69}},\ \bibinfo {pages} {174504} (\bibinfo {year}
  {2004}{\natexlab{b}})}\BibitemShut {NoStop}%
\bibitem [{\citenamefont {Pugach}\ \emph {et~al.}(2017)\citenamefont {Pugach},
  \citenamefont {Safonchik}, \citenamefont {Champel}, \citenamefont
  {Zhitomirsky}, \citenamefont {L{\"a}hderanta}, \citenamefont {Eschrig},\ and\
  \citenamefont {Lacroix}}]{Pugach_APL2018}%
  \BibitemOpen
  \bibfield  {author} {\bibinfo {author} {\bibfnamefont {N.~G.}\ \bibnamefont
  {Pugach}}, \bibinfo {author} {\bibfnamefont {M.}~\bibnamefont {Safonchik}},
  \bibinfo {author} {\bibfnamefont {T.}~\bibnamefont {Champel}}, \bibinfo
  {author} {\bibfnamefont {M.~E.}\ \bibnamefont {Zhitomirsky}}, \bibinfo
  {author} {\bibfnamefont {E.}~\bibnamefont {L{\"a}hderanta}}, \bibinfo
  {author} {\bibfnamefont {M.}~\bibnamefont {Eschrig}}, \ and\ \bibinfo
  {author} {\bibfnamefont {C.}~\bibnamefont {Lacroix}},\ }\href {\doibase
  10.1063/1.5000315} {\bibfield  {journal} {\bibinfo  {journal} {Appl. Phys.
  Lett.}\ }\textbf {\bibinfo {volume} {111}},\ \bibinfo {pages} {162601}
  (\bibinfo {year} {2017})}\BibitemShut {NoStop}%
\bibitem [{\citenamefont {Waintal}\ and\ \citenamefont
  {Brouwer}(2002)}]{Waintal_PRB2002}%
  \BibitemOpen
  \bibfield  {author} {\bibinfo {author} {\bibfnamefont {X.}~\bibnamefont
  {Waintal}}\ and\ \bibinfo {author} {\bibfnamefont {P.~W.}\ \bibnamefont
  {Brouwer}},\ }\href {\doibase 10.1103/PhysRevB.65.054407} {\bibfield
  {journal} {\bibinfo  {journal} {Phys. Rev. B}\ }\textbf {\bibinfo {volume}
  {65}},\ \bibinfo {pages} {054407} (\bibinfo {year} {2002})}\BibitemShut
  {NoStop}%
\bibitem [{\citenamefont {Halterman}\ and\ \citenamefont
  {Alidoust}(2016)}]{Halterman_SupSciTech2016}%
  \BibitemOpen
  \bibfield  {author} {\bibinfo {author} {\bibfnamefont {K.}~\bibnamefont
  {Halterman}}\ and\ \bibinfo {author} {\bibfnamefont {M.}~\bibnamefont
  {Alidoust}},\ }\href {http://stacks.iop.org/0953-2048/29/i=5/a=055007}
  {\bibfield  {journal} {\bibinfo  {journal} {Supercond. Sci. Technol.}\
  }\textbf {\bibinfo {volume} {29}},\ \bibinfo {pages} {055007} (\bibinfo
  {year} {2016})}\BibitemShut {NoStop}%
\bibitem [{\citenamefont {Anderson}(1959)}]{anderson_jphyschemsol1959}%
  \BibitemOpen
  \bibfield  {author} {\bibinfo {author} {\bibfnamefont {P.}~\bibnamefont
  {Anderson}},\ }\href {\doibase https://doi.org/10.1016/0022-3697(59)90036-8}
  {\bibfield  {journal} {\bibinfo  {journal} {J. Phys. Chem. Solids}\ }\textbf
  {\bibinfo {volume} {11}},\ \bibinfo {pages} {26 } (\bibinfo {year}
  {1959})}\BibitemShut {NoStop}%
\bibitem [{\citenamefont {Gor'kov}\ and\ \citenamefont
  {Rusinov}(1964)}]{Rusinov_JETP1964}%
  \BibitemOpen
  \bibfield  {author} {\bibinfo {author} {\bibfnamefont {L.}~\bibnamefont
  {Gor'kov}}\ and\ \bibinfo {author} {\bibfnamefont {A.}~\bibnamefont
  {Rusinov}},\ }\href
  {http://www.jetp.ac.ru/cgi-bin/e/index/e/19/4/p922?a=list} {\bibfield
  {journal} {\bibinfo  {journal} {Sov. Phys.--JETP 19, 922.[Zh. Eksp. Teor.
  Fiz. 46, 1363.]}\ } (\bibinfo {year} {1964})}\BibitemShut {NoStop}%
\bibitem [{\citenamefont {Jacobsen}\ \emph {et~al.}(2016)\citenamefont
  {Jacobsen}, \citenamefont {Kulagina},\ and\ \citenamefont
  {Linder}}]{Jacobsen_SciRep2016}%
  \BibitemOpen
  \bibfield  {author} {\bibinfo {author} {\bibfnamefont {S.~H.}\ \bibnamefont
  {Jacobsen}}, \bibinfo {author} {\bibfnamefont {I.}~\bibnamefont {Kulagina}},
  \ and\ \bibinfo {author} {\bibfnamefont {J.}~\bibnamefont {Linder}},\ }\href
  {https://www.nature.com/articles/srep23926} {\bibfield  {journal} {\bibinfo
  {journal} {Sci. Rep.}\ }\textbf {\bibinfo {volume} {6}},\ \bibinfo {pages}
  {23926} (\bibinfo {year} {2016})}\BibitemShut {NoStop}%
\bibitem [{\citenamefont {Ouassou}\ \emph {et~al.}(2017)\citenamefont
  {Ouassou}, \citenamefont {Jacobsen},\ and\ \citenamefont
  {Linder}}]{Jacobsen_PRB2017}%
  \BibitemOpen
  \bibfield  {author} {\bibinfo {author} {\bibfnamefont {J.~A.}\ \bibnamefont
  {Ouassou}}, \bibinfo {author} {\bibfnamefont {S.~H.}\ \bibnamefont
  {Jacobsen}}, \ and\ \bibinfo {author} {\bibfnamefont {J.}~\bibnamefont
  {Linder}},\ }\href {\doibase 10.1103/PhysRevB.96.094505} {\bibfield
  {journal} {\bibinfo  {journal} {Phys. Rev. B}\ }\textbf {\bibinfo {volume}
  {96}},\ \bibinfo {pages} {094505} (\bibinfo {year} {2017})}\BibitemShut
  {NoStop}%
\bibitem [{\citenamefont {Freeman}\ \emph {et~al.}(2018)\citenamefont
  {Freeman}, \citenamefont {Zholud}, \citenamefont {Dun}, \citenamefont
  {Zhou},\ and\ \citenamefont {Urazhdin}}]{Freeman_PRL2018}%
  \BibitemOpen
  \bibfield  {author} {\bibinfo {author} {\bibfnamefont {R.}~\bibnamefont
  {Freeman}}, \bibinfo {author} {\bibfnamefont {A.}~\bibnamefont {Zholud}},
  \bibinfo {author} {\bibfnamefont {Z.}~\bibnamefont {Dun}}, \bibinfo {author}
  {\bibfnamefont {H.}~\bibnamefont {Zhou}}, \ and\ \bibinfo {author}
  {\bibfnamefont {S.}~\bibnamefont {Urazhdin}},\ }\href {\doibase
  10.1103/PhysRevLett.120.067204} {\bibfield  {journal} {\bibinfo  {journal}
  {Phys. Rev. Lett.}\ }\textbf {\bibinfo {volume} {120}},\ \bibinfo {pages}
  {067204} (\bibinfo {year} {2018})}\BibitemShut {NoStop}%
\bibitem [{\citenamefont {Kurt}\ \emph {et~al.}(2002)\citenamefont {Kurt},
  \citenamefont {Loloee}, \citenamefont {Eid}, \citenamefont {Pratt},\ and\
  \citenamefont {Bass}}]{Kurt_APL_2002}%
  \BibitemOpen
  \bibfield  {author} {\bibinfo {author} {\bibfnamefont {H.}~\bibnamefont
  {Kurt}}, \bibinfo {author} {\bibfnamefont {R.}~\bibnamefont {Loloee}},
  \bibinfo {author} {\bibfnamefont {K.}~\bibnamefont {Eid}}, \bibinfo {author}
  {\bibfnamefont {W.~P.}\ \bibnamefont {Pratt}}, \ and\ \bibinfo {author}
  {\bibfnamefont {J.}~\bibnamefont {Bass}},\ }\href {\doibase
  10.1063/1.1528737} {\bibfield  {journal} {\bibinfo  {journal} {Appl. Phys.
  Lett.}\ }\textbf {\bibinfo {volume} {81}},\ \bibinfo {pages} {4787} (\bibinfo
  {year} {2002})}\BibitemShut {NoStop}%
\bibitem [{\citenamefont {Nguyen}\ \emph {et~al.}(2016)\citenamefont {Nguyen},
  \citenamefont {Ralph},\ and\ \citenamefont {Buhrman}}]{Nguyen_PRL2016}%
  \BibitemOpen
  \bibfield  {author} {\bibinfo {author} {\bibfnamefont {M.-H.}\ \bibnamefont
  {Nguyen}}, \bibinfo {author} {\bibfnamefont {D.~C.}\ \bibnamefont {Ralph}}, \
  and\ \bibinfo {author} {\bibfnamefont {R.~A.}\ \bibnamefont {Buhrman}},\
  }\href {\doibase 10.1103/PhysRevLett.116.126601} {\bibfield  {journal}
  {\bibinfo  {journal} {Phys. Rev. Lett.}\ }\textbf {\bibinfo {volume} {116}},\
  \bibinfo {pages} {126601} (\bibinfo {year} {2016})}\BibitemShut {NoStop}%
\bibitem [{\citenamefont {Chen}\ and\ \citenamefont
  {Zhang}(2015)}]{Chen_PRL2015}%
  \BibitemOpen
  \bibfield  {author} {\bibinfo {author} {\bibfnamefont {K.}~\bibnamefont
  {Chen}}\ and\ \bibinfo {author} {\bibfnamefont {S.}~\bibnamefont {Zhang}},\
  }\href {\doibase 10.1103/PhysRevLett.114.126602} {\bibfield  {journal}
  {\bibinfo  {journal} {Phys. Rev. Lett.}\ }\textbf {\bibinfo {volume} {114}},\
  \bibinfo {pages} {126602} (\bibinfo {year} {2015})}\BibitemShut {NoStop}%
\bibitem [{\citenamefont {Garello}\ \emph {et~al.}(2013)\citenamefont
  {Garello}, \citenamefont {Miron}, \citenamefont {Avci}, \citenamefont
  {Freimuth}, \citenamefont {Mokrousov}, \citenamefont {Bl{\"u}gel},
  \citenamefont {Auffret}, \citenamefont {Boulle}, \citenamefont {Gaudin},\
  and\ \citenamefont {Gambardella}}]{Garello_nnano_2013}%
  \BibitemOpen
  \bibfield  {author} {\bibinfo {author} {\bibfnamefont {K.}~\bibnamefont
  {Garello}}, \bibinfo {author} {\bibfnamefont {I.~M.}\ \bibnamefont {Miron}},
  \bibinfo {author} {\bibfnamefont {C.~O.}\ \bibnamefont {Avci}}, \bibinfo
  {author} {\bibfnamefont {F.}~\bibnamefont {Freimuth}}, \bibinfo {author}
  {\bibfnamefont {Y.}~\bibnamefont {Mokrousov}}, \bibinfo {author}
  {\bibfnamefont {S.}~\bibnamefont {Bl{\"u}gel}}, \bibinfo {author}
  {\bibfnamefont {S.}~\bibnamefont {Auffret}}, \bibinfo {author} {\bibfnamefont
  {O.}~\bibnamefont {Boulle}}, \bibinfo {author} {\bibfnamefont
  {G.}~\bibnamefont {Gaudin}}, \ and\ \bibinfo {author} {\bibfnamefont
  {P.}~\bibnamefont {Gambardella}},\ }\href {\doibase 10.1038/nnano.2013.145}
  {\bibfield  {journal} {\bibinfo  {journal} {Nature Nanotech.}\ }\textbf
  {\bibinfo {volume} {8}},\ \bibinfo {pages} {587} (\bibinfo {year}
  {2013})}\BibitemShut {NoStop}%
\bibitem [{\citenamefont {Kurebayashi}\ \emph {et~al.}(2014)\citenamefont
  {Kurebayashi}, \citenamefont {Sinova}, \citenamefont {Fang}, \citenamefont
  {Irvine}, \citenamefont {Skinner}, \citenamefont {Wunderlich}, \citenamefont
  {Nov{\'a}k}, \citenamefont {Campion}, \citenamefont {Gallagher},
  \citenamefont {Vehstedt} \emph {et~al.}}]{Hide_nnano_2014}%
  \BibitemOpen
  \bibfield  {author} {\bibinfo {author} {\bibfnamefont {H.}~\bibnamefont
  {Kurebayashi}}, \bibinfo {author} {\bibfnamefont {J.}~\bibnamefont {Sinova}},
  \bibinfo {author} {\bibfnamefont {D.}~\bibnamefont {Fang}}, \bibinfo {author}
  {\bibfnamefont {A.}~\bibnamefont {Irvine}}, \bibinfo {author} {\bibfnamefont
  {T.}~\bibnamefont {Skinner}}, \bibinfo {author} {\bibfnamefont
  {J.}~\bibnamefont {Wunderlich}}, \bibinfo {author} {\bibfnamefont
  {V.}~\bibnamefont {Nov{\'a}k}}, \bibinfo {author} {\bibfnamefont
  {R.}~\bibnamefont {Campion}}, \bibinfo {author} {\bibfnamefont
  {B.}~\bibnamefont {Gallagher}}, \bibinfo {author} {\bibfnamefont
  {E.}~\bibnamefont {Vehstedt}},  \emph {et~al.},\ }\href {\doibase
  10.1038/nnano.2014.15} {\bibfield  {journal} {\bibinfo  {journal} {Nature
  Nanotech.}\ }\textbf {\bibinfo {volume} {9}},\ \bibinfo {pages} {211}
  (\bibinfo {year} {2014})}\BibitemShut {NoStop}%
\bibitem [{\citenamefont {Satchell}\ and\ \citenamefont
  {Birge}(2018)}]{Satchell_PRB2018}%
  \BibitemOpen
  \bibfield  {author} {\bibinfo {author} {\bibfnamefont {N.}~\bibnamefont
  {Satchell}}\ and\ \bibinfo {author} {\bibfnamefont {N.~O.}\ \bibnamefont
  {Birge}},\ }\href {\doibase 10.1103/PhysRevB.97.214509} {\bibfield  {journal}
  {\bibinfo  {journal} {Phys. Rev. B}\ }\textbf {\bibinfo {volume} {97}},\
  \bibinfo {pages} {214509} (\bibinfo {year} {2018})}\BibitemShut {NoStop}%
\end{thebibliography}%
\end{document}